\documentclass[fleqn,usenatbib]{mnras}

\usepackage[T1]{fontenc}
\usepackage{ae,aecompl}

\usepackage{graphicx}
\usepackage{amssymb}
\usepackage{amsmath}
\usepackage{caption}
\usepackage{subcaption}
\usepackage{epstopdf}

\title[ALMA observations of exocometary CO lines in $\beta$ Pic]{Exocometary gas structure, origin and physical properties around $\beta$ Pictoris through ALMA CO multi-transition observations}
\author[L. Matr\`a et al.]{L.\ Matr\`a$^{1,2}$\thanks{E-mail:
l.matra@ast.cam.ac.uk}, W.\ R.\ F.\ Dent$^{3}$, M.\ C.\ Wyatt$^{1}$, Q. Kral$^{1}$, D. J. Wilner$^{4}$, O. Pani\'c$^{1}$, \newauthor A. M. Hughes$^{5}$, I. de Gregorio-Monsalvo$^{3}$, A. Hales$^{3}$, J.-C. Augereau$^{6,7}$, \newauthor  J. Greaves$^{8}$, A. Roberge$^{9}$ 
\\
$^{1}$Institute of Astronomy, University of Cambridge, Madingley Road, Cambridge CB3 0HA, UK \\
$^{2}$European Southern Observatory, Alonso de C\'ordova 3107, Vitacura, Santiago, Chile \\
$^{3}$ALMA SCO, Alonso de C\'ordova 3107, Vitacura, Santiago, Chile \\
$^{4}$Harvard-Smithsonian Center for Astrophysics, 60 Garden Street, Cambridge, MA 02138, USA \\
$^{5}$Department of Astronomy, Van Vleck Observatory, Wesleyan University, 96 Foss Hill Dr., Middletown, CT 06459, USA \\
$^{6}$Univ. Grenoble Alpes, Institut de Plan\'etologie et d'Astrophysique de Grenoble (IPAG, UMR 5274), 38000 Grenoble, France \\
$^{7}$CNRS, Institut de Plan\'etologie et d'Astrophysique de Grenoble (IPAG, UMR 5274), 38000 Grenoble, France \\
$^{8}$School of Physics \& Astronomy, Cardiff University, 4 The Parade, Cardiff CF24 3AA, UK \\
$^{9}$Exoplanets and Stellar Astrophysics Laboratory, NASA Goddard Space Flight Center, Greenbelt, MD, USA
}

\date{Accepted 2016 September 20. Received 2016 September 01; in original form 2016 July 18.}
\pubyear{2016}

\begin{document}
\label{firstpage}
\pagerange{\pageref{firstpage}--\pageref{lastpage}}
\maketitle

\begin{abstract}
Recent ALMA observations unveiled the structure of CO gas in the 23 Myr-old $\beta$ Pictoris planetary system, a component that has been discovered in many similarly young debris disks.
We here present ALMA CO J=2-1 observations, at an improved spectro-spatial resolution and sensitivity compared to previous CO J=3-2 observations. We find that 1) the CO clump is radially broad, favouring the resonant migration over the giant impact scenario for its dynamical origin, 2) the CO disk is vertically tilted compared to the main dust disk, at an angle consistent with the scattered light warp. We then use position-velocity diagrams to trace Keplerian radii in the orbital plane of the disk. Assuming a perfectly edge-on geometry, this shows a CO scale height increasing with radius as $R^{0.75}$, and an electron density (derived from CO line ratios through NLTE analysis) in agreement with thermodynamical models.
Furthermore, we show how observations of optically thin line ratios can solve the primordial versus secondary origin dichotomy in gas-bearing debris disks. As shown for $\beta$ Pictoris, subthermal (NLTE) CO excitation is symptomatic of H$_2$ densities that are insufficient to shield CO from photodissociation over the system's lifetime. 
This means that replenishment from exocometary volatiles must be taking place, proving the secondary origin of the disk.
In this scenario, assuming steady state production/destruction of CO gas, we derive the CO+CO$_2$ ice abundance by mass in $\beta$ Pic's exocomets to be at most $\sim$6\%, consistent with comets in our own Solar System and in the coeval HD181327 system.

\end{abstract}


\section{Introduction}
The circumstellar environment of $\beta$ Pictoris, a nearby \citep[$19.44\pm0.05$ pc,][]{vanLeeuwen2007}, young \citep[$23\pm3$ Myr,][]{Mamajek2014} main sequence A6 star \citep{Gray2006}, has been a continuous source of discoveries since more than three decades ago, when first circumstellar gas through optical absorption lines \citep{Slettebak1975} and then dust through an infrared excess from IRAS \citep{Aumann1985} were discovered. Its edge-on geometry was unveiled soon after through the first resolved image of a circumstellar dust disk \citep{Smith1984}, and explained the observed circumstellar gas absorption lines \citep{Hobbs1985}. Later, it was understood that to maintain the low levels of dust present in many nearby main sequence stars such as $\beta$ Pictoris, a replenishment mechanism is needed \citep{Backman1993}. This need for replenishment of second generation dust sets the physical definition of a \textit{debris disk}, and represents the fundamental difference with primordial protoplanetary disks, where the presence of large amounts of gas means the dust does not need replenishment \citep[e.g.][]{Wyatt2015}. 

Considered one of the archetypes of debris disks, the presence of gas has given $\beta$ Pictoris particular attention. The observed atomic absorption lines were seen to have both a stable component, at a radial velocity similar to that of the star, and a variable component, seen at different radial velocities \citep[e.g.][]{Slettebak1983, Kondo1985, Lagrange1987}. The latter feature was attributed to evaporating cometary bodies approaching the star on eccentric orbits \citep{Ferlet1987, Beust1990, Kiefer2014}. In addition, \textit{Hubble Space Telescope} (HST) observations showed that the gas composition is rather peculiar, with an extreme overabundance of carbon compared to other elements. This carbon acts as a braking agent and  explains how the observed metallic gas levels can be maintained through braking against stellar radiation pressure \citep{Roberge2006, Fernandez2006}. Though generally absorption studies against the stellar continuum are the most sensitive, they are limited in that they only probe the gas column along the line of sight to the star. Recent studies have therefore been focusing on emission lines, which on the other hand can be used to trace the overall morphology of the gas disk. Firstly resolved observations of metallic atoms in the optical/UV \citep{Olofsson2001, Brandeker2004, Nilsson2012}, and subsequently observations of ionised carbon and oxygen in the far-IR \citep{Cataldi2014, Kral2016a} showed that the bulk of the atomic gas does not originate from the infalling cometary bodies at a few stellar radii, but from a more extended gas disk in Keplerian rotation at several tens of AU around the central star.

Molecular gas has been more difficult to detect, with only upper limits on H$_2$ and OH \citep{Martin-Zaidi2008, Vidal-Madjar1994}. The presence of CO gas along the line of sight to the star was first marginally detected in absorption by the \textit{International Ultraviolet Explorer} \citep{Deleuil1993} and then confirmed through HST observations \citep[e.g.][]{Vidal-Madjar1994, Roberge2000}. However, detection of its rotational transitions at millimetre wavelengths proved impossible with single dish telescopes, despite very long integration times \citep{Liseau1998}. It is only through recent interferometric observations with the Atacama Large Millimeter/sub-millimeter Array (ALMA), bringing a drastic improvement in sensitivity and angular resolution, that detection of the J=3-2 transition at 345 GHz has been made possible \citep{Dent2014}. The data revealed for the first time the spatial distribution of CO, presenting a clump of emission at 85 AU on the SW side of the disk, which is co-located with both a radial peak in the dust millimetre emission and a SW dust clump similarly observed at mid-IR wavelengths \citep{Telesco2005}. This was interpreted as evidence for a common production location for both second-generation debris dust and second-generation CO. The clumpy azimuthal structure was then attributed to enhanced collision rates between icy bodies at specific azimuthal locations, which could be the dynamical evidence of either a giant impact between Mars-sized bodies \citep{Jackson2014} or the migration of a yet unseen planet sweeping icy planetesimals into resonance \citep{Wyatt2003, Wyatt2006}.

We here present new ALMA follow-up observations of the $\beta$ Pictoris system. In this work, we focus on CO J=2-1 emission at 230 GHz, observed at an improved sensitivity and spatial resolution of $\sim5.5$ AU, and analyse it together with archival 345 GHz ALMA observations of the J=3-2 transition. We postpone the analysis of the dust continuum and upper limits on SiO emission as the subject of forthcoming work.
In Section \ref{sect:obs}, we describe the observations, calibration and imaging procedures, whereas in Section \ref{sect:res} we analyse the radial, vertical and azimuthal structure of both CO transitions as well as the ratio between the two. In Section \ref{sect:mod}, we model the resolved line ratios using the non-local thermodynamic equilibrium methods developed in \citet{Matra2015}, showing how they can be used to probe undetectable species in the disk such as electrons and H$_2$, and measure the mass and optical thickness of CO gas. Finally, in Section \ref{sect:disc}, we discuss how our analysis impacts the current understanding of gas in the $\beta$ Pictoris system itself as well as other gas-bearing debris disks.

\section[]{Observations}
\label{sect:obs}
\subsection{ALMA Band 6}
We observed the $\beta$ Pictoris disk with ALMA during its Cycle 2 (project code 2012.1.00142.S) using band 6 receivers. Observations were performed using both the 12-m array and the Atacama Compact Array (ACA). The 12-m observations were carried out in two antenna configurations; for the most compact one, observed in December 2013, a mosaic strategy was used with two pointings at $\pm$5$\arcsec$ from the stellar location along the disk midplane, whereas for the most extended one, observed in August 2015, a single-pointing strategy was adopted. ACA observations were also carried out in single-pointing mode during October 2013.
The on-source times were 28, 114 and 50 minutes respectively for the 12-m (compact and extended) and ACA observations.
The spectral setup of the correlator consisted of four spectral windows; of these, two were centred around 218.5 and 232.5 GHz and set in time division mode (TDM) to achieve maximum bandwidth ($\sim$2 GHz each) for continuum observations. The remaining two were set in frequency division mode (FDM) to target the $^{12}$CO J=2-1 line (at rest frequency 230.538 GHz) and the SiO J=5-4 line (at 217.105 GHz) with a channel spacing of 244.141 and 488.281 kHz, respectively. The corresponding velocity resolutions are 0.64 and 1.35 km/s (after Hanning smoothing). Across all observations, either Uranus, Ganymede or J0519-454 were used as amplitude calibrators, whereas J0519-4546 was used as phase calibrator and either J0538-4405 or J0334-4008 as bandpass calibrators.

The data calibration was carried out using the CASA software version 4.3.0, including appropriate weighting between different observations and/or array configurations. We note that the flux calibration is estimated to be accurate within 10\% (ALMA Cycle 2 Technical Handbook). The calibrated visibilities from the ACA and the 12-m datasets (covering baselines from 9 to 1574 m) were then concatenated, achieving a u-v coverage which gives us sensitivity to structure on scales between 0\farcs3 and 27\arcsec. To image the CO line, we first subtracted the continuum from the combined visibility dataset using the \textit{uvcontsub} task within CASA. Then, we imaged velocity channels within $\pm$10 km/s from the radial velocity of the star \citep[20.0$\pm$0.7 km/s in the heliocentric reference frame,][]{Gontcharov2006} using the CLEAN algorithm \citep{Hogbom1974}. Natural weighting of the visibilities was applied, resulting in a synthesised beam of size 0\farcs30$\times$0\farcs26 and position angle (PA, East of North) of -83\fdg9, corresponding to 5.8$\times$5.1 AU at the known distance to the system.

As one of the main goals of this work is to directly compare this new Band 6 CO J=2-1 dataset with the archival CO J=3-2 Band 7 dataset, we also produced a final CO J=2-1 data cube at the degraded spatial and spectral resolution of the archival J=3-2 dataset; details of the procedure are described in Appendix \ref{app:3}.

\subsection{ALMA Band 7}

In addition to the new Band 6 observations, we retrieved archival ALMA Band 7 CO J=3-2 data obtained within Cycle 0 (project code 2011.0.00087.S), which are described in \citet{Dent2014}. In summary, a two-point mosaic strategy analogous to the Band 6 compact configuration observations was applied, although without use of the ACA (baselines ranging from 16 to 382 m). For consistency, we repeated the data reduction and imaging procedure the same way as done for the Band 6 data, using the same version of CASA. This resulted in a final synthesised beam of size 0\farcs69$\times$0\farcs55 (13.4$\times$10.7 AU) and PA 65\fdg4, with a channel width of 488.281 kHz (corresponding to a 0.85 km/s resolution at 345.796 GHz). As for the Band 6 dataset, the uncertainty in the flux calibration is estimated to be 10\% (ALMA Cycle 0 Technical Handbook).

\begin{figure*}
\begin{subfigure}{0.95\textwidth}
\vspace{-5mm}
 \hspace{-15mm}
  \includegraphics*[scale=1.0]{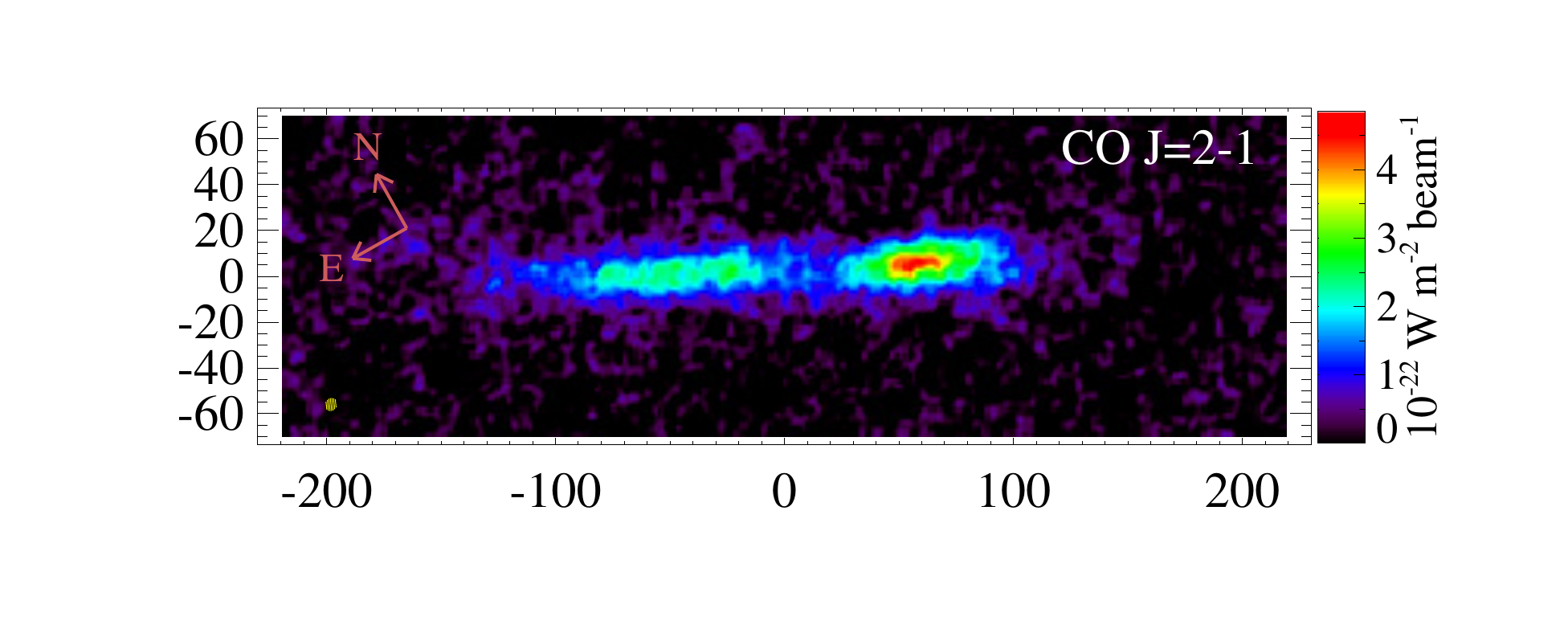}
\vspace{-5mm}
\end{subfigure} \\
\begin{subfigure}{0.95\textwidth}
\vspace{-34mm}
 \hspace{-15mm}
  \includegraphics*[scale=1.0]{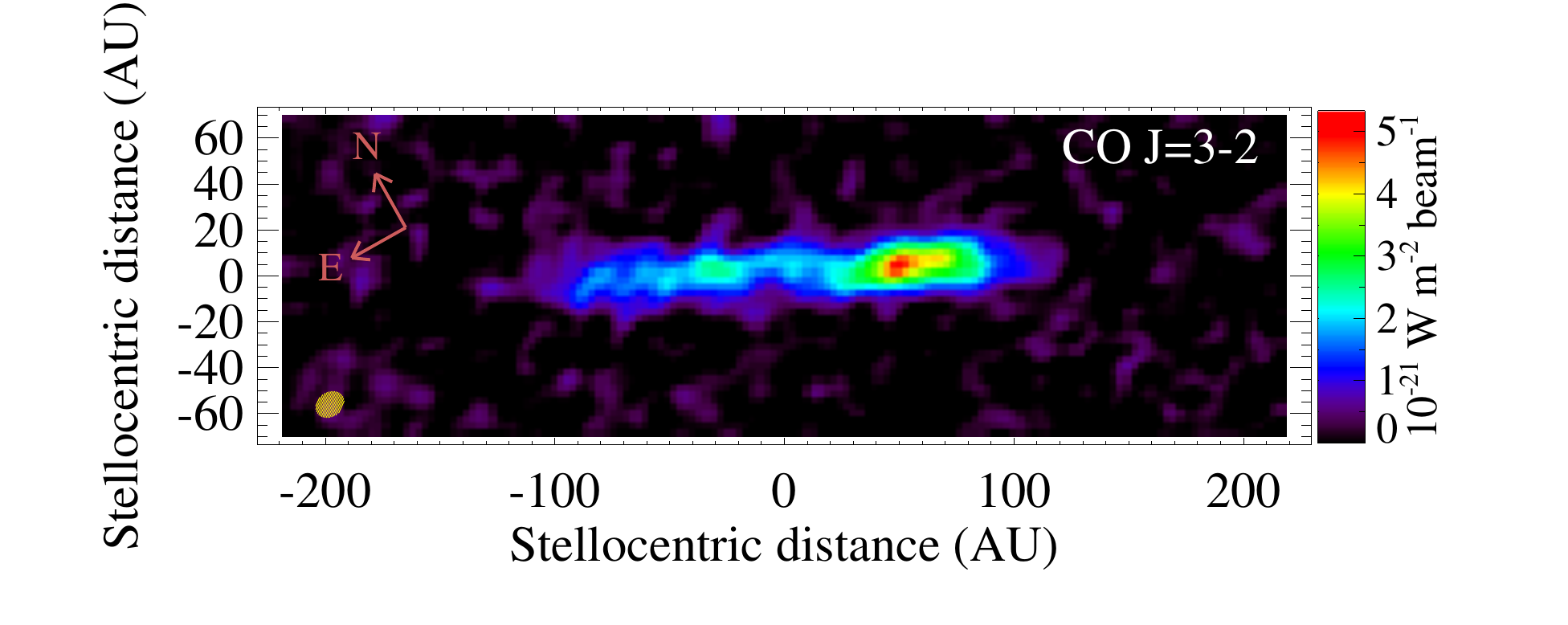}
\vspace{-10mm}
\end{subfigure}
\caption{CO J=2-1 and J=3-2 spectrally integrated (moment-0) CLEAN images of the $\beta$ Pic disk, at their original synthesized resolution (natural weighting). The images have been rotated by the position angle of the main disk observed in scattered light (29.3$^{\circ}$). This way, we can define the $x_{\rm sky}$ axis as the direction along the disk on-sky midplane (positive towards the SW), and the $y_{\rm sky}$ axis as the direction perpendicular to it (positive towards the NW). The origin of the axis is set to the stellar location, and the restoring beam is shown as the yellow ellipse in the bottom left corner.}
\label{fig:mom0}
\end{figure*}

\section[]{Results}
\label{sect:res}

Fig.\ \ref{fig:mom0} shows CO J=2-1 and J=3-2 moment-0 CLEAN images of the $\beta$ Pictoris disk at their original spatial resolution, obtained by spectrally integrating over  heliocentric velocities between 14 and 26 km/s ($\pm 6$ km/s in the reference frame of the star).
The 1-$\sigma$ noise levels on the moment-0 images are 2.7 and 24 $\times$10$^{-23}$ W m$^{-2}$ beam$^{-1}$ (or 3.5 and 21 mJy km/s), respectively, yielding a peak detection of CO integrated line emission at a SNR per beam of 17 and 21 for the 2-1 and 3-2 transitions, respectively.
The total line flux was measured on the primary-beam-corrected moment-0 images by spatially integrating over a region that contains all significant disk emission, but small enough to avoid significant noise contamination. The integrated fluxes measured  are $(3.5\pm0.4) \times10^{-20}$ W m$^{-2}$ for the J=2-1 transition, and $(6.7\pm0.7) \times 10^{-20}$ W m$^{-2}$ for the J=3-2 transition, indicating an integrated 3-2/2-1 line ratio of $1.9\pm0.3$. Absolute errors on integrated line fluxes take into account both the noise level in the moment-0 images and the flux calibration uncertainty; we then summed the relative errors on both fluxes in quadrature (under the assumption that they are Gaussian in nature) to obtain the uncertainty on the line ratio.

\subsection{Spectrally-integrated radial structure}

\begin{figure}
\begin{subfigure}{0.45\textwidth}
 \hspace{-4mm}
  \includegraphics*[scale=0.27]{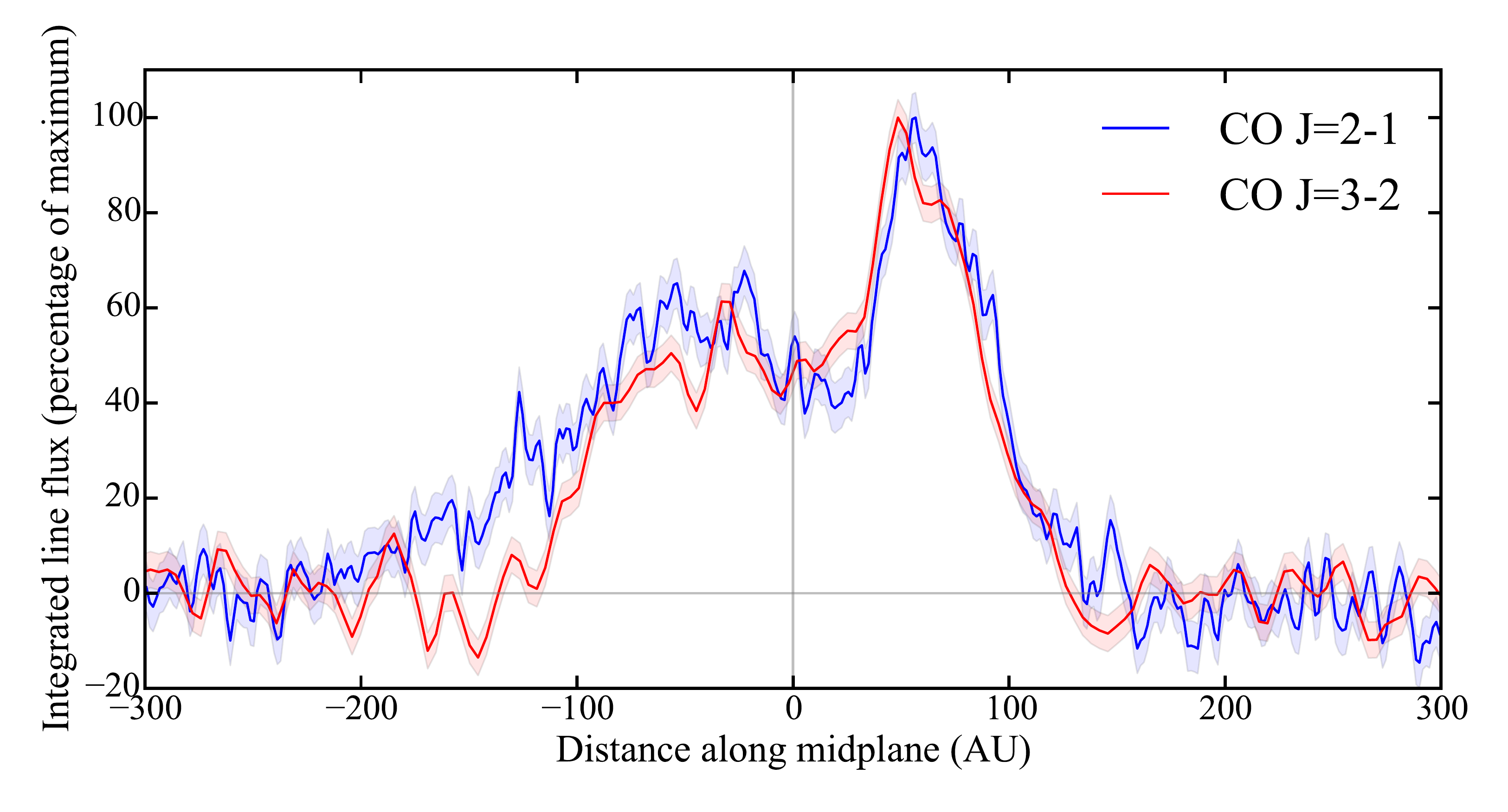}
\vspace{-5mm}
\end{subfigure} 
\caption{Radial distribution of CO line flux obtained from spatially integrating the moment-0 images in Fig.\ \ref{fig:mom0} along the perpendicular to the main disk midplane $y_{\rm sky}$, where $\left|y_{\rm sky}\right|<20$ AU. In order to highlight differences in the projected radial structure between the two datasets, the profile for each image was normalised to its maximum.}
\label{fig:mom0prof}
\end{figure}

In Fig.\ \ref{fig:mom0prof} we present the radial profile of CO emission along the $\beta$ Pic disk midplane ($x_{\rm sky}$ axis in Fig.\ \ref{fig:mom0}). This was derived from the moment-0 images by vertically integrating disk emission from pixels within $\pm$20 AU from the midplane (i.e. with $\left|y_{\rm sky}\right|<20$ AU). 
The J=2-1 dataset confirms the presence of a clump of CO on the SW side of the disk \citep[previously discovered in J=3-2 by][]{Dent2014} at a projected separation of 60 AU from the star, and faint J=2-1 emission detected out to larger distances on the NE side than J=3-2 emission, which will become even clearer when looking at the CO velocity structure (Sect. \ref{sect:velstruct}). This is attributable to both the lower sensitivity of the Band 7 dataset, but also to an intrinsic decrease in the J=3-2/J=2-1 line ratio at increasing disk radii.

\subsection{Spectrally-integrated vertical structure}
\label{vertstruct}

\begin{figure*}
\begin{subfigure}{0.45\textwidth}
\vspace{-3mm}
 \hspace{-9mm}
  \includegraphics[scale=0.44]{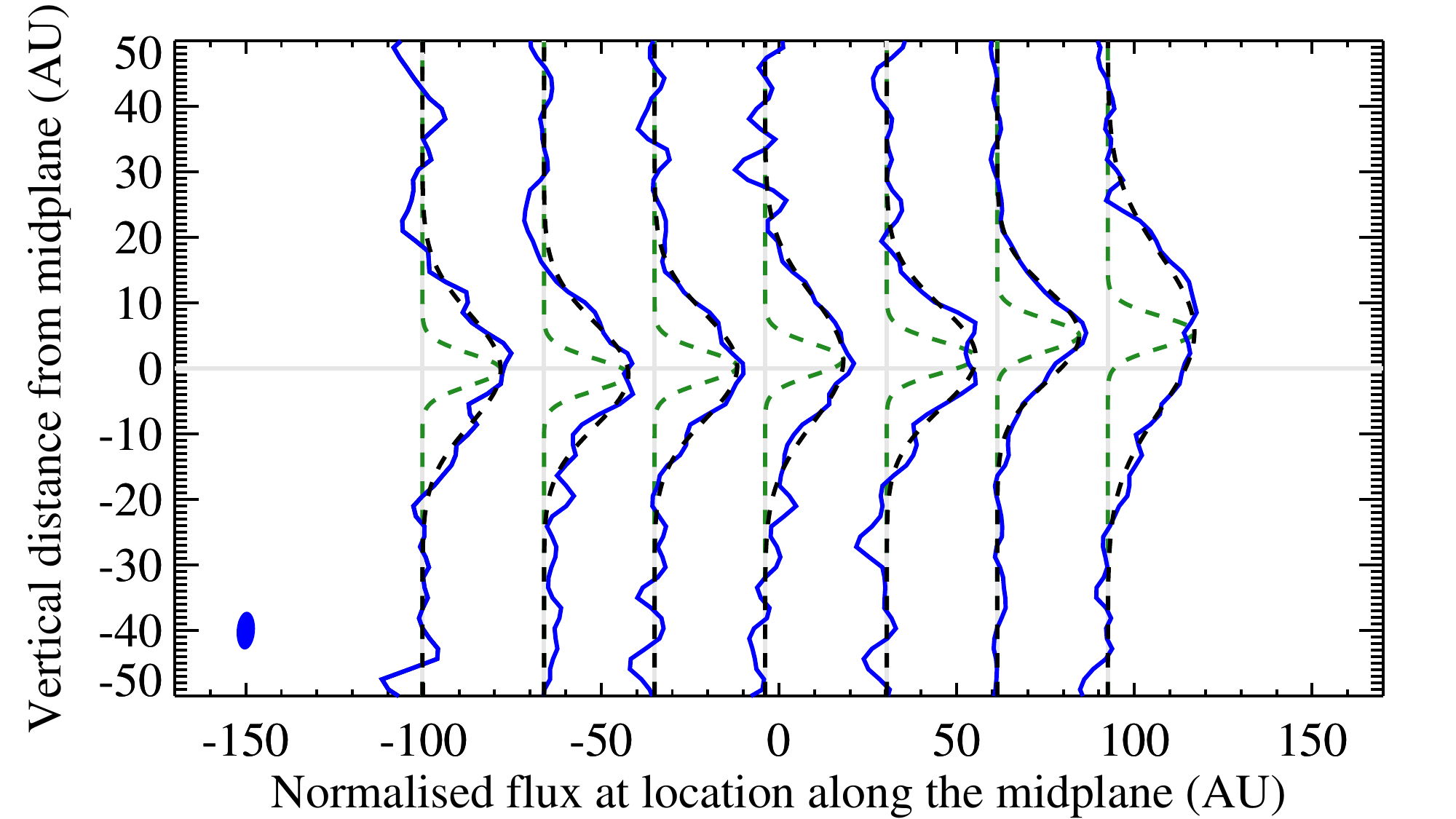}
\vspace{-2mm}
\end{subfigure}
\begin{subfigure}{0.45\textwidth}
\vspace{-3mm}
 \hspace{-2mm}
  \includegraphics[scale=0.44]{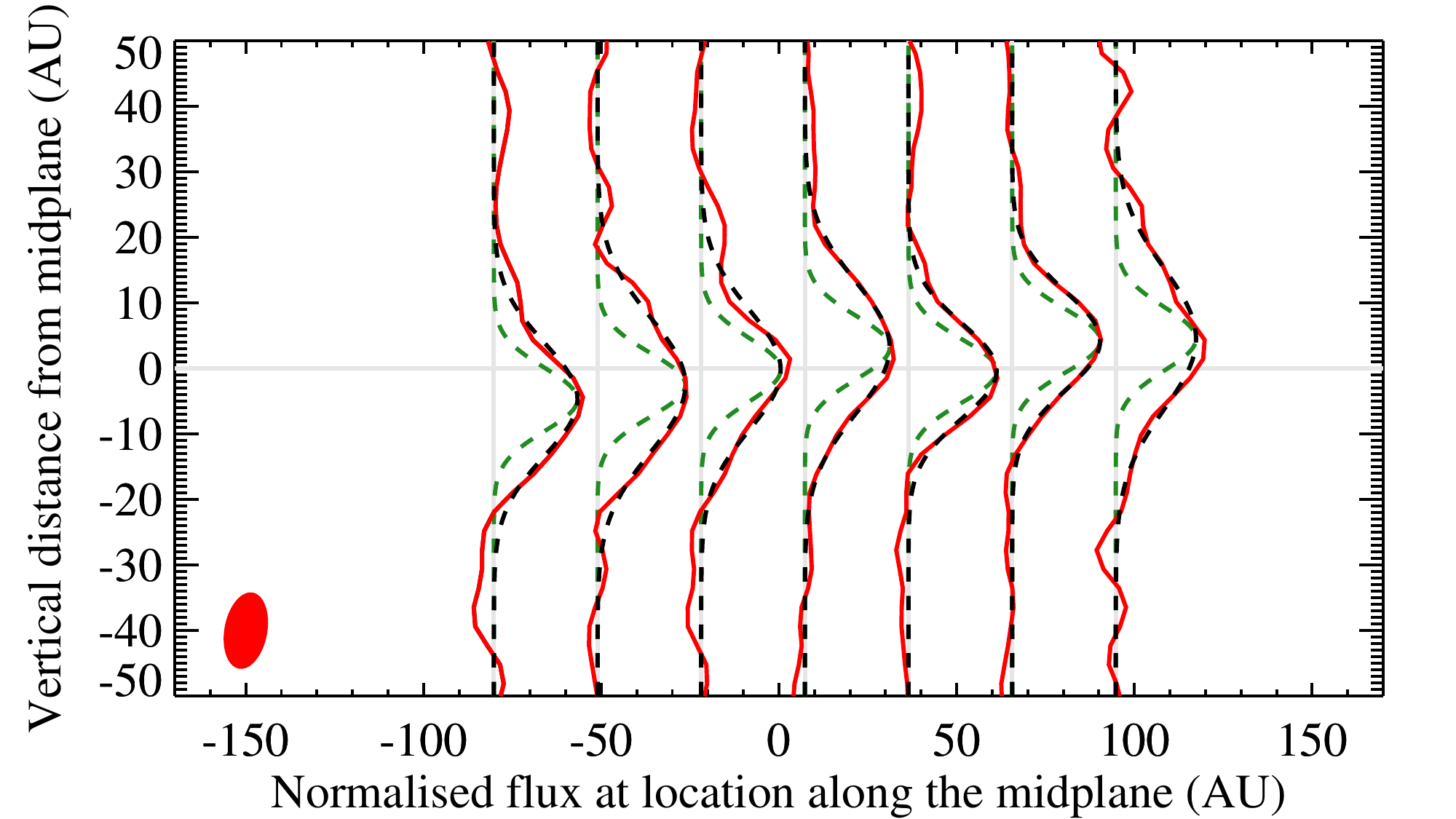}
\vspace{-2mm}
\end{subfigure} \\
\begin{subfigure}{0.45\textwidth}
\vspace{-3mm}
 \hspace{-7mm}
  \includegraphics[scale=0.43]{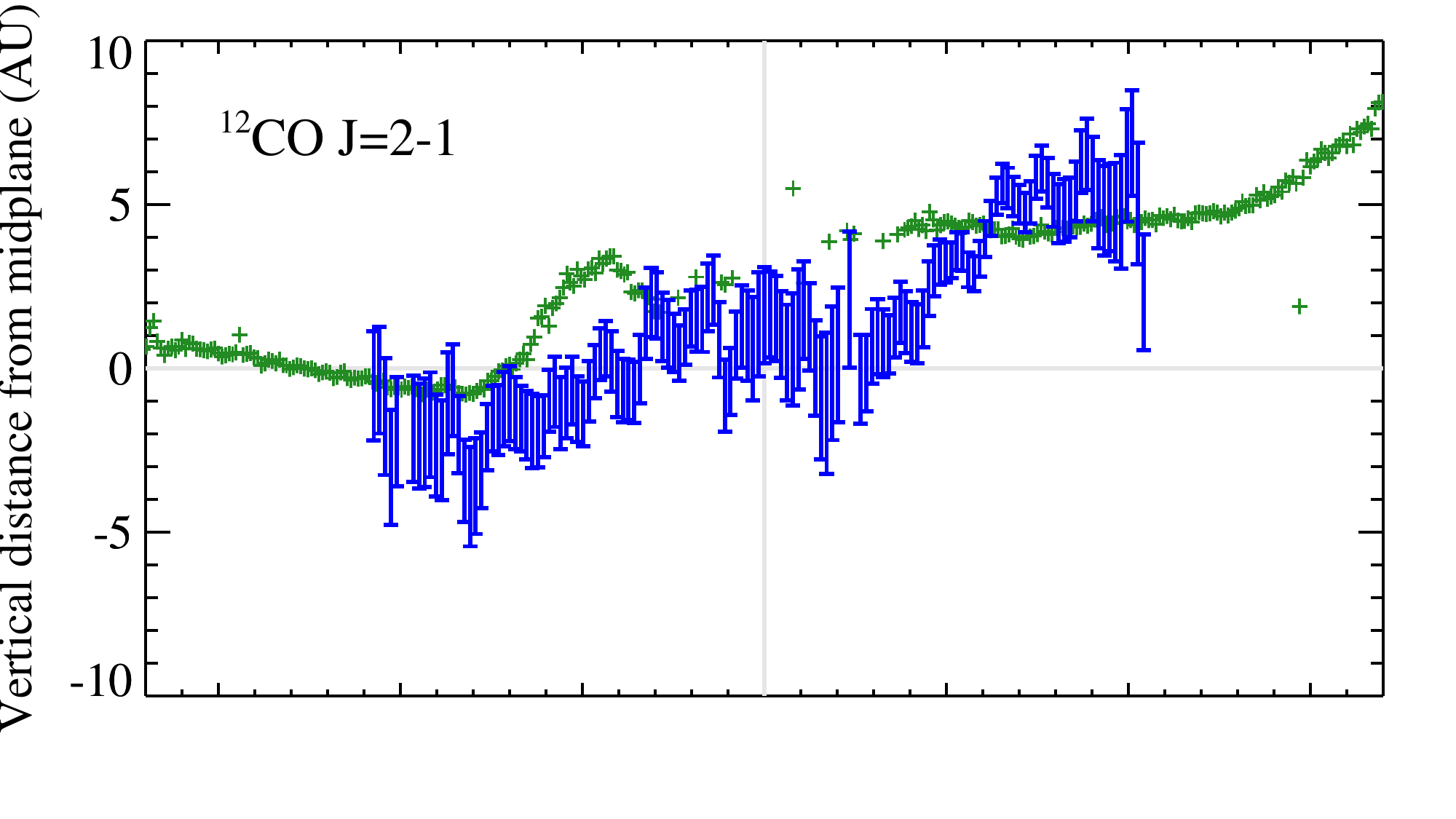}
\vspace{-2mm}
\end{subfigure} 
\begin{subfigure}{0.45\textwidth}
\vspace{-3mm}
 \hspace{0mm}
  \includegraphics[scale=0.43]{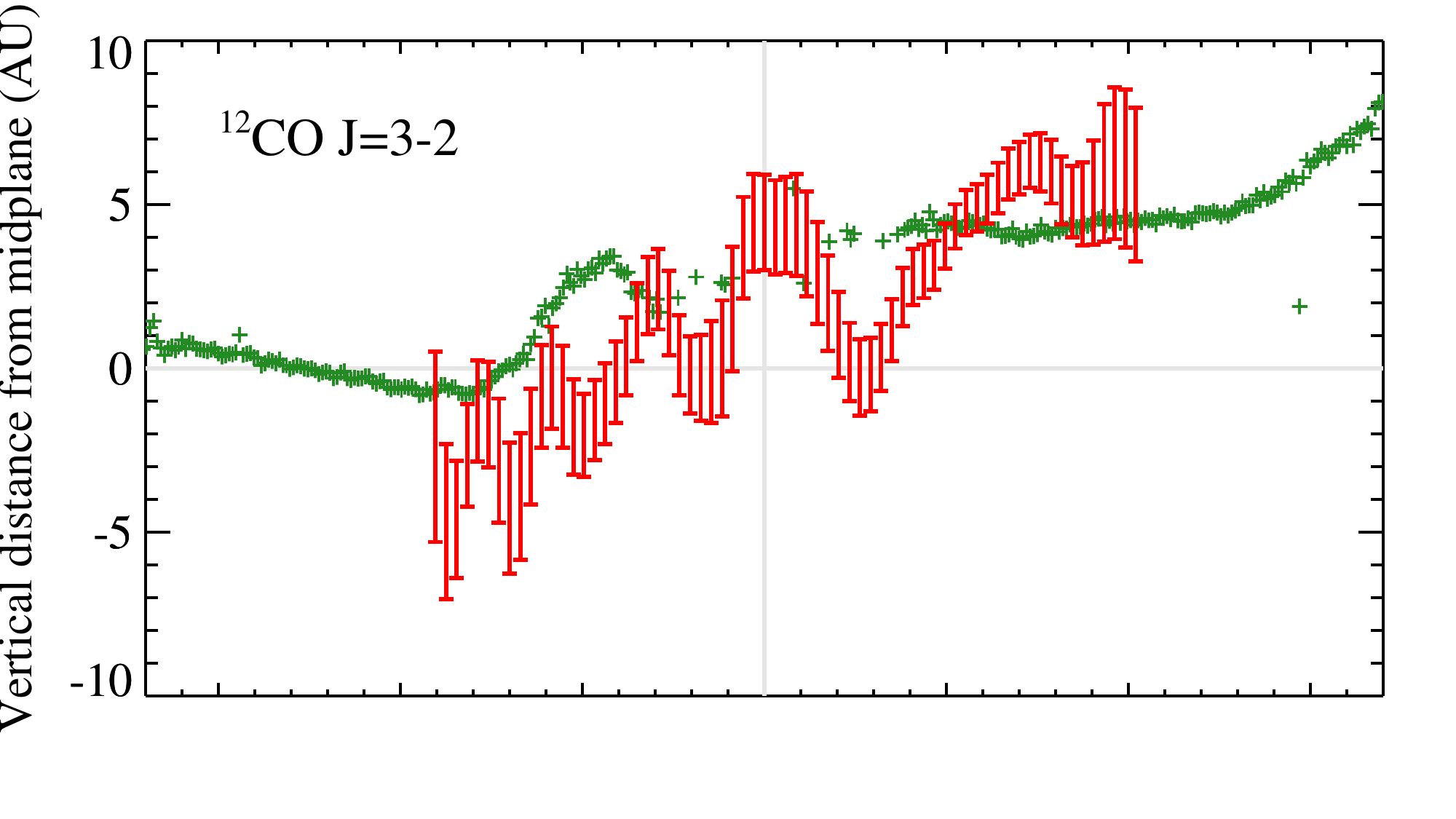}
\vspace{-2mm}
\end{subfigure} \\
\begin{subfigure}{0.45\textwidth}
\vspace{-10mm}
 \hspace{-7mm}
  \includegraphics*[scale=0.43]{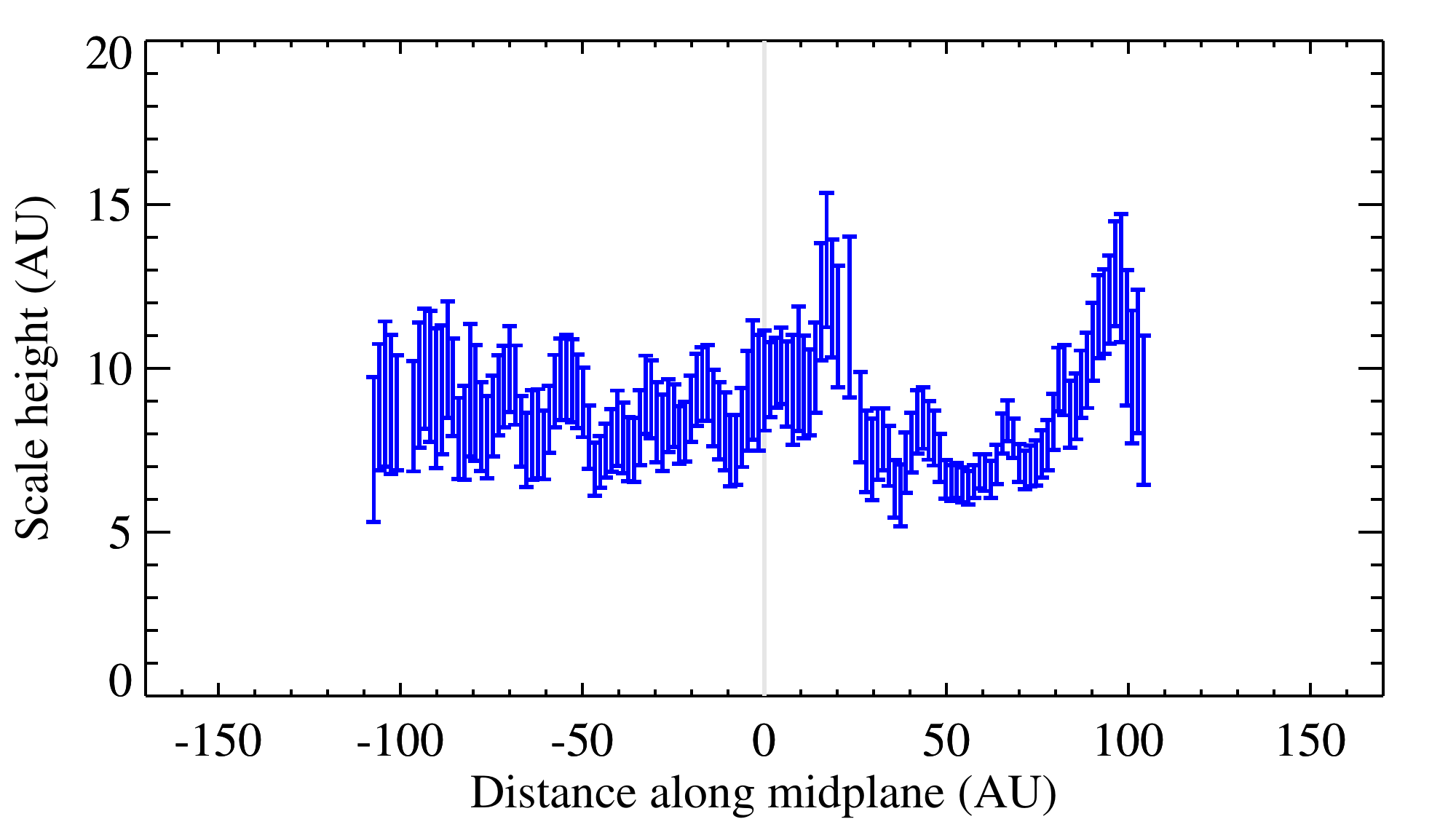}
\vspace{-2mm}
\end{subfigure} 
\begin{subfigure}{0.45\textwidth}
\vspace{-10mm}
 \hspace{0mm}
  \includegraphics*[scale=0.43]{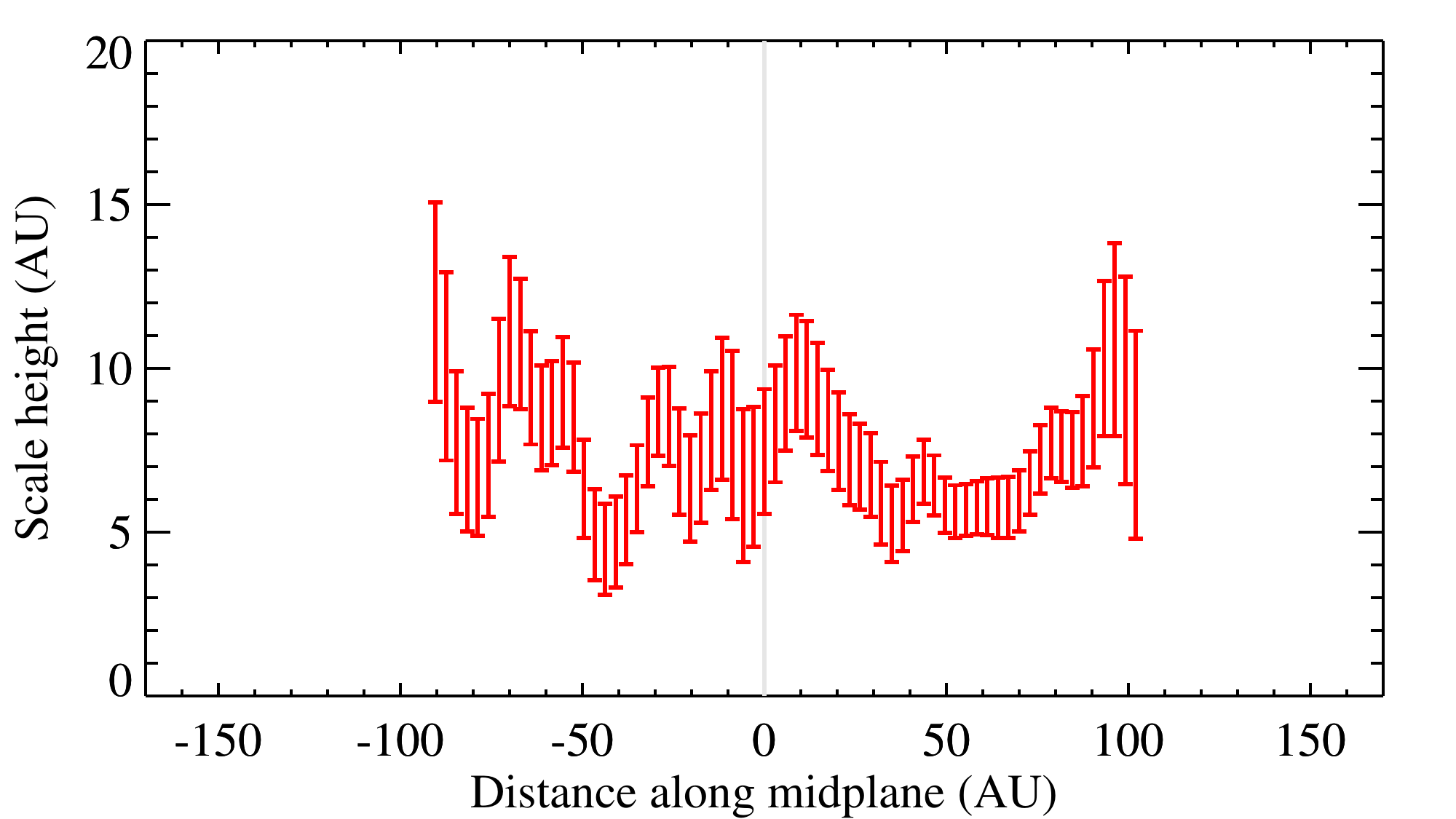}
\vspace{-2mm}
\end{subfigure}
\caption{Vertical structure of the CO disk for the J=2-1 line (blue, left column) and the J=3-2 line (red, right column), from the moment-0 images (Fig. \ref{fig:mom0}).
\textit{Top:} Fitting procedure to the disk's vertical emission profiles as a function of distance along the midplane. Coloured lines represent the measured flux as a function of distance along the vertical ($y_{\rm sky}$) axis for different midplane locations $x_{\rm sky}$. The black dashed lines are the best-fits derived, whereas the green dashed lines (as well as the filled ellipses) represent the instrumental resolution, centred on the best-fit Gaussian peaks $y_{\rm obs}$. \textit{Centre:} Disk spine, i.e. the vertical offset of Gaussian fits $y_{\rm obs}$ from the top panel as a function of midplane location $x_{\rm sky}$. The green crosses represent the same spine derived for the dust emission from HST observations \citep{Apai2015}. \textit{Bottom:} Scale height $H$ as a function of distance along the midplane, derived from the standard deviation $\sigma_{\rm obs}$ of best-fit Gaussians in the top panel.}
\label{fig:vertprofs}
\end{figure*}

In order to probe the vertical structure in the disk, we analyse its \textit{spine}, which we define as the locus of the centres of best-fit vertical Gaussians measured at different locations along the disk midplane. We remind the reader that the disk images shown in Fig.\ \ref{fig:mom0} have been rotated by the position angle (PA) of the main disk observed in scattered light, corresponding to 29\fdg3$^{+0.2}_{-0.3}$ \citep{Lagrange2012}. In scattered light, this differs from the PA of the inner part of the disk, which appears tilted when compared to the main disk observed in the outer regions \citep[e.g.][]{Apai2015}. 

Fig.\ \ref{fig:vertprofs} (top row) illustrates the procedure employed. We apply vertical cuts in the $y_{\rm sky}$ direction for each location $x_{\rm sky}$ along the midplane where emission greater than 5$\sigma$ is detected. The measured (normalised) flux versus $y_{\rm sky}$ is shown as red and blue solid lines. The green dashed Gaussian profiles have width equal to the FWHM of the restoring beam projected onto the $y_{\rm sky}$ direction (11.4 and 5.7 AU for J=3-2 and J=2-1, respectively), and indicate that the disk is resolved vertically for both CO lines. We then employ 1D Gaussians to fit the vertical profiles and obtain the observed best-fit vertical location of the Gaussian peak, which we define as $y_{\rm obs}$ (see fitting procedure and error determination in Appendix \ref{app:2}). Repeating the process at several midplane locations yields the locus of vertical Gaussian centres $y_{\rm obs}$ at different $x_{\rm sky}$ locations, i.e. the disk spine. This appears significantly tilted with respect to the PA of the main disk midplane, presenting an extra anticlockwise rotation \citep[as was noted by][]{Dent2014} by a tilt angle dPA. The latter is similar to the tilt angle observed for the scattered light inner disk \citep[$\sim$4$^{\circ}$,][]{Apai2015}, and is most pronounced at the location of the clump. 

Accurate measurement of the true warp angle from this sky-projected tilt is challenging. To begin with, the disk spines derived from both CO and scattered light are not well represented by a straight line, meaning that measurement of the sky-projected tilt itself is highly sensitive to the disk radii between which the straight line is drawn. Secondly, even assuming a perfectly edge-on disk, there is no reason to believe that the warp axis lies in the plane of the sky; any azimuthal displacement of the warp axis in the orbital plane would cause a substantial difference between the true warp angle and the observed projected sky tilt. To further investigate any potential azimuthal dependence on this tilt, we repeat the same analysis as a function of radial velocity in Sect. \ref{sect: structedgeon}.

As well as the vertical displacement from the midplane, the same Gaussian fits on the moment-0 images yield vertical standard deviations $\sigma_{\rm obs}$ as a function of midplane location. Assuming that the disk is edge-on and that its true vertical structure can be represented by a Gaussian (of standard deviation defined as the scale height $H$), we can derive $H$ by simply deconvolving $\sigma_{\rm obs}$ by the observed instrumental resolution projected along the vertical axis, i.e. $H=\sqrt{\sigma_{\rm obs}^2 -\sigma_{\rm res}^2}$. 
The resulting scale height as a function of midplane location is shown in the bottom row of Fig. \ref{fig:vertprofs}. Given the near edge-on geometry of the disk, its rather flat appearance is unsurprising, as the scale height observed on-sky at any midplane location will tend to trace the scale height at the disk's orbital outer radius. As such, in Sect. \ref{sect: structedgeon} we add the velocity information from the data cube to retrieve the scale height dependence on the orbital radius and explore its relation to the disk temperature.

\subsection{Velocity information}
\label{sect:velstruct}

In order to understand the 3D gas disk kinematics and extract its vertical and azimuthal structure, we analyse the CO dataset through position-velocity (PV) diagrams, i.e. maps of different quantities as a function of both position along the disk midplane ($x_{\rm sky}$ axis in Fig. \ref{fig:mom0}) and radial velocity $v_{\rm rad}$. In particular, we aim to link the $\left(x_{\rm sky}, v_{\rm rad}\right)$ PV location in these diagrams to a $\left(x,y\right)$ location in the orbital plane of the disk. This can be done for a given inclination $i$ if we assume that the gas disk is infinitely thin vertically and in Keplerian rotation (see Appendix \ref{app:1} for details). In Sect. \ref{sect: structedgeon}, we begin by analysing the sky-projected vertical structure along $y_{\rm sky}$ in PV space to derive constraints on the disk vertical and azimuthal structure, assuming a perfectly edge-on disk ($i=90^{\circ}$). In Sect. \ref{sect: structnonedgeon}, on the other hand, we interpret the same sky-projected vertical structure along $y_{\rm sky}$ in PV space purely as azimuthal structure for a disk close to, but not perfectly edge-on ($i<90^{\circ}$). Finally, in Sect. \ref{sect:reslinerat} we carry out a PV diagram comparison between the new CO J=2-1 and the archival J=3-2 dataset.

\subsubsection{CO 3D structure in the edge-on assumption}
\label{sect: structedgeon}
\begin{figure}
\begin{subfigure}{0.47\textwidth}
\vspace{-8mm}
 \hspace{-15mm}
  \includegraphics*[scale=0.56]{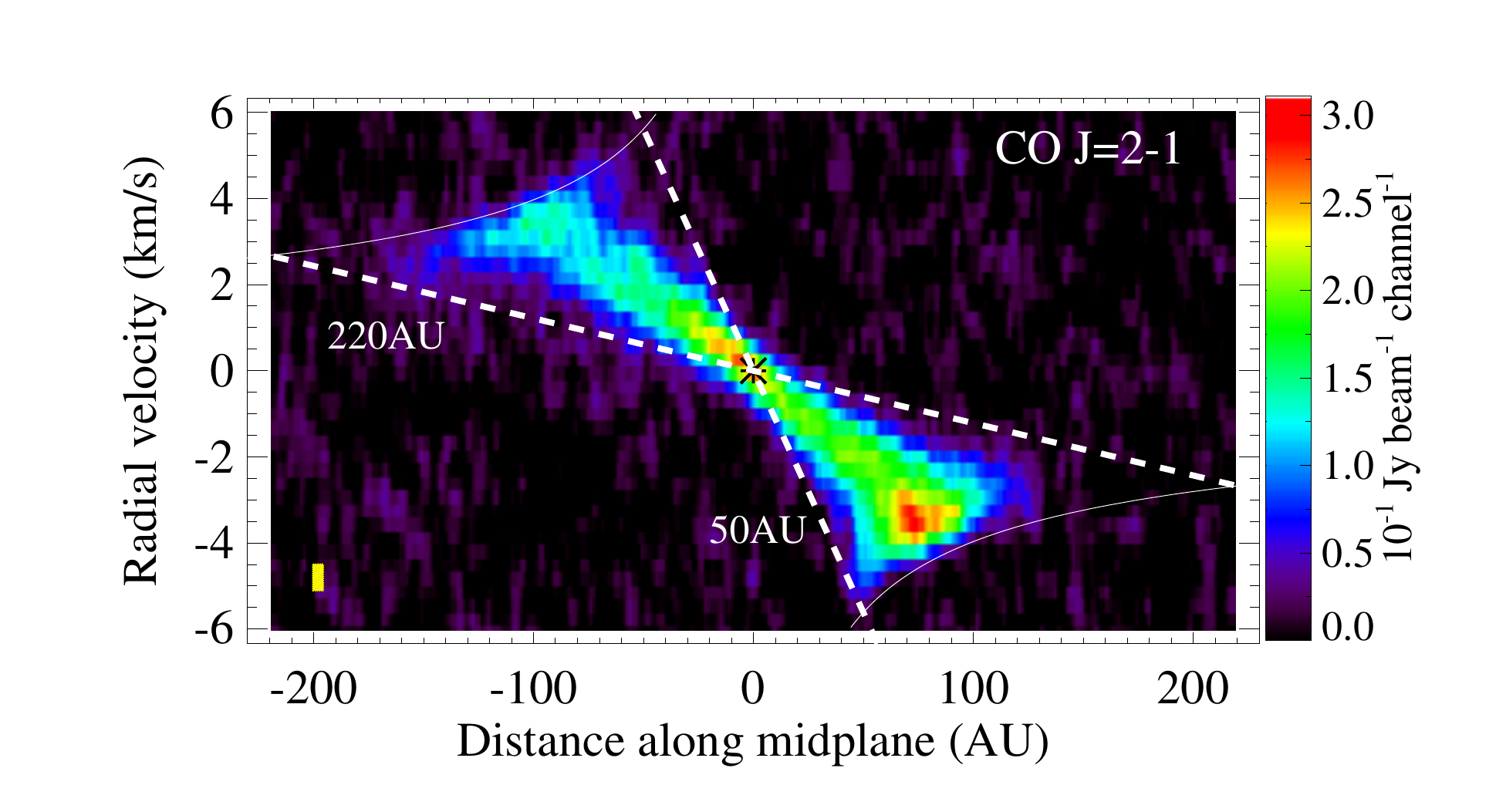}
\vspace{0mm}
\end{subfigure} \\
\begin{subfigure}{0.47\textwidth}
\vspace{-23.8mm}
 \hspace{-15mm}
  \includegraphics*[scale=0.56]{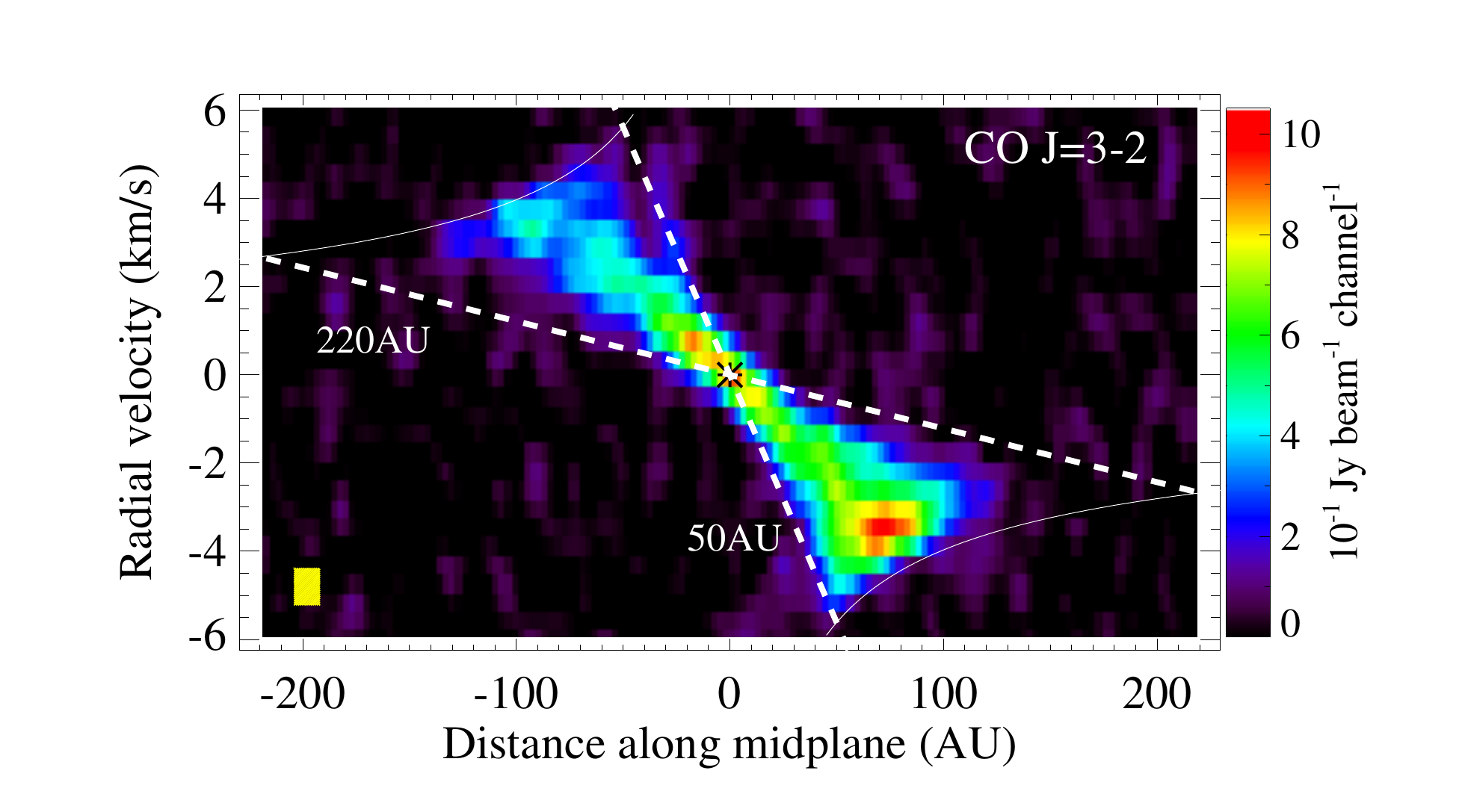}
\vspace{0.0mm}
\end{subfigure}
\vspace{-7mm}
\caption{Position-velocity (PV) diagrams of the $\beta$ Pic disk, showing CO intensity as a function of position along the disk on-sky midplane $x_{\rm sky}$ and radial velocity $v_{\rm rad}$ for each of the two transitions observed. The black asterisk represents the stellar position, while the solid white curves are the maximum radial velocity observable in an edge-on Keplerian disk around a 1.75 M$_{\odot}$ star. The yellow rectangle in the bottom left corner represents the spectro-spatial resolution. The dashed white lines represent different radii $R$ in the orbital plane of the disk, assuming Keplerian rotation (see Appendix \ref{app:1}).}
\label{fig:pvs}
\end{figure}

In Fig.\ \ref{fig:pvs}, we show PV diagrams of CO J=3-2 and J=2-1 line flux obtained by integrating emission vertically (i.e. along the $y_{\rm sky}$ axis in Fig. \ref{fig:mom0}) within 20 AU above and below the disk midplane, including all significant disk emission. 
The spectro-spatial resolution is given by the projection of the restoring beam along the disk midplane $x_{\rm sky}$, combined with the velocity resolution along $v_{\rm rad}$.
The two CO transitions show similar PV structure, consistent with a near edge-on gas disk in Keplerian rotation around a 1.75 M$_{\odot}$ star \citep{Crifo1997}. Diagonal lines on the diagrams represent different radii $R$ in the orbital plane of the disk (Appendix \ref{app:1}), which we here assume to be perfectly edge-on; we can use this to constrain the disk's radial extent in the orbital plane to between $\sim$50 and $\sim$220 AU.  
For both transitions, two flux enhancements are observed; one in the resolved SW clump, approaching us at a velocity $v_{\rm rad}$ between 3 and 4 km/s and sky-projected midplane location $x_{\rm sky}$ between 60 and 90 AU. The other enhancement, less pronounced, is found around the projected stellar location and radial velocity $\left(x_{\rm sky},v_{\rm rad}\right)\sim\left(0,0\right)$. As the latter is not observed towards the star in the moment-0 images, it is likely due to the larger disk volume per velocity channel probed at low compared to higher radial velocities.
Another difference lies in the NE side of the disk, where there is a clear deficit of J=3-2 compared to J=2-1 emission at large radii and low radial velocities (Fig.\ \ref{fig:pvs}). This suggests that J=2-1 emission is detected extending further out in the disk on the NE side compared to J=3-2 emission. 

\begin{figure*}
\begin{subfigure}{0.45\textwidth}
\vspace{-7mm}
 \hspace{-17mm}
  \includegraphics*[scale=0.52]{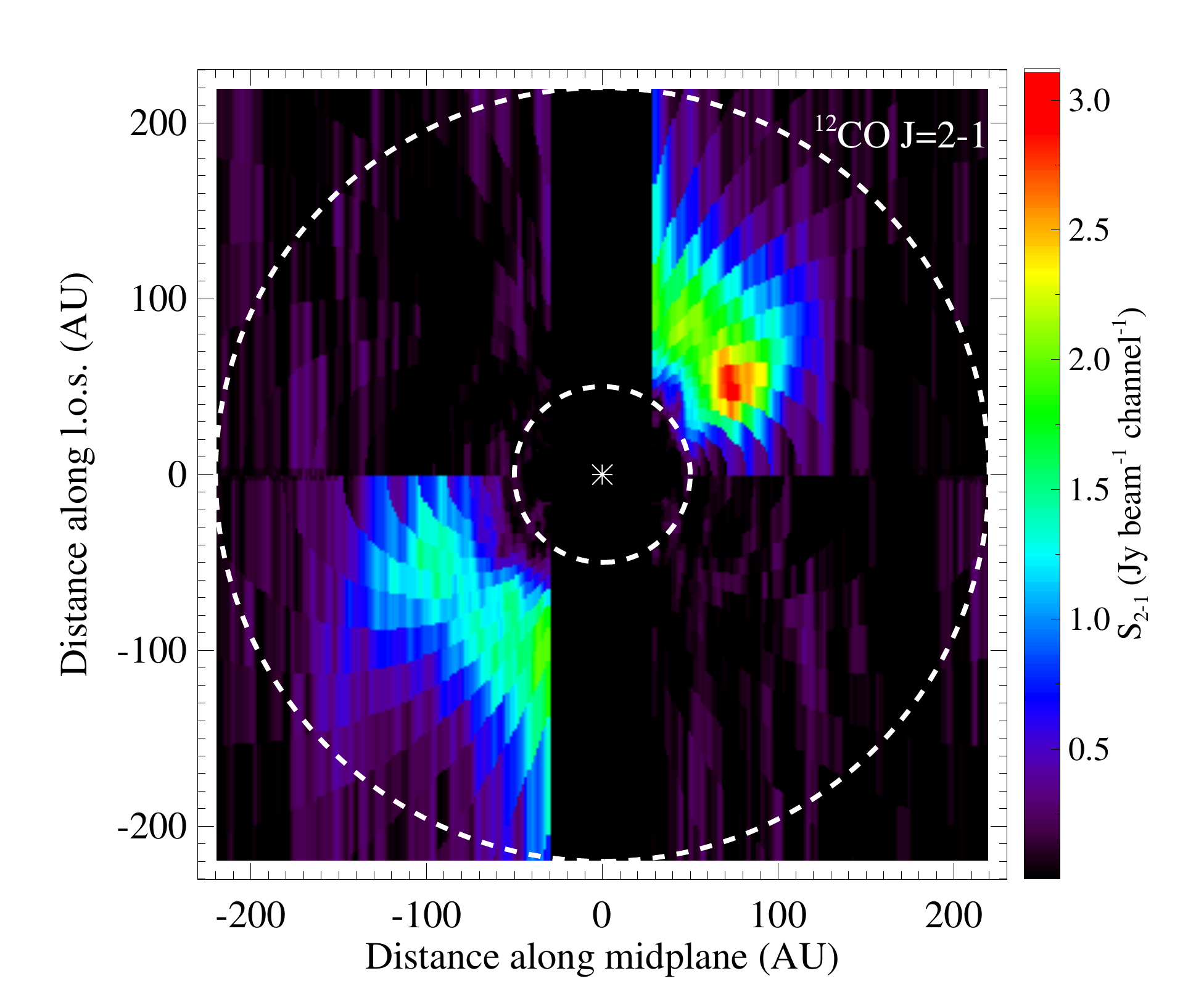}
\vspace{-2mm}
\end{subfigure}
\begin{subfigure}{0.45\textwidth}
\vspace{-7mm}
 \hspace{-3mm}
  \includegraphics*[scale=0.52]{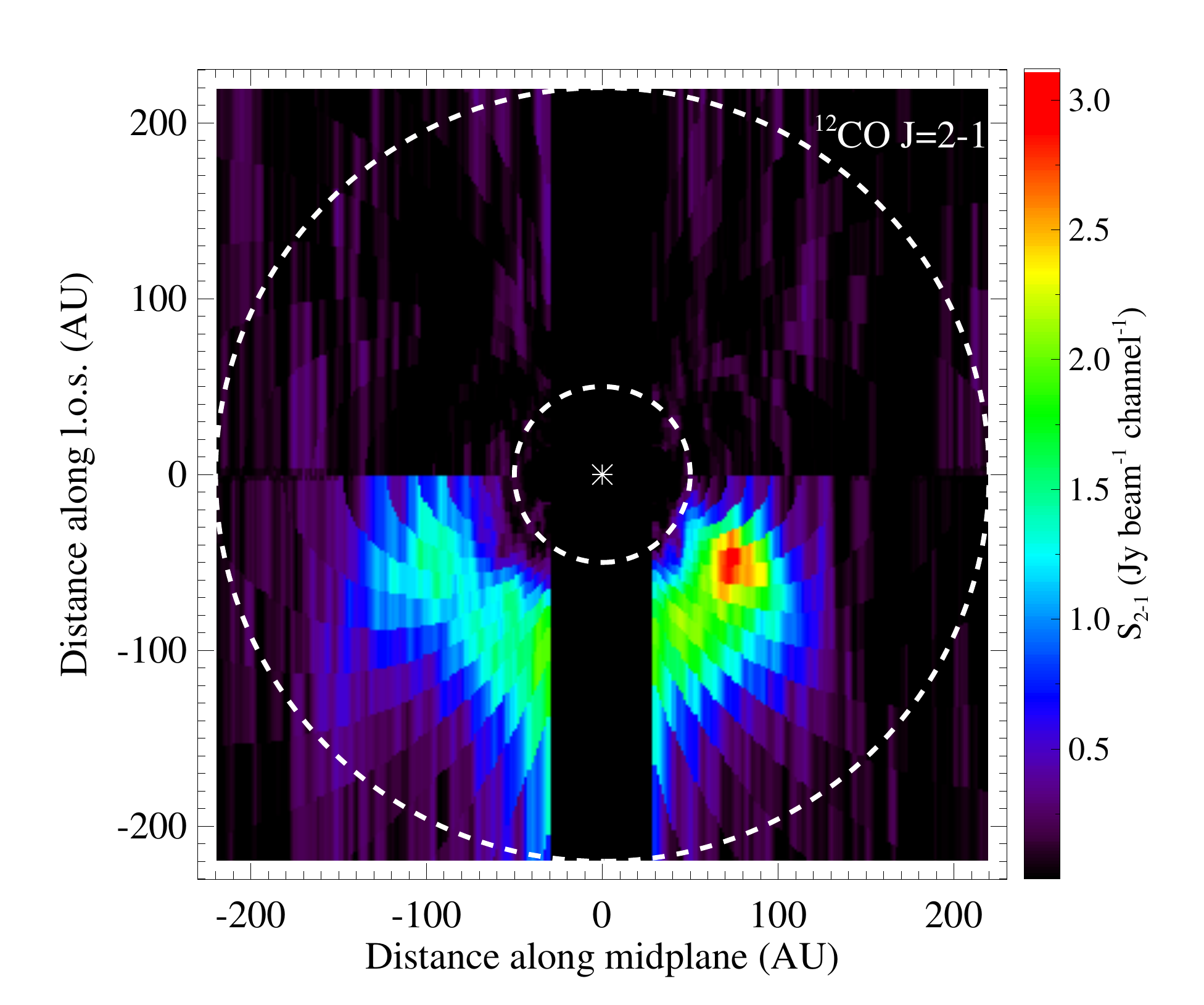}
\vspace{-2mm}
\end{subfigure} \\
\begin{subfigure}{0.45\textwidth}
\vspace{-18.5mm}
 \hspace{-17mm}
  \includegraphics*[scale=0.52]{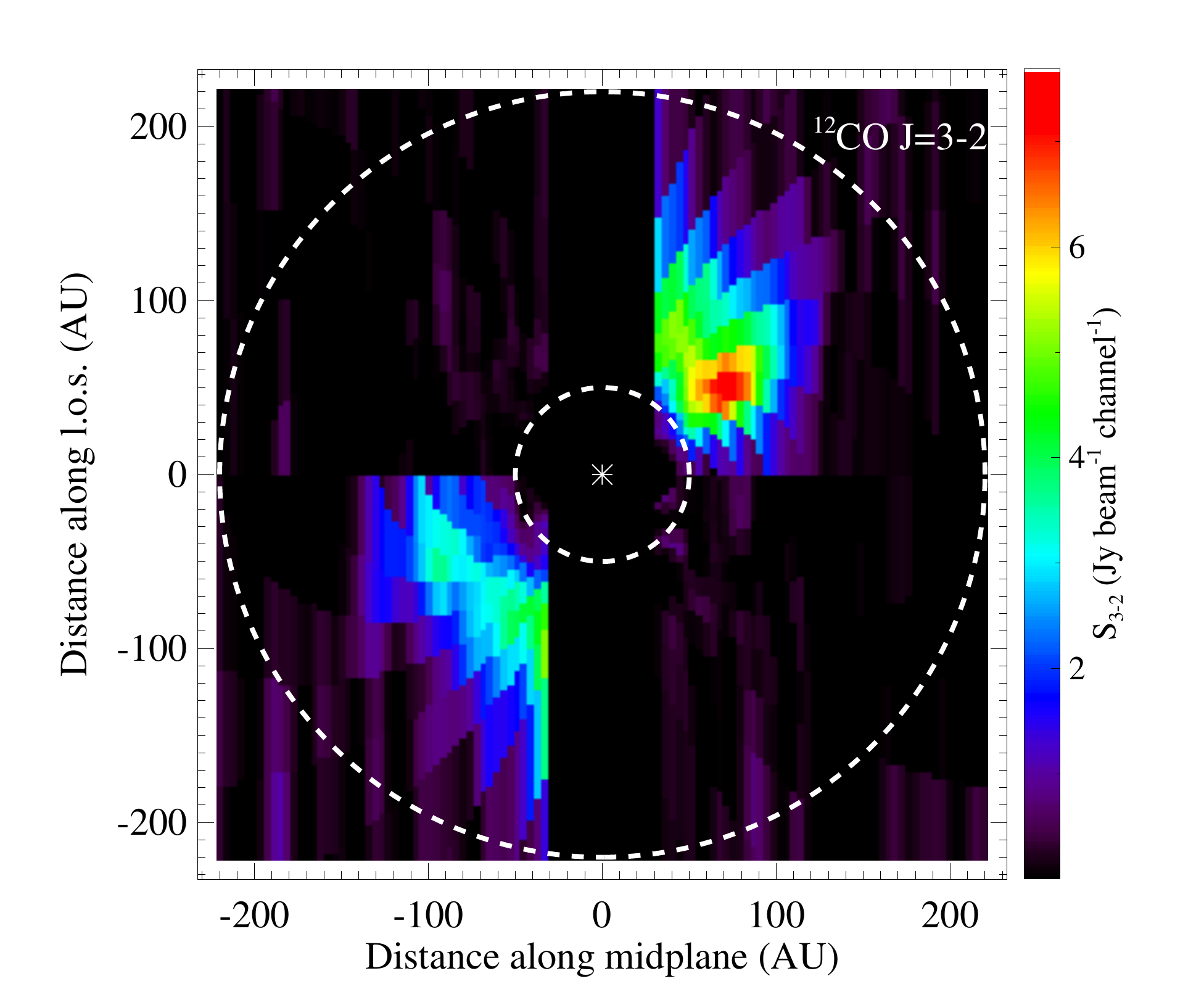}
\vspace{-2mm}
\end{subfigure} 
\begin{subfigure}{0.45\textwidth}
\vspace{-18.5mm}
 \hspace{-3mm}
  \includegraphics*[scale=0.52]{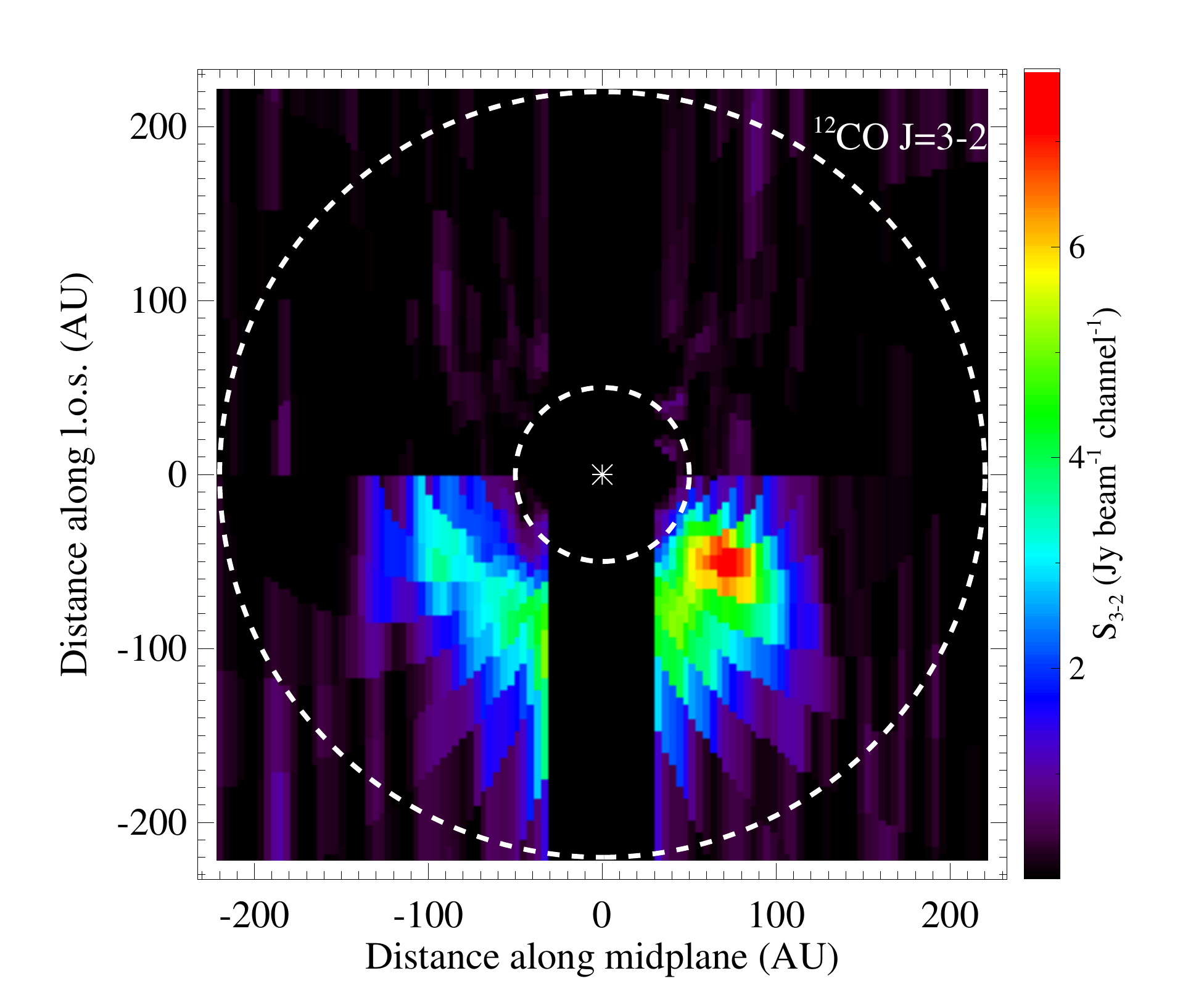}
\vspace{-2mm}
\end{subfigure} 
\vspace{-3mm}
\caption{CO emission in the $\left(x, y\right)$ orbital plane of the disk, derived from the $\left(x_{\rm sky}, v_{\rm rad}\right)$ information in the PV diagrams for the J=2-1 transition (top) and for the J=3-2 transition (bottom), under the assumption of a perfectly edge-on disk. The two columns represent two plausible deprojections of the PV diagrams, justified by the dynamical scenarios proposed in \citet{Dent2014}. The asterisk represents the location of the star, with the inner and outer dashed circles representing deprojected radii of 50 and 220 AU. The 'fishbone' appearance is caused by the finite velocity resolution of the dataset, with any radial velocity tracing a characteristic curve in the displayed $\left(x, y\right)$ space. The central $\pm$30 AU have been masked due to all $y$ locations approaching zero radial velocity along the line of sight to the star, rendering the deprojection highly degenerate.}
\label{fig:deprojectpvs}
\end{figure*}

As mentioned above (again, see Appendix \ref{app:1} for details), for a disk in circular Keplerian rotation, each PV location (i.e. each $\left(x_{\rm sky},v_{\rm rad}\right)$ \textit{spaxel}) in the diagram corresponds to a radius $R$ and hence an $\left(x,\pm y\right)$ location in the orbital plane of the disk. The degeneracy in the sign of $y$, however, means that each spaxel traces two orbital locations, $\left(x, +y\right)$ and $\left(x, -y\right)$. Thus, we do not know how much flux belongs to either of the two locations; we only know that the sum of the two fluxes must equal that of the original spaxel. 
This leads to an infinite number of possible deprojections; Fig.\ \ref{fig:deprojectpvs} shows the two that are most likely physically plausible, as justified by dynamical models \citep[see Sect. \ref{sect:rescl} and][]{Dent2014}. We obtained these by placing \textit{all} of the spaxel emission in the NE side (negative $x_{\rm sky}$) at $-y$ (i.e. in front of the sky plane) and that in the SW (positive $x_{\rm sky}$) at either $+y$ (i.e. behind the sky plane, left column in Fig.\ \ref{fig:deprojectpvs}) or $-y$ (right column in Fig.\ \ref{fig:deprojectpvs}). The direction of rotation is known to be clockwise from the sign of the observed radial velocity. The central $\pm$30 AU from the star are masked as our spectro-spatial resolution is not sufficient to determine orbital locations accurately.
These deprojected images of CO J=3-2 and J=2-1 line emission reveal the azimuthal structure of the disk; on the SW side, the clump peaks at $\phi\sim\pm32^{\circ}$ (measured from the positive $x_{\rm sky}$ direction), with a tail of emission extending to either the front or the back of the star along the line of sight direction. Depending on the system configuration, emission on the NE side can be interpreted either as the continuation of the tail originating from the SW clump (right column), or as a separate dimmer clump at $\phi\sim+212^{\circ}$ with its own tail also extending into the line of sight to the star (left column).

\begin{figure}
\begin{subfigure}{0.47\textwidth}
\vspace{-2mm}
 \hspace{-12mm}
  \includegraphics*[scale=0.35]{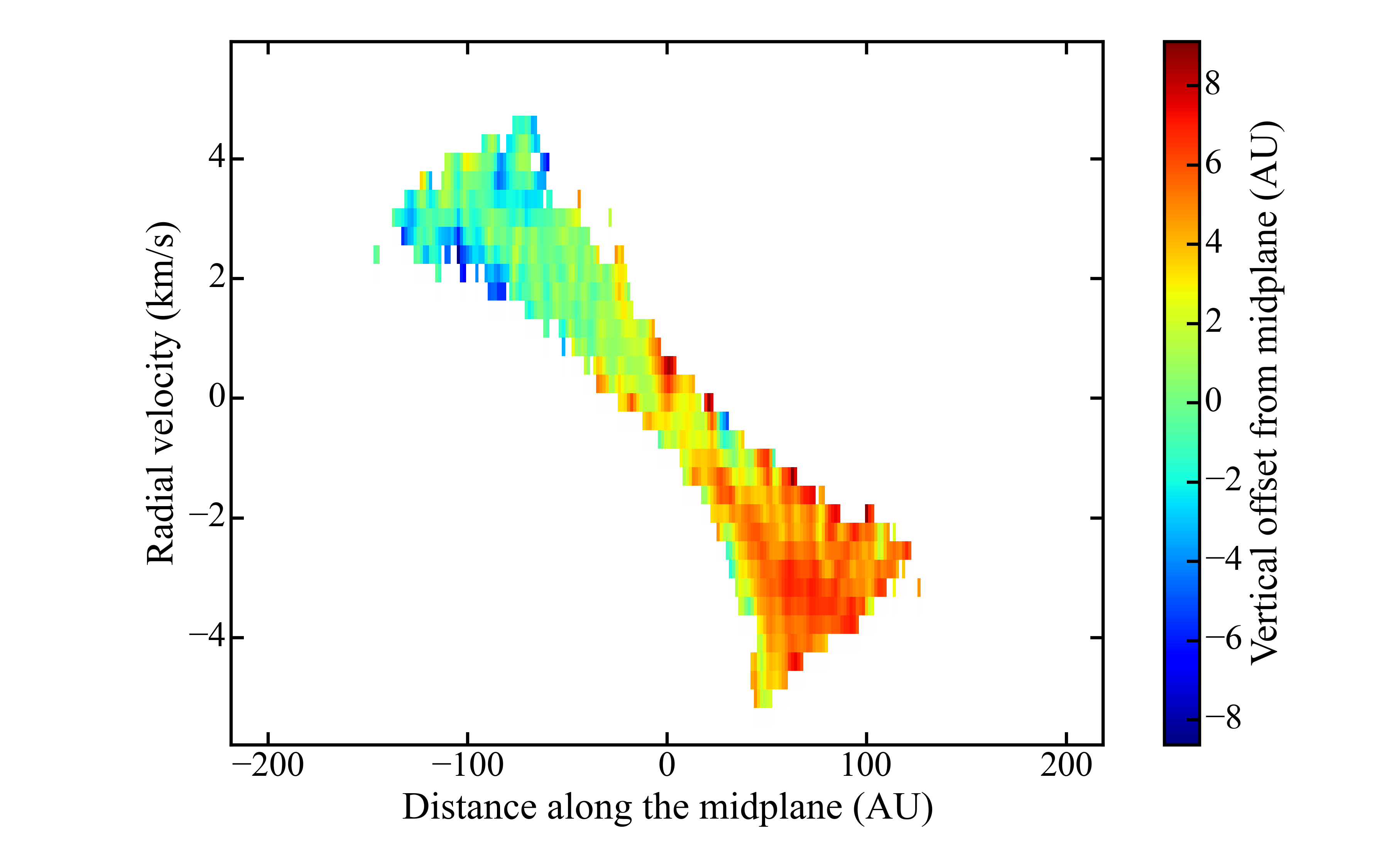}
\vspace{0mm}
\end{subfigure} \\
\begin{subfigure}{0.47\textwidth}
\vspace{-13mm}
 \hspace{-12mm}
  \includegraphics*[scale=0.35]{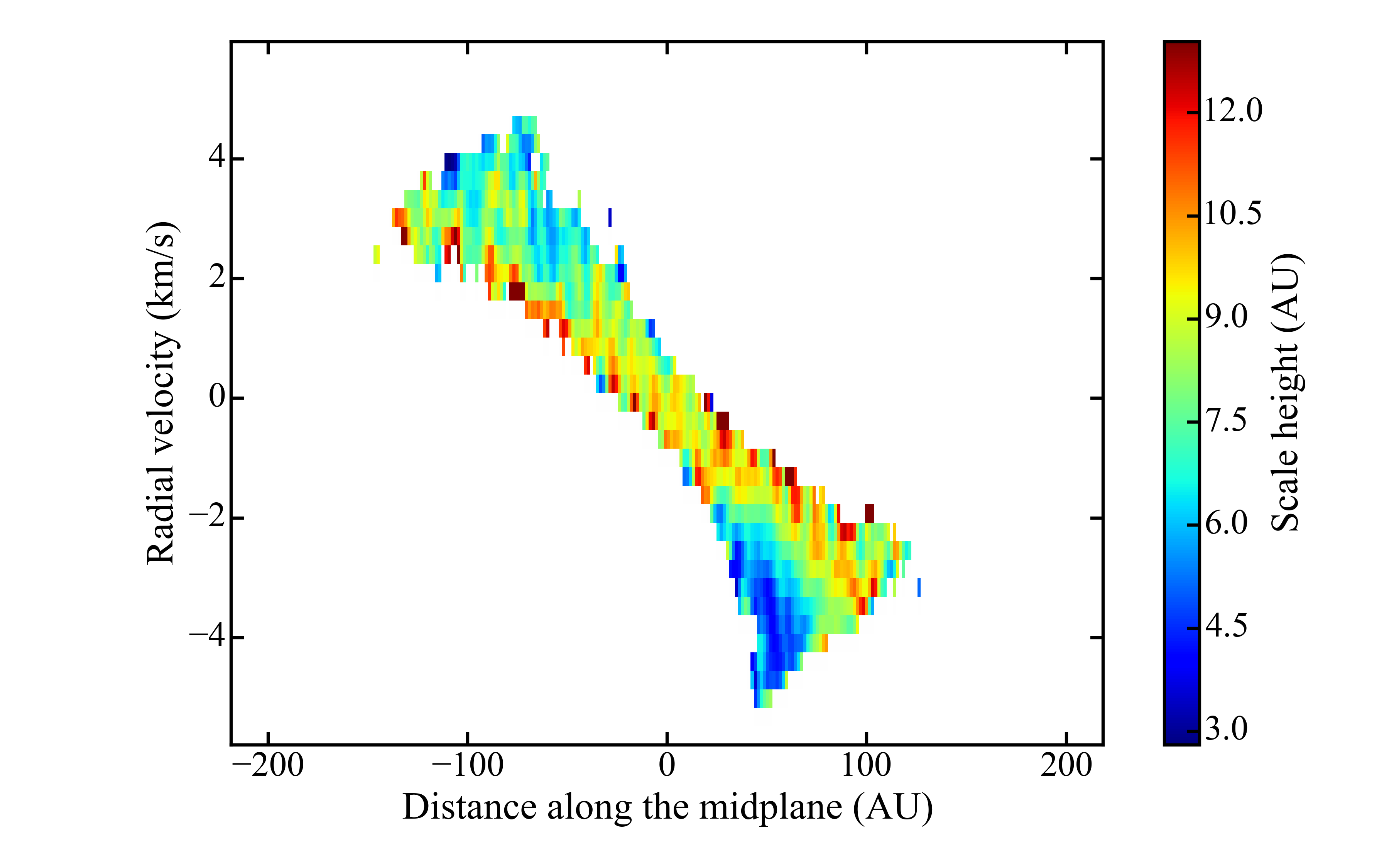}
\vspace{0.0mm}
\end{subfigure}
\vspace{-5mm}
\caption{Position-velocity diagrams of both the CO J=2-1 disk vertical offset $y_{\rm obs}$ with respect to the main dust disk (\textit{top}) and of the disk scale height $H$ (\textit{bottom}), for an assumed perfectly edge-on disk. These measurements were obtained through Gaussian fitting of measured fluxes along cuts perpendicular to the disk midplane (as displayed in Fig. \ref{fig:vertprofs}) repeated for each radial velocity channel.}
\label{fig:pvvert}
\end{figure}

In Section \ref{vertstruct} we analysed the spectrally-integrated disk vertical structure along $y_{\rm sky}$ as a function of position along the disk midplane $x_{\rm sky}$. Given our velocity information, however, it is more insightful to measure the disk spine and width separately for each of the radial velocities $v_{\rm rad}$ at which the disk is detected, using the same procedure as outlined in Appendix \ref{app:2}. This yields a measurement of the sky-projected scale height $H$ and of the disk vertical offset from the main disk midplane $y_{\rm obs}$ as a function of position along the midplane $x_{\rm sky}$ and radial velocity $v_{\rm rad}$. In other words, we obtain a PV diagram (shown for the J=2-1 transition in Fig. \ref{fig:pvvert}).

The anti-clockwise tilt (dPA=PA$_{\rm obs}$-PA$_{\rm main\ disk}$) observed in the moment-0 images (Fig. \ref{fig:vertprofs}) is also present throughout the radial velocity channels, with CO at negative radial velocities (SW) being vertically displaced above CO at positive velocities (NE). Some substructure is observed, with an enhanced vertical displacement at low radial velocities (hence large orbital radii) on the NE side, and a decreased displacement at high radial velocities (hence small orbital radii) on the SW side. In addition, a positive vertical offset of disk emission along the line of sight to the star (enhanced at positive velocities) is present. These local features are marginally significant at the typical 0.5-1 AU 1$\sigma$ uncertainty on vertical offsets in each spaxel; their interpretation in terms of a putative 3D warp structure requires detailed dynamical modelling, and is beyond the scope of this work.

\begin{figure}
\begin{subfigure}{0.47\textwidth}
\vspace{-2mm}
 \hspace{-2mm}
  \includegraphics*[scale=0.29]{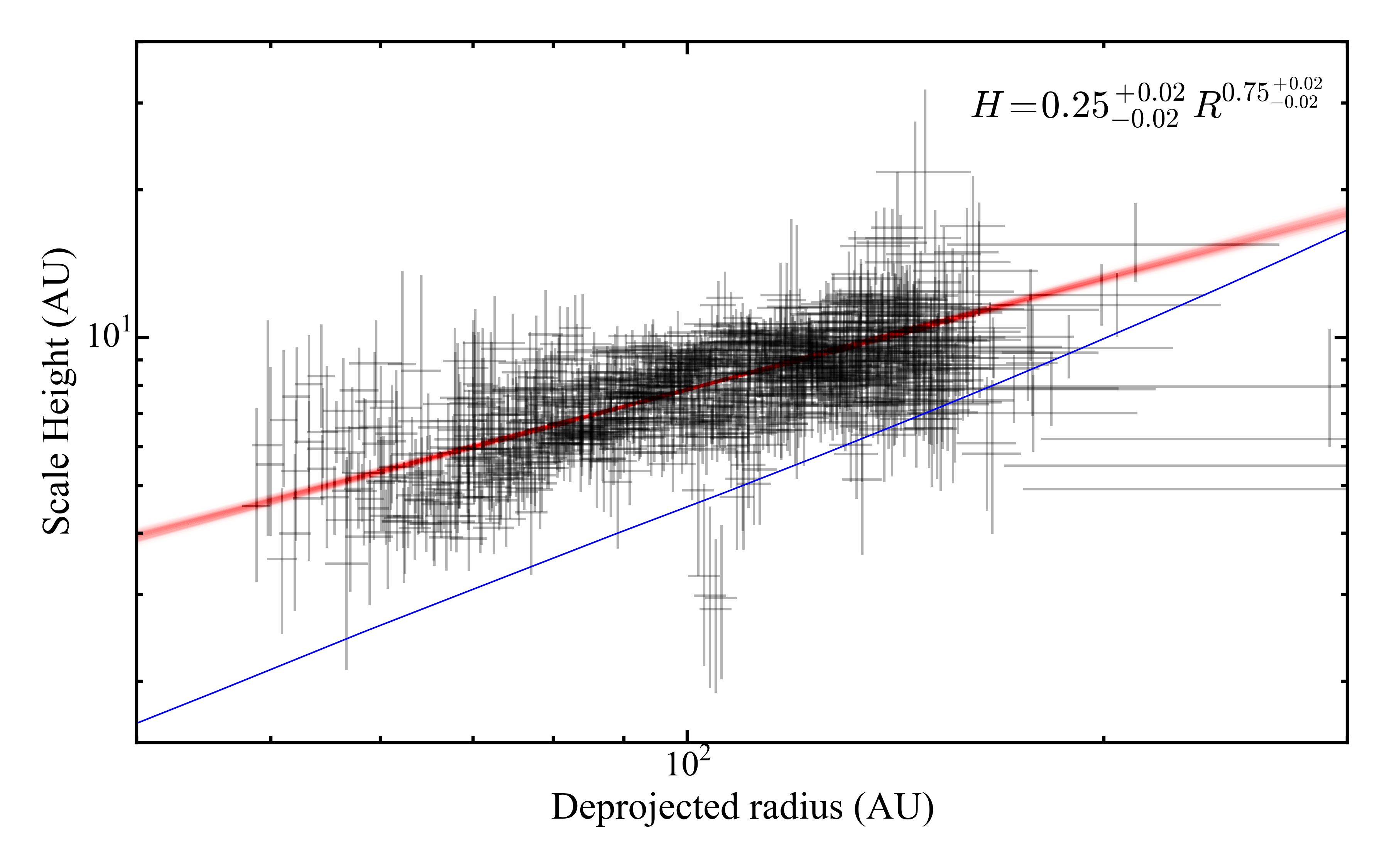}
\vspace{0mm}
\end{subfigure} \\
\begin{subfigure}{0.47\textwidth}
\vspace{-11.7mm}
 \hspace{-2mm}
  \includegraphics*[scale=0.29]{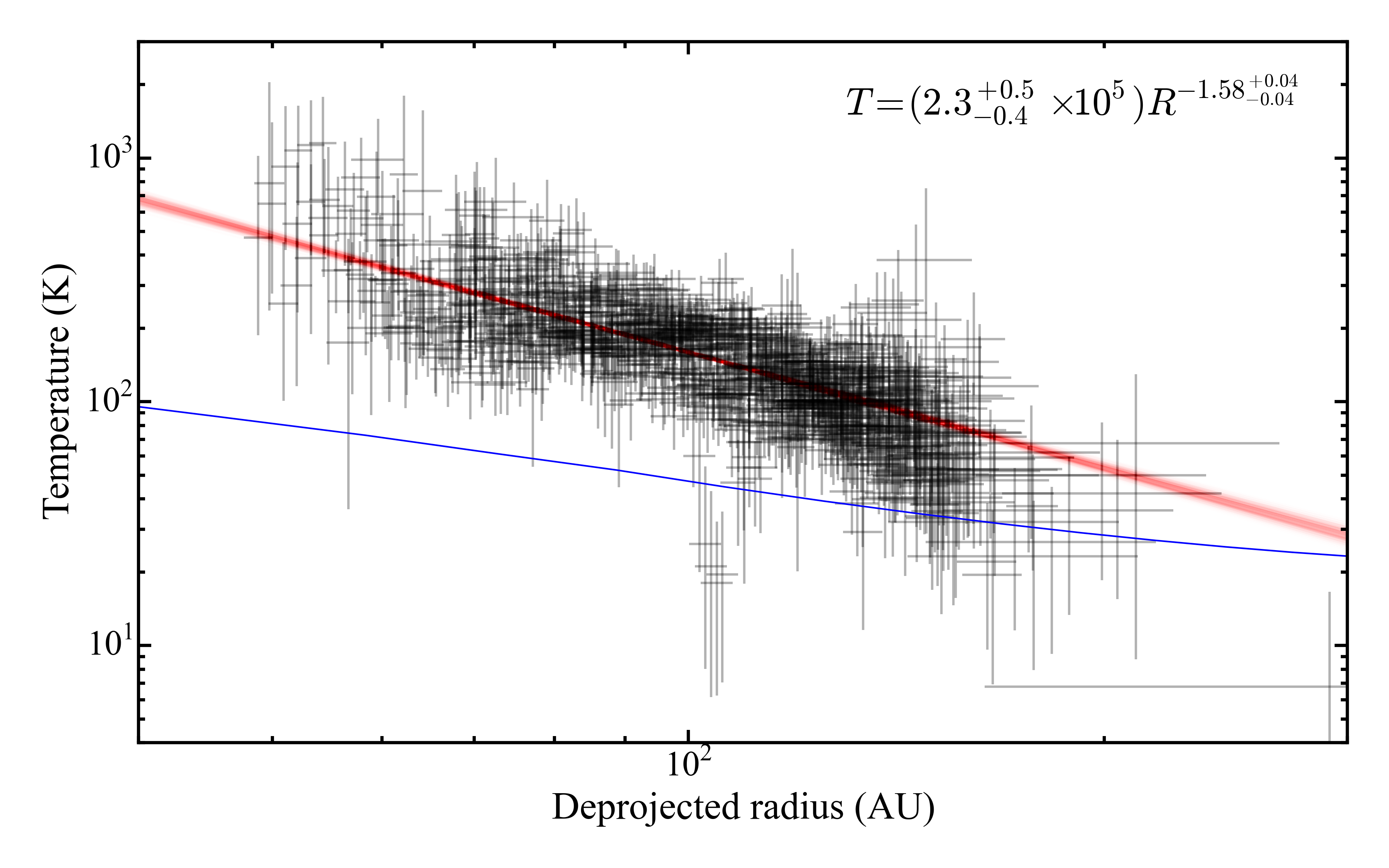}
\vspace{0.0mm}
\end{subfigure}
\vspace{-5mm}
\caption{\textit{Top:} Measured dependence of the CO disk scale height on orbital radius, derived from the PV diagrams in Fig. \ref{fig:pvvert} (bottom) assuming Keplerian rotation. Each point corresponds to a spaxel in the J=2-1 PV diagram. \textit{Bottom:} Radial dependence of the temperature derived from the scale height under the assumption of a vertically isothermal disk.
In both plots, the red lines are constructed by randomly picking 1000 values of our fitting parameters (intercept and slope in log-log space) from their joint posterior probability distribution. The best-fit power law coefficients are displayed in the top right of the diagrams, with their respective 1$\sigma$ uncertainties. The blue lines represent model predictions by \citet{Kral2016a}.}
\label{fig:vert1D}
\end{figure}

The scale height also presents significant PV structure in the form of a gradient from low values at high radial velocities to high values at lower radial velocities (Fig. \ref{fig:pvvert}, bottom). For the assumed perfectly edge-on configuration, and given that assuming Keplerian rotation different diagonal lines represent different orbital radii, this gradient is representative of an increase in the disk scale height as a function of orbital radius. Indeed, if we assign an orbital radius $R$ to each PV location (Appendix \ref{app:1}) and assume the disk scale height to be azimuthally symmetric, in Fig. \ref{fig:vert1D} (top) we show that the scale height $H$ scales as
\begin{equation}
H=7.0^{+0.6}_{-0.6} \times \left(\frac{R}{85 \rm AU}\right)^{0.75^{+0.02}_{-0.02}}.
\end{equation}
The error bars (1$\sigma$) in Fig. \ref{fig:vert1D} (top) represent the uncertainty in the derivation of both the scale height $H$ and the orbital radius $R$. The error on the scale height was calculated from MCMC fits as described in Appendix \ref{app:2}, whereas the error on the orbital radius $R$ was propagated from the uncertainty on the PV location (assumed to be equal to the size of a spaxel). 

Assuming the CO disk to be vertically isothermal and in hydrostatic equilibrium, the scale height can then be used to trace the disk temperature at a certain orbital radius through $T=\frac{GM_{\ast}\mu}{kN_A} \frac{H^2}{R^3}$, where $G$ is the gravitational constant, $M_{\ast}$ is the mass of the star, $\mu$ is the mean molecular mass of the gas, $k$ is Boltzmann's constant, and $N_A$ is Avogadro's number. We here assume the gas to be dominated by the carbon and oxygen atoms released from CO photodissociation, giving a mean molecular mass $\mu=14$ (see Sect. \ref{sect:atomdom}).
We can then use our measured dependence of the scale height on radius to derive the radial dependence of the gas temperature in the $\beta$ Pic disk (Fig. \ref{fig:vert1D}, bottom). This decreases as a function of radius, scaling as
\begin{equation}
T=\left(2.1^{+0.4}_{-0.4} \times 10^{2}\right) \times \left(\frac{R}{85 \rm AU}\right)^{-1.58}
\end{equation}
where $T$ is in K and $R$ is in AU. No temperature increase/decrease is seen at the clump location compared to the rest of the disk, which is in line with the expectation that the CO, after release, should quickly collide and couple to the rest of the atomic disk (again, see discussion in Sect. \ref{sect:atomdom}). 

\subsubsection{CO 3D structure for a non-edge-on configuration}
\label{sect: structnonedgeon}
In the previous section, we neglected a potential deviation of the disk inclination from perfectly edge-on, which has been proposed before through modelling observations of optical and near-IR scattered light from the dust disk \citep[e.g.][]{Ahmic2009, Milli2014}. If the disk is inclined from edge-on, the vertical structure becomes coupled with the azimuthal structure. On one hand, this means that our derivation of the disk scale height and temperature in the previous section may be in part biased by the edge-on assumption. On the other hand, we can interpret the vertical displacement $y_{\rm obs}$ of the disk from the midplane (i.e. the disk spine) purely as an effect of azimuthal structure seen at an inclination $i<90^{\circ}$.

Using the observed vertical displacement $y_{\rm obs}$ measured at each PV location $\left(x_{\rm sky},v_{\rm rad}\right)$ (see Fig. \ref{fig:pvvert} top) we can determine whether CO emission originates in front or behind the plane of the sky (i.e. at $+$ or $-y$ in Fig. \ref{fig:deprojectpvs}). This is because for a given orbit, characterised here by its inclination from face-on $i$ and its on-sky tilt angle dPA, an on-sky location $\left(x_{\rm sky, orb}, +y_{\rm sky, orb}\right)$ at radial velocity $v_{\rm rad}$ will correspond to either orbital location $\left(x, +y\right)$ behind the sky plane or $\left(x, -y\right)$ in front of the sky plane.
Then, in the presence of a $\pm y$ asymmetry in the orbital plane of the disk, the flux contribution from the two on-sky vertical locations $+y_{\rm sky, orb}$ and $-y_{\rm sky, orb}$ will differ and the vertical centroid $y_{\rm obs}$ of the CO emission will be displaced from the midplane. As described in Appendix \ref{app:4}, we use this vertical displacement to infer the level of CO emission originating from $\left(x_{\rm sky, orb}, +y_{\rm sky, orb}\right)$ as opposed to $\left(x_{\rm sky, orb}, -y_{\rm sky, orb}\right)$ in the sky plane. 

A degeneracy remains in that we do not know which between $+y_{\rm sky, orb}$ and $-y_{\rm sky, orb}$ is in front or behind the plane of the sky. This may be solved using scattered light imaging, since material in front of the star will appear brighter than behind the star, owing to the phase function of the grains being peaked towards and hence favouring forward-scattering angles \citep[e.g.][]{Milli2015}. For $\beta$ Pictoris, scattered light emission above the midplane is seen to be brighter than below the midplane \citep[e.g.][]{Golimowski2006}; that suggests that dust at $+y_{\rm sky}$ lies preferentially in front of the star, whereas dust at $-y_{\rm sky}$ is located behind the star. However, this does not take into account that in the presence of an azimuthal asymmetry there may be an excess of material above the midplane, which would affect the above argument by leading us into interpreting the observed flux excess 
as a phase function effect.

As the method produces a model data cube for a given orbit, we can compare different orbits and hence find how well different $\left(i, \rm dPA\right)$ combinations fit the data (see Appendix \ref{app:4} and $\chi^2$ map in Fig. \ref{fig:chisqmap}). We note once again that this approach does not take into account any intrinsic disk vertical structure (and in particular its dependence on orbital radius) nor any uncertainty in the determination of the vertical displacement $y_{\rm obs}$ from the on-sky midplane. Keeping this in mind, we find that the formal best-fit (i.e. the $\chi^2$ minimum) is at $i\sim88^{\circ}$ and dPA $\sim3^{\circ}$, though we consider all inclinations $\geq86^{\circ}$ and on-sky tilt angles $\leq5^{\circ}$ to provide reasonable fits.

\begin{figure}
\begin{subfigure}{0.45\textwidth}
\vspace{-2mm}
 \hspace{-2mm}
  \includegraphics*[scale=0.40]{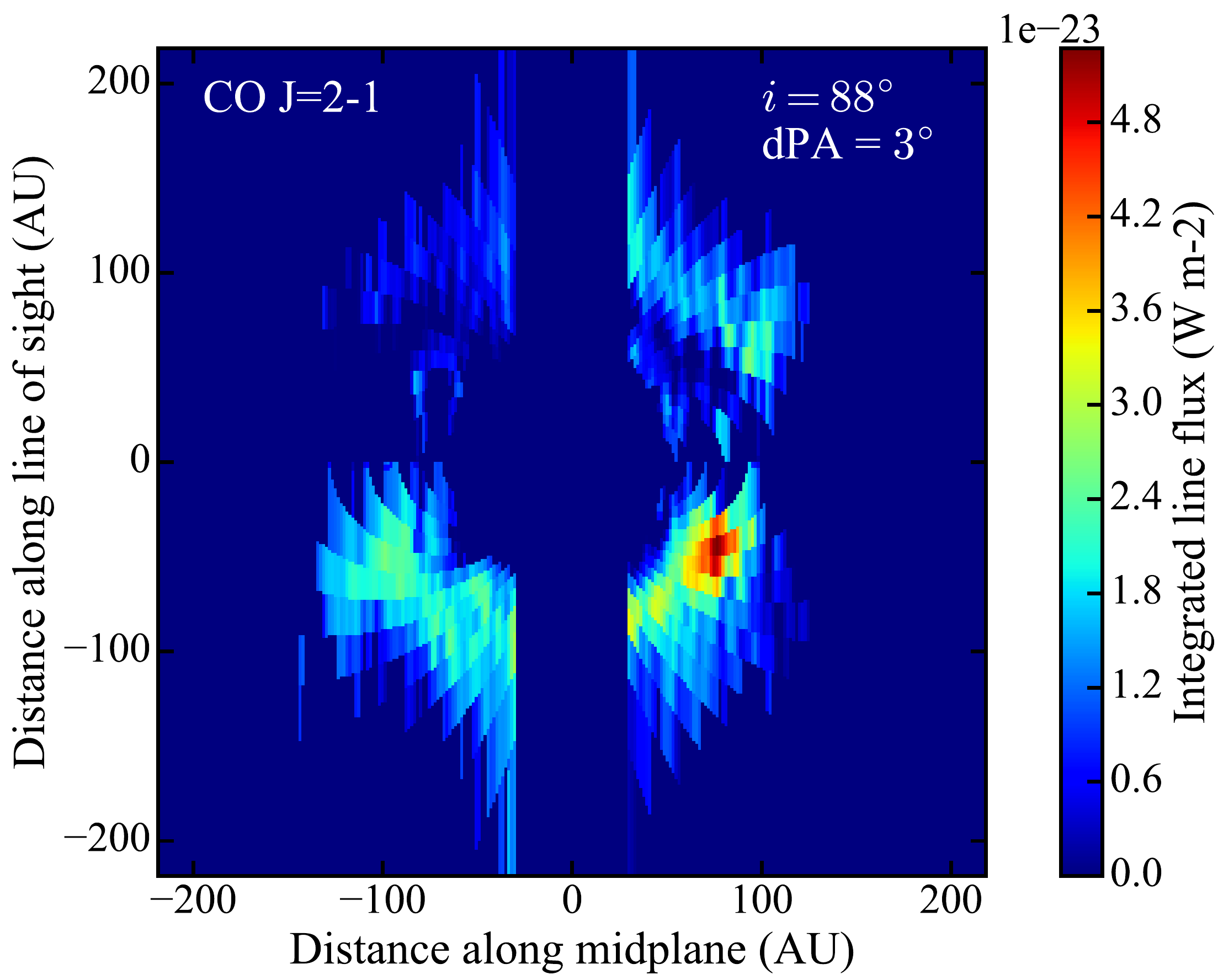}
\end{subfigure} \\
\begin{subfigure}{0.45\textwidth}
\vspace{-0mm}
 \hspace{-2mm}
  \includegraphics*[scale=0.40]{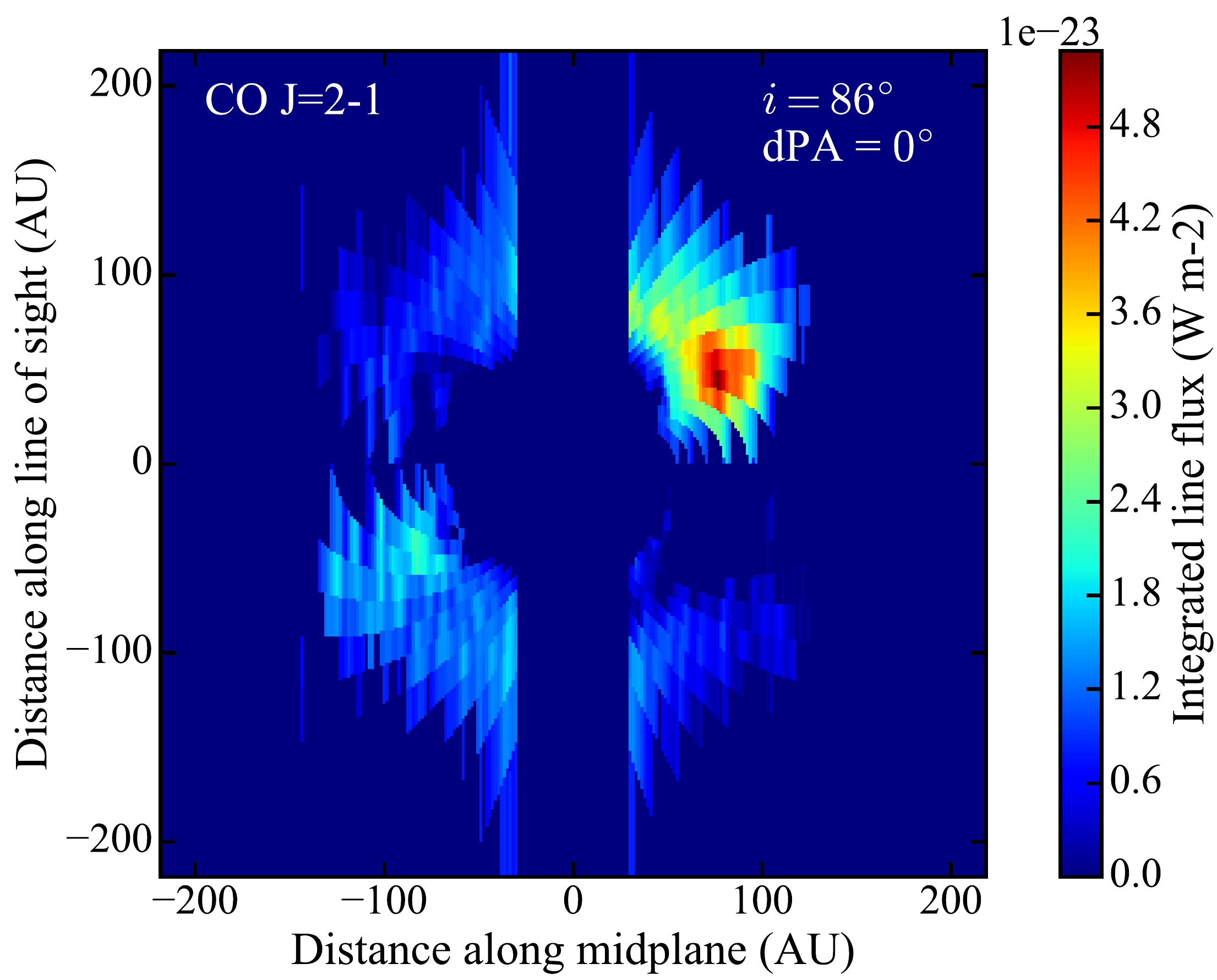}
\end{subfigure}
\caption{Face-on deprojection of CO disk emission, derived from the velocity information in the PV diagrams for the J=2-1 transition, for two cases of disk inclination $i$ and on-sky tilt angle dPA. Flux at each $x$ location has been divided between + and -$y$ locations in the orbital plane displayed here using the vertical displacement $y_{\rm obs}$ of disk emission in the plane of the sky (Fig. \ref{fig:pvvert}, top), using the method described in Appendix \ref{app:4}. Though the $\left(i,\rm dPA\right)$ choice for the deprojection in the top panel is formally the best-fitting in the framework of our simple model (see main text for details), the image in the bottom panel reproduces well the expected morphology from resonant trapping of planetesimals due to an outward migrating planet.}
\label{fig:extradeproj}
\end{figure}

Of course, for any choice of $\left(i, \rm dPA\right)$, we can use our method to produce a map of emission in the orbital plane of the disk, as we now have determined what fraction of emission belongs to $+y$ and to $-y$ for a given $x$. Fig. \ref{fig:extradeproj} shows two examples, for our formal best-fit orbit ($i=88^{\circ}$, dPA = $3^{\circ}$) and in the case of a lower inclination ($i=86^{\circ}$) and no on-sky tilt (dPA $=0^{\circ}$). We report two main findings: 
\begin{itemize}
\item The majority of the flux from the CO clump on the SW side of the disk is located above the sky-projected disk midplane for any reasonable choice of $i$ ($\geq 86$) and dPA ($<5^{\circ}$). This confirms that the flux in the sky-projected clump does originate from a physical clump, located either behind or ahead of the sky plane (due to the aforementioned degeneracy). 
\item The disk deprojection obtained for a lower inclination of $86^{\circ}$ and no dPA qualitatively reproduces well the resonance sweeping scenario proposed in \citet{Dent2014} and based on previous work by \citet{Wyatt2003, Wyatt2006}, with two clumps on opposite sides of the star and their respective trailing tails (see discussion in Sect. \ref{sect:rescl}). Further 3D modelling taking into account at the same time the disk vertical and azimuthal structure as well as its geometry is needed to rule out or confirm this possibility.
\end{itemize}

\subsubsection{CO 3-2/2-1 line ratio}
\label{sect:reslinerat}

\begin{figure*}
\begin{subfigure}{0.45\textwidth}
 \hspace{-24mm}
  \includegraphics*[scale=0.57]{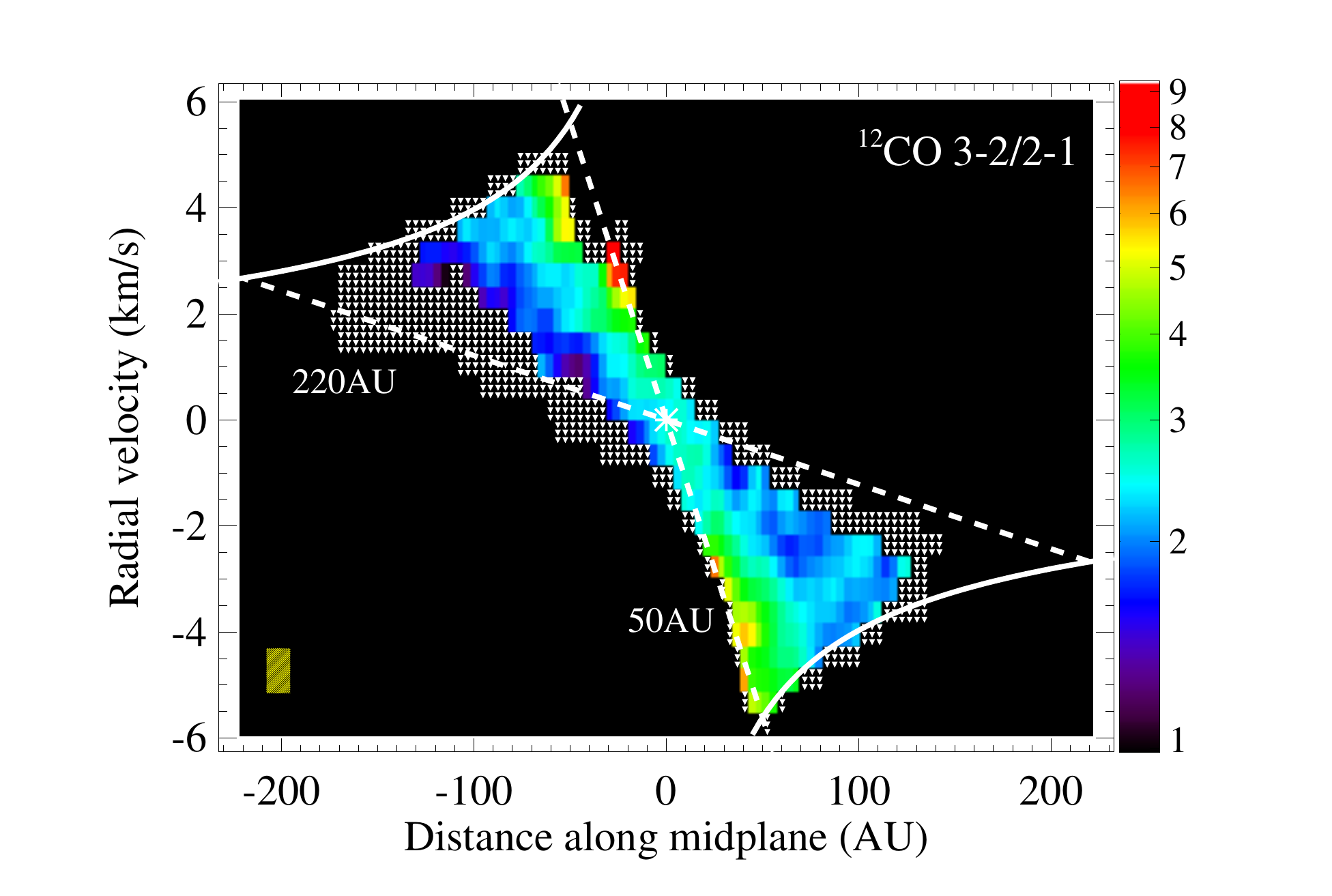}
\vspace{0mm}
\end{subfigure}
\begin{subfigure}{0.45\textwidth}
 \hspace{-8mm}
  \includegraphics*[scale=0.57]{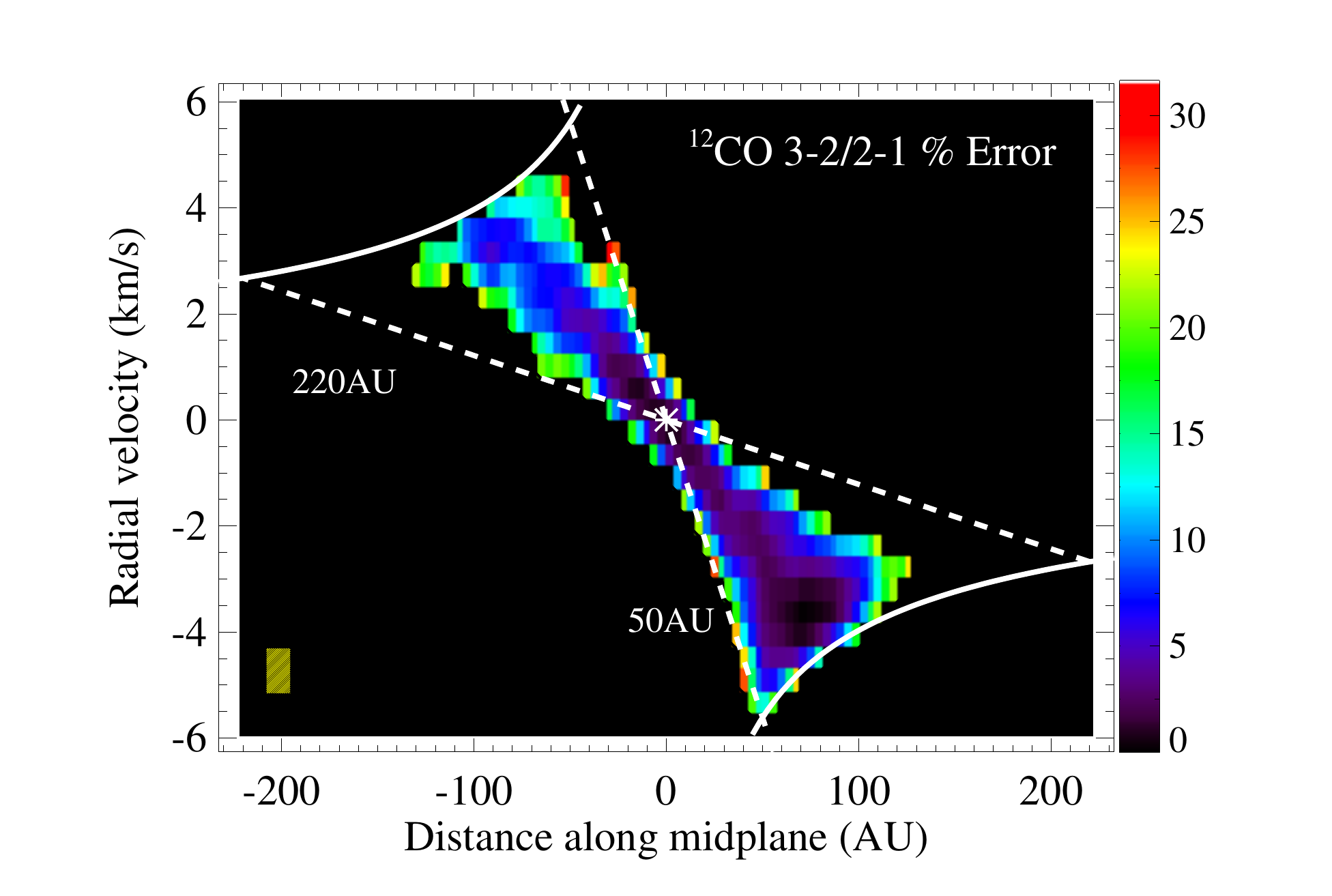}
\vspace{0mm}
\end{subfigure}
\vspace{-5mm}
\caption{\textit{Left:} CO J=3-2/J=2-1 line ratio position-velocity diagram of the $\beta$ Pic disk. Only spaxels where \textit{either} CO transition is detected (at $\ge$4 $\sigma$) are shown; downward arrows indicate spaxels where only the J=2-1 transition was detected. The white asterisk represents the stellar position and velocity, while the solid white curves are the maximum radial velocity observable in an edge-on Keplerian disk around a 1.75 M$_{\odot}$ star. The yellow rectangle in the bottom left corner represents the spectro-spatial resolution. \textit{Right:} Percentage error map of the line ratio PV diagram, where this can only be quantified in spaxels where \textit{both} transitions are detected.}
\label{fig:pvratios}
\end{figure*}

In order to better investigate morphological differences between the two transitions, we combined the J=2-1 and J=3-2 PV diagrams to obtain line ratios in each position-velocity spaxel. As such ratios can be extreme in noise-dominated regions, we facilitate visualisation by only displaying ratios in spaxels where either CO J=3-2 and J=2-1 emission is over 4$\sigma$ (where $\sigma$ is the RMS noise level measured in each PV diagram). A detection at 3-2 but only an upper limit at 2-1 will then lead to a lower limit on the line ratio, and conversely a detection at 2-1 but not at 3-2 will lead to an upper limit on the ratio. Such line ratio PV diagram is shown in Fig.\ \ref{fig:pvratios} (left), with the corresponding uncertainty diagram in Fig.\ \ref{fig:pvratios} (right). At each spaxel, the uncertainty is the quadratic sum of the relative errors on the flux for the two original PV diagrams, where these are given by the RMS levels. We here do not take into account the 10\% flux calibration error on each dataset, since we are here most interested in the presence of gradients rather than in the absolute value of the ratios.

The line ratio PV diagram shows a noticeably different structure to the flux PV diagrams for the individual transitions in Fig.\ \ref{fig:pvs}. In particular, the line ratio shows no evidence of the CO clump nor the NE/SW asymmetry. On the other hand, it shows a significant gradient where the line ratio is seen increasing from low values at low radial velocities to high values at high radial velocities, for any given on-sky midplane location. On the NE side, where individual line fluxes are generally dimmer than in the SW side (see Fig.\ \ref{fig:pvs}), this causes a non-detection of the 3-2 line at low velocities (as illustrated through the upper limits on the ratios). This then explains the differences between the two lines that were already apparent from the individual moment-0 images and PV diagrams (Fig.\ \ref{fig:mom0} and \ref{fig:pvs}), with 2-1 emission lying at larger projected radii than the 3-2 line on the NE side of the disk. The physical significance of this gradient is explored through modelling in Sect. \ref{sect:mod} below.

\section[]{Line ratio modelling}
\label{sect:mod}

Through detailed analysis of resolved ALMA CO J=3-2 and CO J=2-1 line observations, we unveiled the radial, vertical and azimuthal structure of CO line emission in the $\beta$ Pictoris disk. Under the assumption that the system is viewed perfectly edge-on, we interpreted the scale height observed as a function of orbital radius to give us an estimate for the temperature of the gas and its radial dependence. We then relaxed the edge-on assumption and interpreted the observed vertical structure as solely caused by azimuthal structure, in order to break the degeneracy in deprojecting PV diagrams to the disk orbital plane.
Finally, we analysed the CO 3-2/2-1 line ratio PV diagram and discovered a significant gradient with the ratio increasing with radial velocity at any given midplane location.
In this section, we aim to use the non-local thermodynamic equilibrium (NLTE) analysis of CO excitation developed in \citet{Matra2015} to interpret these line observations (particularly resolved line ratios) in terms of the physical conditions, origin and mass of CO in the $\beta$ Pic disk.

\subsection{Second generation: NLTE line ratios as a probe for electron densities}
\label{modelec}

We begin by assuming CO line emission to be optically thin at the wavelengths of the observed transitions, and double-check this assumption later in Sect. \ref{sect:optthick}. In this regime, CO excitation and hence line ratios are only dependent on the kinetic temperature $T_{\rm k}$  and the collisional partner density $n_{\rm coll}$, for the known external millimetre radiation field $J_{\nu}$ at the frequency of the transitions in question. In a scenario where the gas is of second-generation origin, the dominant collisional partners for which collisional rates are available are electrons, which originate from the photoionisation of atoms created by rapid CO photodissociation. In the absence of H$_2$, electrons are more efficient colliders than H$_2$O \citep{Matra2015}, but also \ion{H}{I}, since the latter, despite being potentially $\sim$6 times more abundant than electrons \citep{Kral2016a}, is on average a much less efficient collider. For example, collisional de-excitation of a CO molecule from rotational level 3 to 2, for the given \ion{H}{I}/e$^-$ abundance of 6 happens at a rate that is $(\gamma_{3-2, \ion{H}{I}} n_{\ion{H}{I}}) / (\gamma_{3-2, e^-} n_{\rm e^-}) \sim 50$ times faster for electron collisions than for \ion{H}{I} collisions\footnote{Collisional rate coefficients $\gamma$ were obtained from \citet{Dickinson1975} for electrons and from \citet{Walker2015} for \ion{H}{I}}. Nonetheless, it is useful to keep in mind that other less dominant species may contribute to observed CO collisional excitation, which implies that the electron densities probed here should strictly speaking be considered upper limits.

\begin{figure}
\begin{subfigure}{0.47\textwidth}
\vspace{-4mm}
 \hspace{-5mm}
  \includegraphics*[scale=0.45]{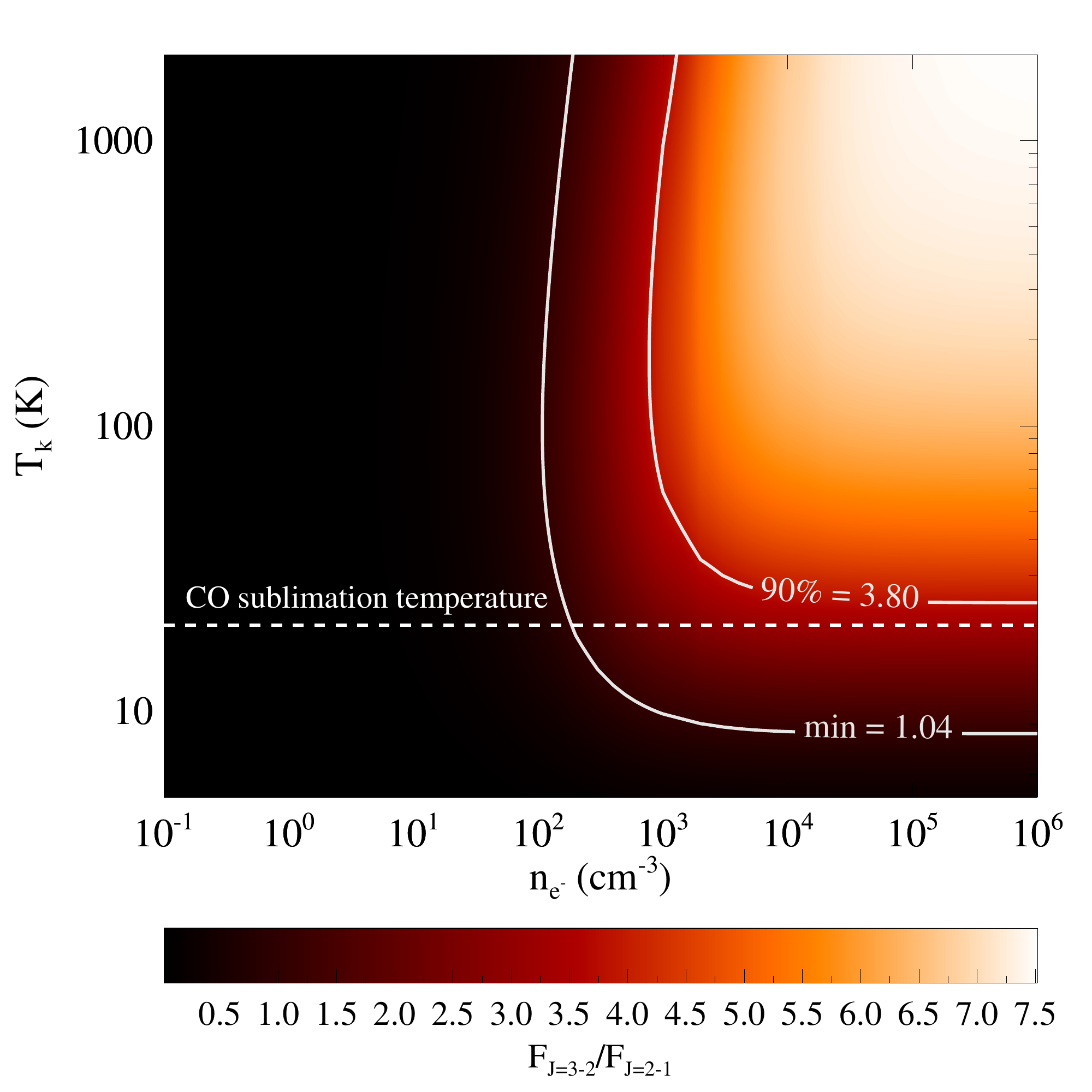}
\vspace{0.0mm}
\end{subfigure} \\
\begin{subfigure}{0.47\textwidth}
\vspace{-4.7mm}
 \hspace{-5mm}
  \includegraphics*[scale=0.45]{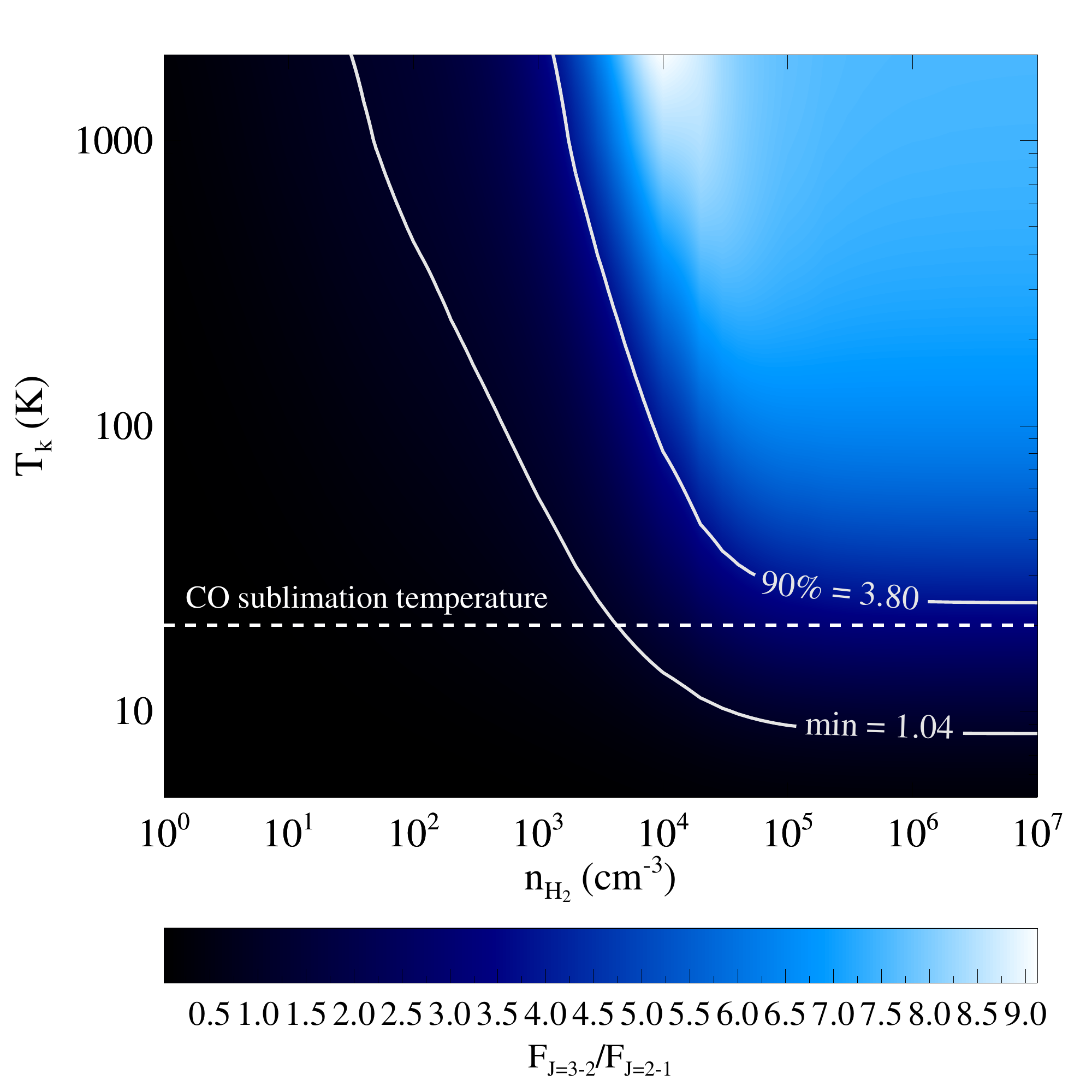}
\vspace{0.0mm}
\end{subfigure}
\vspace{-4mm}
  \caption{Colour maps of the J=3-2/2-1 line ratio expected in an optically thin disk as a function of the density of the dominant collider species and the kinetic temperature of the gas. The dominant collider species is electrons in a secondary origin scenario (top), and H$_2$ in a primordial origin scenario (bottom). Contours represent minimum and $90^{\rm th}$ percentile values observed in our ratio PV diagram (Fig.\ \ref{fig:pvratios})}
\vspace{-6 mm}
\label{fig:ratiotkncoll}
\end{figure}

Fig. \ref{fig:ratiotkncoll} (top) shows the J=3-2 / J=2-1 line ratios expected for CO that is excited by collisions with electrons and by millimetre radiation from the cosmic microwave background (CMB, which dominates over the dust continuum for low CO transitions), as a function of the electron density $n_{\rm e^-}$ and the gas kinetic temperature $T_{\rm k}$. Each ratio traces a line in $\left(T_{\rm k}, n_{\rm e^-}\right)$ space, since these are the only free parameters. 
For $n_{\rm e^-}$, we explore the parameter space between 10$^{-1}$ and 10$^{6}$ cm$^{-3}$, encompassing the transition between the two limiting excitation regimes \citep[LTE and radiation-dominated,][]{Matra2015}. For the kinetic temperature of the gas, we probe a broad range of $T_{\rm k}$ between 5 and 2000 K to illustrate the behaviour of the line ratios. 
Its true value depends on the detailed balance between local heating and cooling processes in the gas disk. Our temperature estimates from the CO scale height (Sect. \ref{sect: structedgeon}) suggest values between 40 and a few hundred K between 50 and 200 AU, whereas detailed thermodynamical modelling of the disk suggests values between 30 and 80K at the same radii \citep{Kral2016a}. In any case, we expect the gas kinetic temperature to be above a few tens of K throughout the disk.

These predictions can then be compared with our measurements, tracing lines overplotted on Fig. \ref{fig:ratiotkncoll} (top). Above kinetic temperatures of $\sim$30 K, the vast majority (up to $\sim$90$^{\rm th}$ percentile) of the line ratios observed in the $\beta$ Pic disk (from spaxel values in Fig. \ref{fig:pvratios}) are very good probes of the electron density, being largely independent of the choice of kinetic temperature. On the other hand, if the electron densities were high enough for CO to reach local thermodynamic equilibrium (LTE, right edge of the colour map), the line ratios would be higher and become solely dependent on the kinetic temperature. While a unique solution for $T_{\rm k}$ and $n_{\rm e^-}$ is only obtainable with two (or more) line ratios, our measurements can confidently constrain the electron density in the disk to around $10^2$-$10^3$ cm$^{-3}$.

This allows us to attribute the gradient of increasing CO line ratio with radial velocity observed in the PV diagrams (Fig. \ref{fig:pvratios}, left) to the electron density distribution in the disk. Since we want to measure the electron density as robustly as possible, we mask the highest 10\% of the line ratios (i.e. those above 3.80), which arise preferentially from high-noise regions of the PV diagram, and have a stronger kinetic temperature dependence.
For the remaining line ratios, this temperature dependence is much weaker, so any assumption we make is not going to affect the result significantly. As such, we assume the temperature radial profile predicted by \citet{Kral2016a}, i.e. a power law with slope of $-0.8$, normalised to a value of 80 K at 50 AU. Using the temperature radial profile derived from our scale height measurement in Fig. \ref{fig:vert1D} (bottom) yields analogous results.

\begin{figure*}
\begin{subfigure}{0.45\textwidth}
 \hspace{-10mm}
  \includegraphics*[scale=0.40]{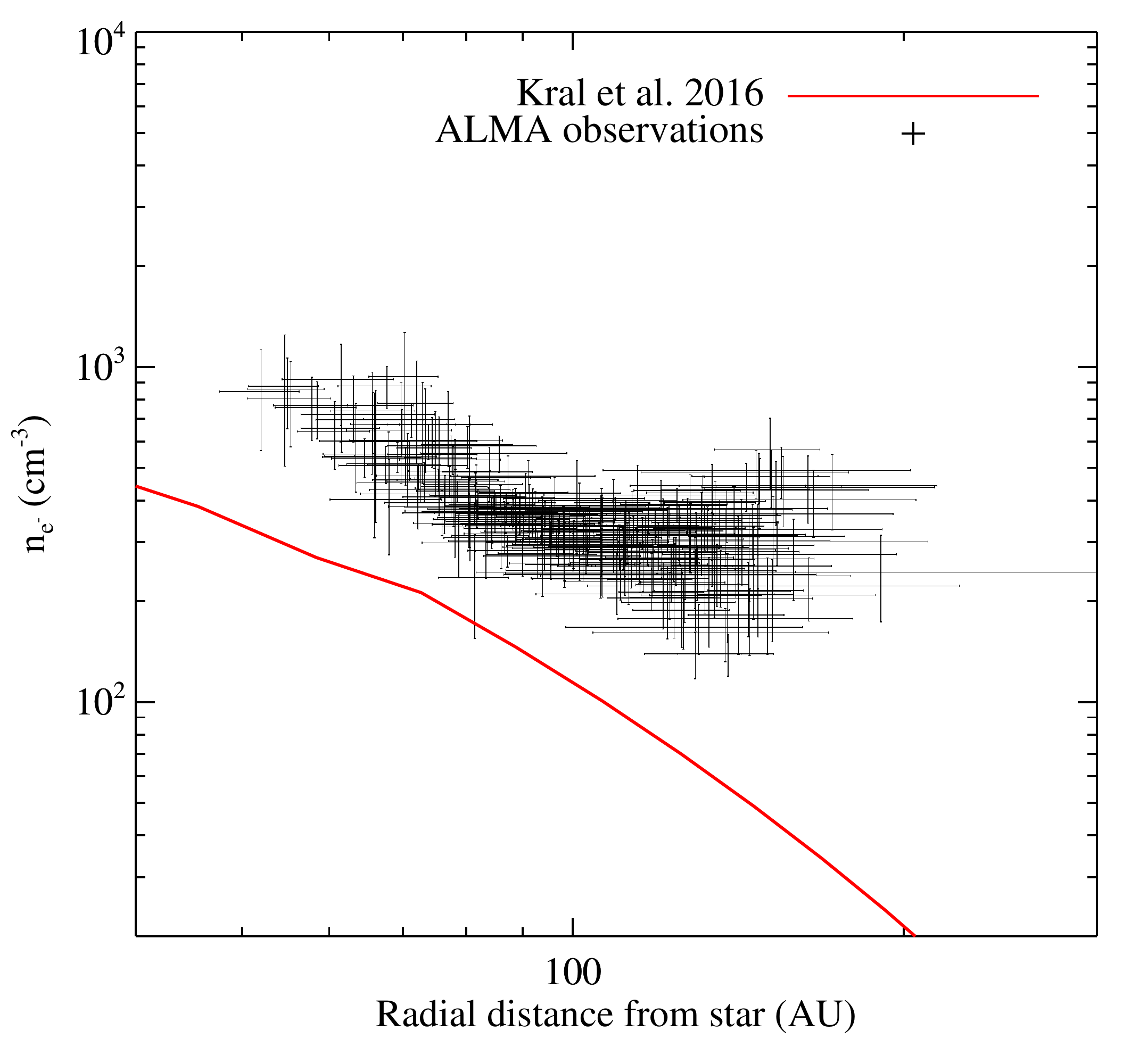}
\vspace{-3mm}
\end{subfigure}
\begin{subfigure}{0.45\textwidth}
 \hspace{-0mm}
  \includegraphics*[scale=0.40]{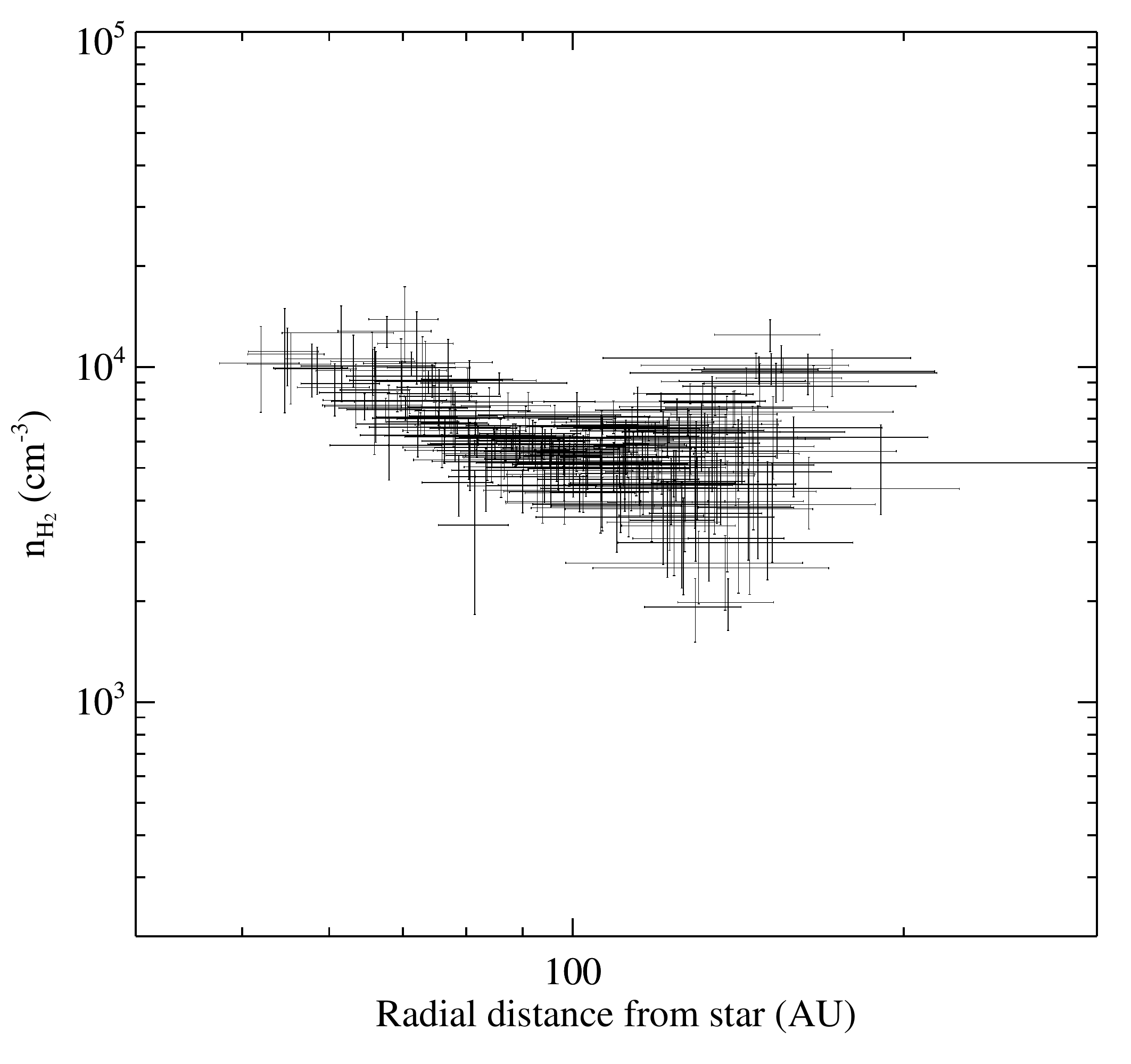}
\vspace{-6.5mm}
\end{subfigure}
  \caption{Electron density (left) and H$_2$ density (right) as a function of deprojected radius derived from the PV ratio diagram in Fig.\ \ref{fig:pvratios} using the NLTE analysis as displayed in Fig.\ \ref{fig:ratiotkncoll}, and assuming temperatures predicted by the hydrodynamical model of \citet{Kral2016a}. Line ratios higher than 3.8 were masked due to their strong dependence on kinetic temperature, which only adds noise to the diagram. For the remaining line ratios, our temperature choice does not influence the resulting electron densities strongly. Error bars in both directions were derived from uncertainties in the line ratio and deprojected radial location. }
\vspace{-6 mm}
\label{fig:neradial}
\end{figure*}

Each location in the PV diagram traces a certain orbital radius within the disk (Appendix \ref{app:1}). This way, we can solve for the radial dependence of the electron density, which is shown in Fig.\ \ref{fig:neradial}, left. The uncertainty on $n_{\rm e^-}$ was propagated from that of the line ratio, assuming a perfectly known $T_{\rm k}$ from the models; the uncertainty on the radius, on the other hand, was calculated from the uncertainty in the $\left(x_{\rm sky}, v_{\rm rad}\right)$ location within a spaxel in the PV diagram. 
Since we are most interested in the radial dependence rather than on the absolute value of the electron density, we have not included sources of systematic error that may shift $all$ of the observational points in our plot (e.g. an error on $M_{\ast}$ or $i$ may cause an overall shift in the radial direction, whereas the ALMA flux calibration systematic may cause an overall shift in the $n_{\rm e^-}$ direction). 

The electron density in the disk shows a power law radial dependence between 40 and 200 AU (i.e. where the CO is detected), with a slope of $\gamma \sim -1$; this is in line with the prediction by the \citet{Kral2016a} model, which we will discuss in more detail in Sect.\ \ref{sect:atomdom}. 

\subsection{Primordial origin: H$_2$ densities from line ratios}
\label{sect:H2}

In the previous section, we derived constraints on the density of electrons using the CO line ratios, where we assumed a gas disk of secondary origin, as already suggested by the CO morphology and short survival timescale \citep{Dent2014}. However, we note that the disk could still contain large amounts of unseen H$_2$, which may itself shield CO and allow it to survive since the protoplanetary phase of evolution. This would imply a primordial origin for the gas disk. 
In such scenario, H$_2$ dominates the disk gas mass and will act as the dominant collider species for CO excitation. 

Then, we can use the same line ratio analysis as in the previous subsection to estimate the H$_2$ density. Fig.\ \ref{fig:ratiotkncoll}, bottom shows the dependence of the 3-2/2-1 line ratios measured on $T_{\rm k}$ and $n_{H_2}$. Though once again for the highest line ratios the temperature dependence becomes stronger and we cannot make an accurate prediction, the large majority of the measured ratios have a weak temperature dependence. We can therefore use them to estimate the H$_2$ density, and use the PV diagram to once again obtain the radial dependence of $n_{H_2}$, shown in Fig.\ \ref{fig:neradial}, right. 
We find that if H$_2$ were to make up the bulk of the gas disk, its density would be between 10$^3$ and 10$^5$ cm$^{-3}$ at radii between 50 and 200 AU. Strictly speaking, these values should be considered upper limits, as the presence of other collider species, here unaccounted for, would lower the derived H$_2$ densities necessary to maintain the observed level of CO excitation. The H$_2$ density levels we derive are much below values observed and expected from theoretical models for protoplanetary disks \citep[e.g.][]{Boneberg2016}, proving that the gas content of the $\beta$ Pic debris disk is intrinsically different in molecular hydrogen content to a typical primordial disk, and is insufficient to allow CO survival over the system's lifetime (see discussion in Sect.\ \ref{sect:dichot}).

\subsection{Optical thickness and total CO mass}
\label{sect:optthick}
Analysing the 3-2/2-1 line ratio also allows us to set improved constraints on the total CO mass in the system. \citet{Dent2014}, having only information on the J=3-2 integrated line flux, obtained a value within a factor 2 of 2.85$\times 10^{-5}$ M$_{\oplus}$, assuming optically thin emission and excitation temperatures $T_{\rm exc}$ between 20 and 85 K. We hereby test both these assumptions and their implications on the derived CO mass. 

For optically thin emission, the excitation temperature corresponds to the kinetic temperature only if LTE applies; in NLTE the discrepancy between the two can be significant, with $T_{\rm exc}<<T_{\rm k}$. In fact, our measured disk-integrated line ratio of $1.9\pm0.3$ corresponds to an excitation temperature $T_{\rm exc}=12\pm4$ K, which may appear low, but is entirely consistent with the expectation of NLTE. In order to calculate the CO mass from a given CO integrated line flux (we here choose the J=2-1), we need to know the fractional population of the upper level of the transition, where this can be obtained for each of the $\left(T_{\rm k},n_{\rm e^-}\right)$ pairs traced by the contour of our disk-integrated line ratio in Fig. \ref{fig:ratiotkncoll} (top). The intrinsic $\left(T_{\rm k},n_{\rm e^-}\right)$ degeneracy implies that a range of CO masses will be possible.

We estimate the range of possible CO masses using Monte-Carlo methods. We assume both integrated line fluxes to follow a Gaussian probability distribution with standard deviation equal to the quoted uncertainty (which includes the absolute flux systematic added in quadrature to the noise level from the cube). For each integrated line flux, we sample this Gaussian 10$^4$ times and for each sample calculate the line ratio, which will itself be sampled 10$^8$ times. For each line ratio we calculate possible $\left(T_{\rm k},n_{\rm e^-}\right)$ pairs (i.e. draw a contour on Fig. \ref{fig:ratiotkncoll}, top) and randomly draw one of these pairs, assuming a uniform underlying $\left(T_{\rm k},n_{\rm e^-}\right)$ distribution. This pair will yield a value for the fractional population of the upper level of the transition in question ($x_2$), which then, combined with a sample of the integrated line flux ($F_{\rm 2-1}$), yields a sample for the CO mass \citep[through Eq. 2 in][]{Matra2015}. Repeating this procedure for the large number of samples drawn from the measured distributions of integrated line fluxes yields a probability distribution for the total CO mass in the disk. The CO mass range obtained is $3.4^{+0.5}_{-0.4} \times 10^{-5}$ M$_{\oplus}$, representing the $(50^{+34}_{-34})^{\rm th}$ percentiles of this distribution.

Low line ratios and hence excitation temperatures, such as observed in $\beta$ Pic, can also be symptomatic of optical thickness of the CO line \citep[e.g.][]{Flaherty2016}. Optical thickness can be tested through observations of CO isotopologues such as $^{13}$CO and C$^{18}$O, assuming interstellar isotopic ratios \citep[e.g.][]{Kospal2013}; there is however no direct measurement of these at millimetre wavelengths in $\beta$ Pic. Another possibility is to estimate it directly through its definition,
\begin{equation}
\tau_{\nu}=\frac{h\nu}{4\pi \Delta\nu}(x_lB_{lu}-x_uB_{ul}) N,
\label{eq:tauthree}
\end{equation}
where $h$ is Planck's constant, $\nu$ is the frequency of the transition in Hz, $\Delta \nu$ is the line width in Hz (a rectangular line profile is assumed), $x_u$ and $x_l$ are the fractional populations of the upper and lower energy level of the transition, respectively, $N$ is the CO column density along the line of sight in m$^{-2}$, and $B_{lu}$ and $B_{ul}$ are the Einstein B coefficients for the upward and downward transition. 
In our case, we measure $x_u$, $x_l$ and $N$ under the optically thin assumption, and use them to verify that the optical thickness is indeed $<1$. We follow this procedure in the PV spaxel where the CO flux is highest, and hence where we would expect the optical thickness to be highest. This corresponds to the location of the clump at a radial velocity of $-3.5$ km/s and midplane location of $75$ AU to the SW of the star. The line ratio measured here is $2.3\pm0.3$; from this, we can derive possible $\left(T_{\rm k},n_{\rm e^-}\right)$ combinations, and corresponding combinations of fractional populations $x_3$ and $x_2$. The column density in the spaxel can then be estimated through a modified version of Eq.\ 2 in \citet{Matra2015},
\begin{equation}
N=\frac{4\pi d^2}{h\nu_{u,l}A_{u,l}\Delta A}\frac{F_{u,l}}{x_{u}},
\label{eq:coldens}
\end{equation} 
where $d$ is the distance to the star, $A_{u,l}$ is the Einstein A coefficient of the transition, $F_{u,l}$ is the integrated line flux observed, and $\Delta A$ is the on-sky area of the line-of-sight column. Applying this column density to Eq.\ \ref{eq:tauthree}, we obtain an optical thickness $\tau_{3-2}=0.27\pm0.05$ for the 3-2 line at the clump location, where the error takes into account both the intrinsic $\left(T_{\rm k},n_{\rm e^-}\right)$ degeneracy and the observational uncertainties on the observed fluxes $F_{3-2}$ and $F_{2-1}$.
This value of the optical depth $\tau_{3-2}$ derived at the clump location can be interpreted as an upper limit to the optical depth, which is likely lower in the rest of the disk, showing that our assumption of optically thin CO emission is a good approximation to the physical conditions in the $\beta$ Pic disk. This is also supported by the absence of the SW clump in the PV diagram of the CO line ratio, as opposed to the CO J=3-2 or J=2-1 PV diagrams; this indicates that CO excitation is independent of the CO column density and hence no optical depth effects are at play.

It is valuable to note that increasing the clump CO density by an order of magnitude would have resulted in CO millimetre emission being optically thick along the line of sight to Earth. A similar argument applies for the vertical optical thickness of CO to UV light, which allows CO to self-shield against photodissociation and hence prolong his survival timescale in the disk \citep[see Sect. \ref{sect:codestr}, and][]{Visser2009}. Despite the low levels observed, if the total CO mass or the clump CO density were only an order of magnitude higher, the CO would have survived longer than an orbital timescale at 85 AU, significantly reducing the azimuthal asymmetry observed and casting significantly more uncertainty on its secondary nature.
This suggests that significant optical thickness in a second generation gas disk can be easily attained for CO production rates slightly higher than observed in $\beta$ Pictoris; as such, observations of optically thick, azimuthally symmetric CO in debris disk systems \citep[e.g. HD141569A and HD21997, ][]{White2016, Kospal2013}, should not be treated as sufficient proof for a primordial disk origin. In such cases, multi-transition observations of optically thin isotopologues are necessary to probe the H$_2$ densities in the disk (Sect. \ref{sect:H2}), and discern between the primordial and secondary origin scenarios (Sect. \ref{sect:dichot}).

\section{Discussion}
\label{sect:disc}

\subsection{Radially resolved CO clump: pointing towards the resonant migration scenario}
\label{sect:rescl}

\citet{Dent2014} brought forward two possible dynamical scenarios to explain the origin of the clumpy morphology in the CO disk. The first (explaining an azimuthal morphology like shown in Fig. \ref{fig:deprojectpvs}, right column) is a recent impact between planetary bodies, releasing dust and planetesimals that once created follow a range of orbits around the star. Crucially, all these orbits have to cross at the original impact location, where they cause enhanced collisional activity \citep{Jackson2014, Kral2015}, which is very confined spatially. This location is fixed in time, so dust and/or gas density enhancement produced should also remain stationary.

The second scenario (explaining an azimuthal morphology like shown in Fig. \ref{fig:deprojectpvs}, left column, or Fig. \ref{fig:extradeproj}, bottom) consists of a planet migrating outwards and trapping planetesimals in its 2:1 and 3:2 resonances \citep{Wyatt2003, Wyatt2006}. The accumulation, or outward sweeping, of these planetesimals at the resonance locations causes once again an increase in collisional activity, causing enhanced dust and/or cometary gas production. Due to the shape of the resonant orbits of the planetesimals, the azimuthal structure predicted consists of two clumps of unequal brightness at $\pm90^{\circ}$ azimuth with respect to planet, which in Fig. \ref{fig:extradeproj} (bottom) would be located at $\sim$+302$^{\circ}$ anticlockwise with respect to the positive $x$ axis. These clumps can be quite extended radially, depending on the eccentricities of the planetesimals that are trapped into resonance. The higher these eccentricities, which correspond to a larger extent of planet migration, the broader the region of enhanced collisional activity \citep[see e.g. Fig. 6 in][]{Wyatt2003}. In addition, the clumps are expected to move, rotating with the orbital period of the planet. 

The improved sensitivity and resolution of the new Band 6 dataset allowed us to detect CO J=2-1 emission between $\sim$50 and $\sim$220 AU. At the azimuthal location of the brightest SW clump spaxel (Fig. \ref{fig:deprojectpvs}), CO emission is detected over a range of at least $\sim$100 AU in radius, where this extent is clearly resolved at the resolution of the observations ($\sim$5.5 AU along the $x$ axis). This can mean two things: either CO is released over a range of radii, favouring eccentric planetesimals trapped in resonance in the planet migration scenario, or CO is released at a point-like location in the giant impact scenario and can spread radially in a timescale much shorter than the orbital timescale. 

In order to understand the dynamics of CO, we need to understand its surrounding environment. Molecules released from cometary ices (e.g. CO itself, see next section) are short lived due to the short photodissociation timescales, and go on to form an accretion disk that is dominated by their long-lived atomic photodissociation products (see Sect. \ref{sect:atomdom}). Therefore, production of fresh CO will happen within the atomic gas disk that is already in place. As such, regardless of the dynamics of its release, any CO produced will rapidly collide and couple to other species in the disk, losing the dynamical 'memory' of its production and following the atomic gas in Keplerian rotation around the star. To verify this, we estimate the collision timescale between CO and \ion{C}{II}, which is created via photoionisation of \ion{C}{I} produced by CO photodissociation. This can be estimated as $\tau_{\rm CO-\ion{C}{II}}=(\sigma nv)^{-1}$, where $n$ is the number density of the \ion{C}{II} gas, $v$ is the relative velocity of the collision, and $\sigma$ is the cross-section of the ion-neutral collision. We estimate the latter through \citep{McDaniel1964, Beust1989}
\begin{equation}
\sigma_{\rm CO-\ion{C}{II}}= \sqrt{\frac{\pi q^2\alpha}{\epsilon_0\mu'}}\frac{1}{v}
\end{equation}
where $q$ is the charge of the ion, $\alpha$ is the polarisability of CO, $\epsilon_0$ is the vacuum permittivity, $\mu'$ is the reduced mass of the system, and $v$ is again the relative impact velocity. For the \ion{C}{II} number density, we adopt the value predicted at 85 AU by \citet{Kral2016a} ($\sim$150 cm$^{-3}$), whereas the relative velocity cancels out through the cross-section expression.

The timescale for CO-\ion{C}{II} collision obtained is 0.19 years, much shorter than both the orbital and photodissociation timescales at 85 AU ($\sim$600 and $\sim$300 years, respectively; see Sect. \ref{sect:codestr}). Given that there are many more gas species present than just \ion{C}{II}, this should be taken as an upper limit on the collision timescale. This means that CO will quickly couple to the surrounding gas, proceeding in Keplerian rotation. It also means that to reproduce the observed radial width, if released at a point-like production location, CO must spread to a width of $\sim100$ AU before colliding with the surrounding atomic gas. For a CO molecule to travel $\sim 50$ AU in less than 0.19 years, under the simplifying assumption of a constant expanding velocity, it would have to be released (and travel) at an implausible velocity of at least $\sim$1200 km/s. Since such high CO velocities are not observed, a point-like production location such as predicted in the giant impact scenario is ruled out.
In turn, this implies that CO needs to be produced at a wide range of radii, leaving resonant sweeping due to outward planet migration as the only dynamical scenario proposed to date that can viably explain the observed CO morphology.

\subsection{The ice content of $\beta$ Pic's exocomets}
\subsubsection{Secondary molecular gas production}
\label{sect:coprod}
The augmented collisional activity produced by a dynamical scenario such as resonant trapping due to planet migration favours release of volatiles from cometary ices, either through direct release from cometary break-up \citep{Zuckerman2012} or UV photodesorption \citep{Grigorieva2007}. Thermal desorption (i.e. sublimation) of CO can be ruled out since we are far within the system's CO snow line, meaning that CO cannot last on the ice surface. On the other hand, as much as 50\% of the original CO ice content can remain trapped in other ices, e.g. water ice \citep{Collings2003}, and can therefore be released with them. Then, in the presence of a collisional cascade, volatile release is favoured through the collisions themselves or through the exposure of fresh ice that is then subject to rapid UV photodesorption. \citet{Grigorieva2007} analyse the latter effect in detail, showing that in the $\beta$ Pic disk the desorption timescale of water ice (and hence CO with it) is shorter than the collisional timescale only for particles of size below $\sim$20 $\mu$m. Hence, in the absence of direct collisional release, the majority of the ice would be retained through the top part of the collisional cascade and released mostly at the bottom of it.

Assuming water and CO ice to be well mixed (which should be a good approximation as long as gas release occurs from undifferentiated bodies of size below tens of km), they will be released into the gas phase at the same production rate through the processes discussed above. For a collisional cascade at steady state, we can estimate the rate at which solid mass (i.e. volatiles plus rock) is being lost from the cascade itself through Eq. 15 and 16 in \citet{Wyatt2008}. We take the same assumptions as in Sect. 4.4 of \citet{Marino2016}, with the additional assumptions of a uniform collision rate in a ring of planetesimals between 50 and 150 AU \citep{Dent2014} with a fractional luminosity $L_{\rm dust}/L_{\ast}$ of $2.1\times 10^{-3}$ around a star of 8.7 $L_{\odot}$ \citep[from best-fit of the SED, ][]{KennedyWyatt2014}. The resulting total mass loss rate from the planetesimal belt is 1.9 M$_{\oplus}$/Myr (or $1.2\times 10^{19}$ kg/year), where most of this release will happen in regions of higher planetesimal mass concentration, since the above loss rate scales with $(L_{\rm dust}/L_{\ast})^2$ (and hence with $M_{\rm belt}^2$). Using the fraction of this solid mass that is composed of CO ice we can obtain the release rate of CO gas, regardless of the choice of release mechanisms discussed above. 

We note that there is a considerable uncertainty on this number, since this calculation relies on assumptions about the unknown planetesimal strength, mean planetesimal eccentricities, and particle velocities \citep[e.g.][]{Marino2016}. In addition, the resonant scenario invoked to explain the clumpy CO morphology would imply a much higher mass loss rate than calculated above, particularly at the clump location. As such, we interpret the value of 1.9 M$_{\oplus}$/Myr as a lower limit estimate to the solid mass loss rate in the disk.

\subsubsection{Secondary molecular gas destruction}
\label{sect:codestr}
On the other hand, the dominant destruction mechanism is photodissociation from the interstellar UV radiation field (ISRF), which will dominate over the stellar UV light at all radii where CO is detected in the $\beta$ Pic disk \citep{Kamp2000}. The photodissociation timescale in an environment that is optically thin to the UV ISRF \citep[][]{Draine1978} is $\sim$120 years \citep[as assumed for $\beta$ Pic in][]{Dent2014}. It is worth noting that the \citet{Kral2016a} model best fits the atomic gas observations for an ISRF 60 times stronger than this, in order to explain the observed high carbon ionisation fraction in the disk. However, a similar ionisation fraction could be attained through effects not taken into account by the model, such as X-ray emission and enhanced UV emission from the central star. These may significantly affect the \ion{C}{I}/\ion{C}{II} levels without altering the CO photodissociation timescale, since X-rays do not induce CO photodissociation and stellar UV will mostly affect the CO-poor but C-rich inner disk regions. As such, we here assume a standard level for the UV ISRF.

The optically thin dust and the low H$_2$ levels (see next section) are not sufficient to attenuate this UV radiation field. However, in the presence of CO column densities larger than $10^{13}$ cm$^{-2}$, the molecule can self-shield leading to much longer photodestruction timescales \citep{Visser2009}. The vertical column density at the clump location (i.e. brightest spaxel in the PV diagram) can be estimated through Eq. \ref{eq:coldens}, where now the area $\Delta A$ of the column in question is the area in the orbital plane (Fig. \ref{fig:deprojectpvs}) corresponding to the spaxel size at that PV diagram location (Fig. \ref{fig:pvs}). The CO column density at the clump location through half of the vertical extent of the disk is $\sim(1.6-1.7) \times 10^{15}$ cm$^{-2}$, leading to a photodissociation timescale that is $\sim$2.5 times longer than for the unshielded case \citep[see Table 6 in][]{Visser2009}. 

In addition to CO self-shielding, we consider UV attenuation in the vertical direction by the \ion{C}{I} ionisation continuum, which needs to be taken into account since it acts at the same UV wavelengths as CO photodissociation. We calculate its effect using Eq. 4 in \citet{Rollins&Rawlings2012} and ionisation cross sections from \citet{vanDishoeck1988}. We estimate the \ion{C}{I} vertical column density at the clump radial location from the \citet{Kral2016a} thermodynamical model, yielding a value of $2.8 \times 10^{15}$ cm$^{-2}$. This results in a photodissociation rate that is $\sim$0.96 times the unshielded rate, showing that at 85 AU vertical shielding of CO by the \ion{C}{I} ionisation continuum is minor compared to CO self-shielding. Using these shielding factors, we estimate a photodissociation timescale in the clump midplane of $\sim$300 years, which together with our measured total mass of CO in the disk yields a CO mass loss rate of 0.11 M$_{\oplus}$/Myr (or $6.8\times 10^{17}$ kg/year).

This is under the simplifying assumption of a photodissociation timescale that is constant across the disk. We note that in reality this timescale will decrease down to the unshielded value as we vertically move away from the midplane. However, most CO mass loss will occur at low heights, where most of the CO mass lies (assuming a vertical Gaussian for the CO number density distribution). This means that in the region where most mass loss occurs, the vertically averaged CO mass loss rate is actually well approximated by the midplane value ($<10\%$ decrease). As well as in the vertical direction, the photodissociation timescale will also decrease away from the clump in the azimuthal direction: if we assume that the factor $\sim5$ drop in flux from the peak of the clump to the end of the tail also represents a factor $\sim5$ drop in CO mass and vertical column density, this corresponds to a factor $\sim1.8$ drop in photodissociation timescale. That means that the mass loss rate will decrease by a factor $5/1.8\sim2.8$ across the CO tail, meaning that the average rate would be $\sim$32\% lower than that measured at the CO clump.

The above considerations only take into account shielding of CO along the vertical direction. In reality, at a given disk location, shielding of interstellar UV radiation over the entire 4$\pi$ solid angle should be considered. This is higher than shielding along the vertical direction alone, as for example the radial optical depth to UV photodissociating radiation is higher than the vertical one, due to a CO column that we can estimate to be $\sim3$ times longer (assuming a Gaussian radial distribution at the clump location), yielding a higher column density in the radial direction than in the vertical direction. This effect implies that the average shielding over the entire 4$\pi$ solid angle is higher, yielding a longer disk-averaged photodissociation timescale, than calculated in the vertical direction alone. Overall, we expect the effect of a non spatially uniform photodissociation timescale and of the increased radial optical depth to change our mass loss rate estimate at most by a factor of a few below and above the quoted value, respectively.

We can then relate our estimate of the photodestruction timescale to the observed azimuthal extent of the CO tail. In the resonant migration picture (e.g. Fig. \ref{fig:extradeproj}, bottom), the two clumps on the SW and NE sides of the disk are rotating at the Keplerian angular velocity of the inner planet that is causing the resonance \citep[assumed to be located at 60 AU, as in][]{Dent2014}, with the CO lagging because of its slower Keplerian angular velocity at 85 AU. The length of the tails can then be calculated as the azimuthal displacement undertaken in one photodissociation timescale at an angular velocity equal to the difference between that of the inner planet at 60 AU and that of gas in Keplerian rotation at 85 AU. Despite the uncertainties in the photodissociation timescale calculation and in the radial location of the planet, we obtain a value of 126$^{\circ}$, which is in good agreement with the deprojection in Fig. \ref{fig:extradeproj}, bottom.
Therefore, a planet located near the inner edge of the disk can feasibly explain the observed CO azimuthal structure for a disk inclination of $\sim$86$^{\circ}$ from face-on.

\subsubsection{Steady state abundances and the impact of CO$_2$}
Unless we are witnessing the system shortly after a stochastic event, which is unlikely, we can assume CO gas to be in steady state, with its production rate matching the destruction rate. Since the production rate of CO is the total mass loss rate from the collisional cascade (derived in Sect. \ref{sect:coprod}) multiplied by the CO mass fraction in the ice, we can obtain the latter by dividing the destruction rate (derived in Sect. \ref{sect:codestr}) by the production rate. Given that the production rate is a lower limit, we obtain an upper limit of 6\% on the CO mass abundance in $\beta$ Pic's exocomets. However, we note that additional CO can be produced through the CO$_2$ molecule, which has a similar ice abundance as CO in Solar System comets \citep[e.g.][]{Mumma2011}. Whether CO$_2$ ice can survive as ice in the $\beta$ Pic disk depends on the temperature of the solid bodies, which in turn depends on their size, composition and radial location. The timescale for a CO$_2$ molecule to sublimate from a comet surface is shorter than an orbital timescale for temperatures $>$50-55 K \citep{Sandford1990}, which matches with the equilibrium temperature of a blackbody at the radial location of the clump (85 AU). Therefore, we predict that in the inner disk regions all surface CO$_2$ will have disappeared, leaving some of it trapped in other ices (e.g. water ice). This, like CO, may then be released through collisions or H$_2$O photodesorption. In the outer regions, on the other hand, CO$_2$ ice can survive on the surface of grains, and may be released again through collisions or photodesorption of CO$_2$ ice itself. Near the radial location of the clump, sublimation of surface CO$_2$ itself may also be contributing to produce CO$_2$ gas.

CO is rapidly created via photodissociation of CO$_2$ gas \citep[e.g. ][]{vanDishoeck1988} and/or photodesorption of CO$_2$ ice \citep[e.g. ][]{Oberg2009}. 
The photodissociation timescale of CO$_2$ gas is considerably faster than that of CO in $\beta$ Pic. In an optically thin medium, using calculated cross sections \citep{Hudson1971, LewisCarver1983}, the interstellar radiation field as expressed in Eq. 3 in \citet{vanDishoeck1988}, and the stellar radiation field obtain from fit to the SED, we calculate a CO$_2$ photodissociation timescale of 0.35 years at 85 AU, $\sim$100 times shorter than the CO photodissociation timescale. While we are not aware of whether CO$_2$ can self-shield significantly, we suggest that to first approximation this can be ignored, since too high a column density would have to be produced in a very short timescale for CO$_2$ to be able to shield itself. Additionally, UV photodesorption of CO$_2$ ice can also directly contribute to CO release, since experiments show that 20-50\% of CO$_2$ ice will be desorbed directly as CO \citep{Oberg2009}.
This strengthens our assumption that any CO$_2$ gas released is rapidly turned into CO, meaning that the CO gas observed can be interpreted to derive a collective CO+CO$_2$ cometary abundance, which we measure to be at most 6\%.

\subsubsection{Comparison with other cometary ices}
In the Solar System, an ice to rock ratio of $\sim$1 is generally assumed, and the (CO+CO$_2$)/H$_2$O abundance ratio observed ranges between 2.4 and 60\% \citep{Mumma2011}, yielding a CO+CO$_2$ mass abundance in the comets of 2.6 to 27.2\%. Therefore, the 6\% upper limit mass abundance of CO+CO$_2$ in $\beta$ Pictoris is consistent with Solar System cometary compositions, suggesting a broadly similar carbon reservoir within the cometary ices of both planetary systems, despite the differences in age and stellar type. On the other hand, compared to the only other extrasolar system where the presence of secondary gas has been confirmed \citep[around the coeval star HD181327,][]{Marino2016}, $\beta$ Pic's exocomets have a CO+CO$_2$ content that is at most $\sim$20 times higher (for the same assumptions in the calculation). While the exact spatial distribution of CO in HD181327 is unknown, the observations suggest that it should be co-located with the dust ring at a radius of $\sim$86 AU from the star, very similar to the radial location of the clump in the $\beta$ Pic disk. As the two stars belong to the same moving group, we assume that they formed from the same molecular cloud, which may suggest they should retain similar molecular reservoirs (which this remains observationally unconstrained). Different abundances, however, may arise if different molecule formation and/or freeze-out mechanisms have been at play in the original protoplanetary disks, which may in turn relate to the different stellar hosts.
However, any such conclusion is highly dependent on the assumptions taken in our cometary abundance calculation, as pointed out before. Though the same assumptions on unknown quantities have been made for both disks, their values may actually differ significantly between the two systems, altering our conclusions on the relative CO+CO$_2$ ice abundance. Nonetheless, it is a valuable result that the abundances obtained give reasonable estimates compared to the Solar System, and shows that a new era is beginning where we can start setting Solar System comets into the wider context of extrasolar cometary bodies.

\subsection{An atom-dominated secondary disk}
\label{sect:atomdom}

Given the short photodissociation timescales, molecules in low gas mass environments such as secondary gas-bearing debris disks are rapidly destroyed to form atoms.
For the $\beta$ Pic disk, a recent model has been put forward by \citet[][]{Kral2016a} that is consistent with all far-IR/sub-mm atomic line observations of the system. The model traces the thermodynamical evolution of the atoms released by CO photodissociation, and shows how these rapidly spread viscously from the production point (i.e. the CO clump) to form an axisymmetric accretion disk. Because of the ionisation of carbon caused by the strong stellar and interstellar UV field, the disk is required to have a relatively high viscosity $\alpha>0.1$, which may be explained by the onset of the magneto-rotational instability \citep[MRI, ][]{Kral&Latter2016}. Carbon ionisation is the dominant heating mechanism in the disk, and produces a considerable amount of electrons, which then act as an efficient collider species to excite the CO molecule and produce the line ratios observed by ALMA (see Sect. \ref{modelec}). 

Our results in Fig.\ \ref{fig:neradial} show that the radial dependence of the electron density predicted by this model is consistent with that derived in this work. While on the other hand the absolute value of the electron density might appear underpredicted by a factor of a few, we caution the reader that our derivation of the electron density from the observations assumes electrons to be the \textit{only} collisional partners that contribute to the excitation of CO. Other species might contribute as well, but the lack of calculated cross sections for collisions between CO and any species that are abundant in the disk (e.g. atomic carbon and oxygen) prevents us from being able to make quantitative predictions \citep[see discussion in][]{Matra2015}. In general, the influence of additional collisional partners would lower our measured $n_{\rm e^-}$, meaning that the model is still consistent with our observations.

Another test for the model is our estimate of the disk scale height (Sect. \ref{sect: structedgeon}).
The gas temperature at radii where the CO is detected was predicted by the model to have a value of $\sim$52 K at 85 AU and decrease as $T\sim R^{-0.8}$, which for a vertically isothermal gas dominated by C and O ($\mu$=14) translates into a scale height of 3.6 AU at 85 AU, increasing as $H\sim R^{1.1}$. As estimated in Sect. \ref{sect:rescl}, CO should rapidly collide and couple with the atomic gas, meaning that we would expect a similar scale height for the CO and atomic disks. However, our results show that CO has a scale height that is twice larger than predicted by the model (7 AU vs 3.6 AU at 85 AU radius) and that it has a shallower radial increase (as $\sim R^{0.75}$ vs $\sim R^{1.1}$). This difference cannot be explained by the CO disk being decoupled from the atomic disk, since the increase in molecular mass would make the discrepancy even larger. As well as a geometrical effect due to a non-negligible inclination of the disk to the line of sight, which is degenerate with the disk's vertical structure, two physical reasons can be invoked to explain this discrepancy: 
\begin{itemize}
\item The gas is not vertically isothermal. This is to be expected if the gas becomes vertically optically thick to \ion{C}{I}-photoionising UV radiation, which is the main heating process in the disk \citep{Kral2016a}. Then, the upper disk layers would be warmer than the disk midplane, which would naturally produce a wider vertical distribution than in the vertically isothermal case. This is also suggested by the vertical distribution of \ion{Fe}{I} and \ion{Ca}{II} gas, which is in fact even broader than that inferred for CO here \citep[the scale height value derived for \ion{Fe}{I} is $\sim$16 AU at 85 AU radius, for a near exponential vertical profile;][]{Nilsson2012}.
\item The excitation of CO, and hence the electron density, is not vertically uniform. In our analysis, we assumed the observed vertical flux distribution (after deconvolution by the instrumental resolution) to be a good tracer of the underlying CO mass distribution. However, if a vertical gradient in electron density is present in the disk, which once again would be expected if \ion{C}{I} photoionisation preferentially takes place in the upper layers, the flux vertical distribution observed would appear broader than the true CO mass distribution, skewing our scale height measurement to higher values.
\end{itemize}
Furthermore, if the vertical optical thickness to \ion{C}{I}-photoionising UV radiation decreases as a function of radius, the effects above would be stronger in the disk inner regions compared to the outer regions. This would then naturally explain the shallower radial dependence of the scale height measured here.

\subsection{Solving the primordial vs secondary origin dichotomy in gas-bearing debris disks}
\label{sect:dichot}
Just like $\beta$ Pic, an increasing number of young, $<$100 Myr-old debris disks has recently been detected in molecular and/or atomic gas lines. This phenomenon had only been observed for disks around A stars until recently, when CO in the debris belt around the late-F star HD181327 (member of the same moving group, and hence coeval with, $\beta$ Pictoris) was detected by ALMA \citep{Marino2016}.
To explain these discoveries, a dichotomy has emerged between the primordial and secondary gas origin scenarios, where the disk is viewed, respectively, as protoplanetary disk where gas dispersal has been delayed \citep[as proposed for HD21997 and HD141569,][]{Kospal2013, Flaherty2016} or as a debris disk where second-generation gas is released by planetesimals through a steady-state collisional cascade and/or a relatively recent stochastic event.

In Sect.\ \ref{sect:H2} we showed how optically thin CO line ratios can be used to probe the density of collisional partners in a gas-bearing debris disk. As discussed in \citet{Matra2015}, at least three optically thin lines are needed to completely solve the collider density-kinetic temperature degeneracy; these do not have to be from the same species, as long as the species are located in the same disk regions and therefore are subject to similar collider densities and temperatures. 
From Fig.\ \ref{fig:ratiotkncoll} we can deduce that if the collider density is high enough (towards the right in the plots), LTE applies and the line ratios become dependent solely on the kinetic temperature, meaning we are unable to probe the collider density in the disk. Nonetheless, we can set a lower limit to it, which will correspond to few times the critical density $n_{\rm crit}$ of the colliders, since by definition LTE applies if $n_{\rm coll}\gg n_{\rm crit}$. If however collider densities are lower, we are in the NLTE regime and we can estimate the collider density directly, either through observations of 3 lines or by making reasonable assumptions on the kinetic temperature, as we did in the case of $\beta$ Pic.

This method is powerful as it can be used to test the presence of otherwise undetectable collider species, such as H$_2$ gas, in the system. Since this species dominates the disk mass in a protoplanetary disk, its presence should be dominant in gas-bearing debris disks of primordial origin. H$_2$ is an extremely volatile molecule and unlike heavier species does not freeze-out onto grains \citep[e.g.][and references therein]{Sandford1993}, meaning that it cannot survive protoplanetary disk dispersal by freezing out on cometary bodies. Therefore, its presence is not expected at all in debris disk gas of secondary origin. However, if we assume a disk to be primordial, H$_2$ will dominate collisions with CO, and in turn the CO line ratio analysis should consistently yield H$_2$ densities typical of protoplanetary disks. In the case of $\beta$ Pic, under this assumption, we derive H$_2$ densities of order $10^3-10^5$ cm$^{-3}$. 

The only direct measurement of the bulk H$_2$ content in a protoplanetary disk comes from the detection of HD in the TW Hydrae protoplanetary disk \citep{Bergin2013}. This detection, combined with detailed models of the disk's physical and chemical structure, yielded midplane H$_2$ densities of order $10^7-10^{10}$ cm$^{-3}$. These are considerably higher than the values we derived for $\beta$ Pic, and crucially are well in the LTE regime of CO excitation. Thus, detection of optically thin CO lines that are excited subthermally (i.e. are out of LTE) like in $\beta$ Pic is already proof that we are looking at environments that are H$_2$-depleted by several orders of magnitude. A simple estimate of the vertical column density can then be obtained by assuming a uniform column of length 15 AU (equivalent to about twice the vertical scale height at the clump location in the $\beta$ Pic disk, Sect. \ref{sect: structedgeon}) and our upper limit of $10^5$ cm$^{-3}$ on the H$_2$ density. This yields a column density of $\sim$10$^{19}$ cm$^{-2}$, which is not sufficient to shield the CO from UV photodestruction over the age of the system \citep{Visser2009}. Therefore, even in the presence of low H$_2$ levels, CO cannot have survived since the protoplanetary phase of evolution, and must therefore be replenished through exocometary volatiles. This provides additional and now definitive confirmation of the secondary origin of gas in the $\beta$ Pictoris disk. Furthermore, we here propose this analysis of multi-transition observations of optically thin, subthermally excited CO as a fundamental way to indirectly probe the H$_2$ density in other gas-bearing debris disks, hence solving the primordial versus secondary origin dichotomy.

\section{Summary and Conclusions}
In this work, we presented new ALMA Band 6 observations of the CO J=2-1 line in the $\beta$ Pictoris debris disk, and combined them with archival Band 7 observations of the CO J=3-2 line to derive the 3D morphology, excitation conditions and total mass of the CO gas. We reach the following conclusions:

\begin{itemize}
\item We confirm the presence of the CO clump discovered by \citet{Dent2014}, peaking at radius of 85 AU and an azimuth of $\pm$32$^{\circ}$ in the orbital plane of the disk (SW on-sky). At the 0$\farcs$3 resolution of the new dataset, we conclude that the clump is radially extended, spanning $\sim$100 AU in orbital radius. Since CO should rapidly couple to the atomic gas disk already in place and quickly proceed in Keplerian rotation around the star, it does not have time to spread radially and must be produced at a broad range of radii. This rules out a giant impact between planetary bodies as a possible dynamical scenario to explain the clumpy morphology, leaving resonant trapping by outward planet migration as the only viable mechanism proposed so far. This scenario is also consistent with the observed azimuthal structure for an assumed disk inclination of $\sim$86$^{\circ}$, and can be tested by looking for the clump's predicted orbital motion.
\item The CO disk is vertically tilted, i.e. its PA presents an extra anticlockwise rotation compared to the main disk observed in scattered light. The degree of misalignment is in line with that inferred for the warp observed in scattered light imaging of the disk, suggesting a common origin between the CO clump and the scattered light warp.
\item Under the assumption of a vertically isothermal disk, the scale height measured is a function of orbital radius (increasing as $H \sim R^{0.75}$ with a value of 7 AU at the clump radial location). As observed for metallic atoms, this implies temperatures that are higher than predicted by thermodynamical models \citep{Kral2016a}. Aside from a geometrical effect due to our edge-on assumption, this discrepancy can be attributed to the fact that the disk is likely not vertically isothermal and/or to a potential higher electron abundance in the disk surface layers, where both effects are expected if the disk is vertically optically thick to C-photoionising UV photons. 
\item The CO J=3-2/J=2-1 line ratio shows a clear gradient with radial velocity in the PV diagrams, which is consistent with a power law decrease with orbital radius. Through NLTE modelling, we attribute this to a radial decrease in the electron density in the disk, in line with model predictions. No enhancement at the clump location nor a significant NE/SW asymmetry are observed, confirming that CO lies in an azimuthally symmetric, atom-dominated secondary disk.
\item Repeating the NLTE analysis above under the assumption of an H$_2$-dominated primordial disk, we derive H$_2$ densities that are many orders of magnitude lower than expected for protoplanetary disks. These imply column densities that are too low to shield CO and allow it to survive over the age of the system, providing further proof that CO needs to be replenished and hence the gas disk must be of secondary origin. We propose such measurement of H$_2$ densities from subthermally excited, optically thin CO lines as a fundamental way to solve the primordial vs secondary dichotomy in other gas-bearing debris disks.
\item Using both CO transitions, we refine the total CO mass measurement to $3.4^{+0.5}_{-0.4} \times 10^{-5}$ M$_{\oplus}$. We find that the vertical CO and \ion{C}{I} column densities are sufficient to partially shield CO from UV light, prolonging its survival timescale to $\sim$300 years, which is still much shorter than both the system age (hence requiring replenishment from cometary ices) and the orbital period (hence producing the observed clumpy morphology and NE/SW asymmetry). Assuming CO gas to be in steady state, being released from ices through the collisional cascade and then rapidly photodestroyed, we estimate the CO+CO$_2$ mass abundance in the comets to be at most 6\%. This is still consistent with cometary abundances measured in both our own Solar System and in coeval HD181327 system.

\end{itemize}

\section*{Acknowledgements}

The authors would like to acknowledge D\'aniel Apai for providing the disk spine from HST observations, and Grant Kennedy for providing the SED fitting parameters. LM acknowledges support by STFC and ESO through graduate studentships and, together with MCW and QK, by the European Union through ERC grant number 279973. Work of OP is funded by the Royal Society Dorothy Hodgkin Fellowship, and AMH gratefully acknowledges support from NSF grant AST-1412647. This paper makes use of the following ALMA data: ADS/JAO.ALMA\#2012.1.00142.S and ADS/JAO.ALMA\#2011.0.00087.S. ALMA is a partnership of ESO (representing its member states), NSF (USA) and NINS (Japan), together with NRC (Canada), NSC and ASIAA (Taiwan), and KASI (Republic of Korea), in cooperation with the Republic of Chile. The Joint ALMA Observatory is operated by ESO, AUI/NRAO and NAOJ.

\bibliographystyle{mnras}
\bibliography{lib_bpic}


\appendix

\section{Comparing the ALMA datasets}
\label{app:3}

In order to allow for direct comparison between the CO J=2-1 and J=3-2 data cubes, both need to have matching pixel and channel size, as well as spatial and velocity resolution. We therefore re-CLEANed the Band 6 dataset to achieve the coarser pixel size, spatial beam and channel size of the Band 7 dataset. 
While the CLEAN task within CASA allows us to choose the channel size of the restored data cube, it does so through simple spectral rebinning, and as such does not change the velocity \textit{resolution} of the dataset (or in other words, its spectral response function). The native spectral response function of ALMA data is a Hanning-smoothed sinc function of width (resolution) equal to twice the native channel width. This function is well approximated by a Gaussian of equal FWHM (ALMA Helpdesk, private communication). Therefore, as well as rebinning of channels at the CLEANing stage, convolution by a Gaussian of FWHM equal to the quadratic difference between the target resolution and the native resolution was necessary to allow spectral comparison of the datasets.

The stellar position from \textit{Hipparcos} was corrected for proper motion separately between datasets, and the final images rotated by the position angle of the main scattered light dust disk \citep[29.3$^{\circ}$,][]{Lagrange2012}, aligning the disk midplane to the horizontal on-sky spatial axis $x_{\rm sky}$ (see Fig.\ \ref{fig:mom0}). 
The astrometric accuracy of an ALMA dataset is based on the phase tracking accuracy of the array; this depends on several factors, including weather, observing band, accuracy in the position of each antenna, distance of the source to the phase calibrator, and frequency of phase calibration measurements. We quantify this as advised in the ALMA Knowledgebase\footnote{\url{https://help.almascience.org/index.php?/Knowledgebase/Article/View/319/0/what-is-the-astrometric-accuracy-of-alma}}, and assume a 1$\sigma$ RMS astrometric accuracy of  $\sim$0$\farcs$037 for the Band 6 and $\sim$0$\farcs$029 for the Band 7 dataset. 

In order to remove potential systematics in the ALMA phase tracking, we recenter the data cubes using the location of the star, which is itself detected in both continuum datasets. Since emission at its location includes a contribution from the disk itself, which may bias our procedure, we determined its position through global fitting of disk+star models in each dataset, whose detailed description will appear in forthcoming work \citep{Matrainprep}. In the Band 6 dataset the best-fit stellar offset from the image centre was found to be ($-0\farcs0084$, $0\farcs034$) along the midplane and perpendicular to it, respectively; this offset was ($0\farcs24$, $0\farcs12$) for the Band 7 dataset, which was likely affected by systematics during the Science Verification Cycle 0 campaign of ALMA.

\section{Fitting the disk vertical structure}
\label{app:2}
The Gaussian fitting procedure employed to derive the vertical structure of the disk takes into account the following uncertainties:
\begin{itemize}
\item The uncertainty $\Delta S_{\rm CO}$ on the observed line flux at each pixel, i.e. the RMS of the images in Fig.\ \ref{fig:mom0}.
\item The uncertainty $\Delta y'_{\rm sky}$ on the the pixel location along the direction which is perpendicular to the disk midplane. Let us define the \textit{rotated} pixel reference frame to be centred on the stellar position, with $x'_{\rm sky}$ along the disk midplane and $y'_{\rm sky}$ perpendicular to it, and with both having a positive sign towards the SW direction. The position in this $\left(x'_{\rm sky},y'_{\rm sky}\right)$ reference frame was obtained from the original \textit{unrotated} reference frame (also centred on the star, with $x_{\rm sky}$ positive towards East and $y_{\rm sky}$ positive towards North), through an anticlockwise rotation by an angle of $90^{\circ}-\theta_{\rm PA}$. Thus, $y'_{\rm sky}$ can be expressed as a function of $x_{\rm sky}$, $y_{\rm sky}$ and $\theta_{\rm PA}$ simply through
\begin{equation}
y'_{\rm sky}=sin(90^{\circ}-\theta_{\rm PA})x_{\rm sky}+cos(90^{\circ}-\theta_{\rm PA})y_{\rm sky}
\end{equation}
This means that uncertainties in the PA (29\fdg3$^{+0.2}_{-0.3}$), $x_{\rm sky}$ and $y_{\rm sky}$
\ will propagate as an uncertainty in y' through
\begin{equation}
\begin{split}
\Delta y_{\rm sky}'^2=sin^2(90^{\circ}-\theta_{\rm PA})(\Delta x_{\rm sky})^2+cos^2 (90^{\circ}-\theta_{\rm PA}) (\Delta y_{\rm sky})^2+ \\
+ (-x_{\rm sky} cos (90^{\circ}-\theta_{\rm PA}) +y_{\rm sky} sin (90^{\circ}-\theta_{\rm PA}))^2(\Delta \theta_{\rm PA})^2
\end{split}
\end{equation}
The uncertainties $\Delta x_{\rm sky}$ and $\Delta y_{\rm sky}$ are given by the astrometric accuracy of the instrument (see Appendix \ref{app:3}), summed in quadrature with the uncertainty on the star's proper motion ($\sim$0\farcs004).
\end{itemize}
The fit was carried out using the affine invariant Markov-Chain Monte-Carlo (MCMC) Ensemble sampler from \citet{GoodmanWeare2010}, implemented through the Python emcee package \citep{Foreman-Mackey2013}. The likelihood function of the data given a model 1D Gaussian was defined through the $\chi^2$ distribution, taking into account uncertainties in both flux and vertical position ($\Delta S_{\rm CO}$ and $\Delta y'_{\rm sky}$). The Gaussian vertical centre $y_{\rm obs}$, FWHM and peak $\rm \rm max(S_{\rm CO, model})$ were left as free parameters, and uniform, uninformative priors set as $|y_{\rm obs}|  <$ 20 AU, $0<\rm max(S_{\rm CO, model})<2.0\times \rm max(S_{\rm CO, observed})$  and FWHM$_{\rm beam}<$FWHM$<80$AU, where FWHM$_{\rm beam}$ is the resolution of the data projected onto the perpendicular to the disk midplane. For each vertical cut, the posterior probability distribution of the Gaussian parameters, given the data, was sampled using 300 walkers through 500 steps. The resulting best-fit Gaussians are shown in the top panel of Fig.\ \ref{fig:vertprofs} as the black dashed lines. Their best-fit vertical centres $y_{\rm obs}$ and FWHM are plotted as a function of midplane location in the central and bottom panel of Fig.\ \ref{fig:vertprofs}. The error bars show the 1$\sigma$ error obtained from the posterior probability distribution of $y_{\rm obs}$ as sampled through the MCMC.

\section{From PV diagrams to orbital plane}
\label{app:1}
Assuming gas in a flat circular disk in Keplerian rotation around the star, the magnitude of its orbital velocity will be given by $\sqrt \frac{GM_{*}}{R}$, where $G$ is the gravitational constant, $M_{*}$ is the stellar mass, and $R$ is the radial distance from the star; its direction will be perpendicular to the radius vector. In Cartesian coordinates (centred on the star), the orbital velocity along the $y$ axis can then be expressed as $v_{y}=\sqrt \frac{GM_{*}}{R}\cos\theta$, where $\theta$ is the angle between the $x$ axis and the radius vector. If we then rotate the disk about the $x$ axis by an angle $i$ corresponding to its inclination to the plane of the sky, its velocity perpendicular to the plane to the sky (i.e. its radial velocity), will be given by $v_{\rm rad}=\sqrt \frac{GM_{*}}{R}\cos\theta\sin(i)$. By definition, in the orbital plane $x=R\cos\theta$, leading to $v_{\rm rad}=\sqrt \frac{GM_{*}}{R^3}\sin(i)x$. Since $x$ is the axis of rotation when transforming from the orbital to the sky plane, the $x$ component of both position and velocity remains unchanged (i.e. $x=x_{\rm sky}$). This has two important consequences: 
\begin{enumerate}
\item Since the PV diagram is by definition a map in $\left(x_{\rm sky},v_{\rm rad}\right)$ space, knowing the inclination angle $i$ means that each PV location traces a certain disk radius $R$. In fact, for a given inclination, each orbital radius defines the slope of a unique straight line in PV space (see Fig.\ \ref{fig:pvs} and \ref{fig:pvratios}). This can be directly used as a probe of the radial extent of the disk. However, the accuracy in measuring $R$ is limited by the finite spectro-spatial resolution of the instrument; this is particularly important at $|v_{\rm rad}|\sim0$ (i.e., along the line of sight to the star), where it becomes impossible to disentangle contributions from different orbital radii.
\item As $R$ and $x$ in the orbital plane are now known, a $y$ position can immediately be derived through $y=\pm\sqrt{R^2-x^2}$. Due to the sign, however, there remains a degeneracy; this can only be solved if the disk is \textit{not} edge-on, as in that case we have an extra observable, the vertical on-sky location, that allows to break the degeneracy. As applies for $R$, derivation of $y$ is severely limited near the line of sight to the star.
\end{enumerate}

Therefore, assuming a flat, circular Keplerian disk and an inclination angle, PV diagrams can be used to map disk emission in its orbital plane (see e.g. Fig.\ \ref{fig:deprojectpvs}). 

\section{Breaking the deprojection degeneracy using the observed vertical structure}
\label{app:4}
Combining the methods outlined in Appendix \ref{app:2} and \ref{app:1} allow us to obtain, for a choice of inclination $i$ and on-sky tilt angle dPA, parameters of the observed best-fit sky-projected vertical Gaussian (centre $y_{\rm obs}$, standard deviation $\sigma_{\rm obs}$ and peak vertical flux $F_{\rm peak}$) at each ($x,\pm y$) location in the disk orbital plane. If the disk is perfectly edge-on ($i=90^{\circ}$), the degeneracy in the sign of $y$ cannot be broken, as emission coming from in front and behind the sky plane will lie at the same sky-projected vertical location $+y_{\rm sky, orb}=-y_{\rm sky, orb}=0$. If the disk is sufficiently inclined to the line of sight ($i<90^{\circ}$), however, emission coming from behind and in front the sky plane will lie at different vertical locations on the sky plane, on opposite sides of the midplane (i.e. $+y_{\rm sky, orb}\neq-y_{\rm sky, orb}$). 
If ($i\ll90^{\circ}$), the distance between the two will be large enough for them to be resolved, if both the disk physical vertical thickness and the instrumental resolution allow. If $i\rightarrow90^{\circ}$, as is the case for $\beta$ Pictoris, the two sides will not be vertically resolved. 

\begin{figure}
\vspace{-3mm}
 \hspace{-4mm}
  \includegraphics*[scale=0.27]{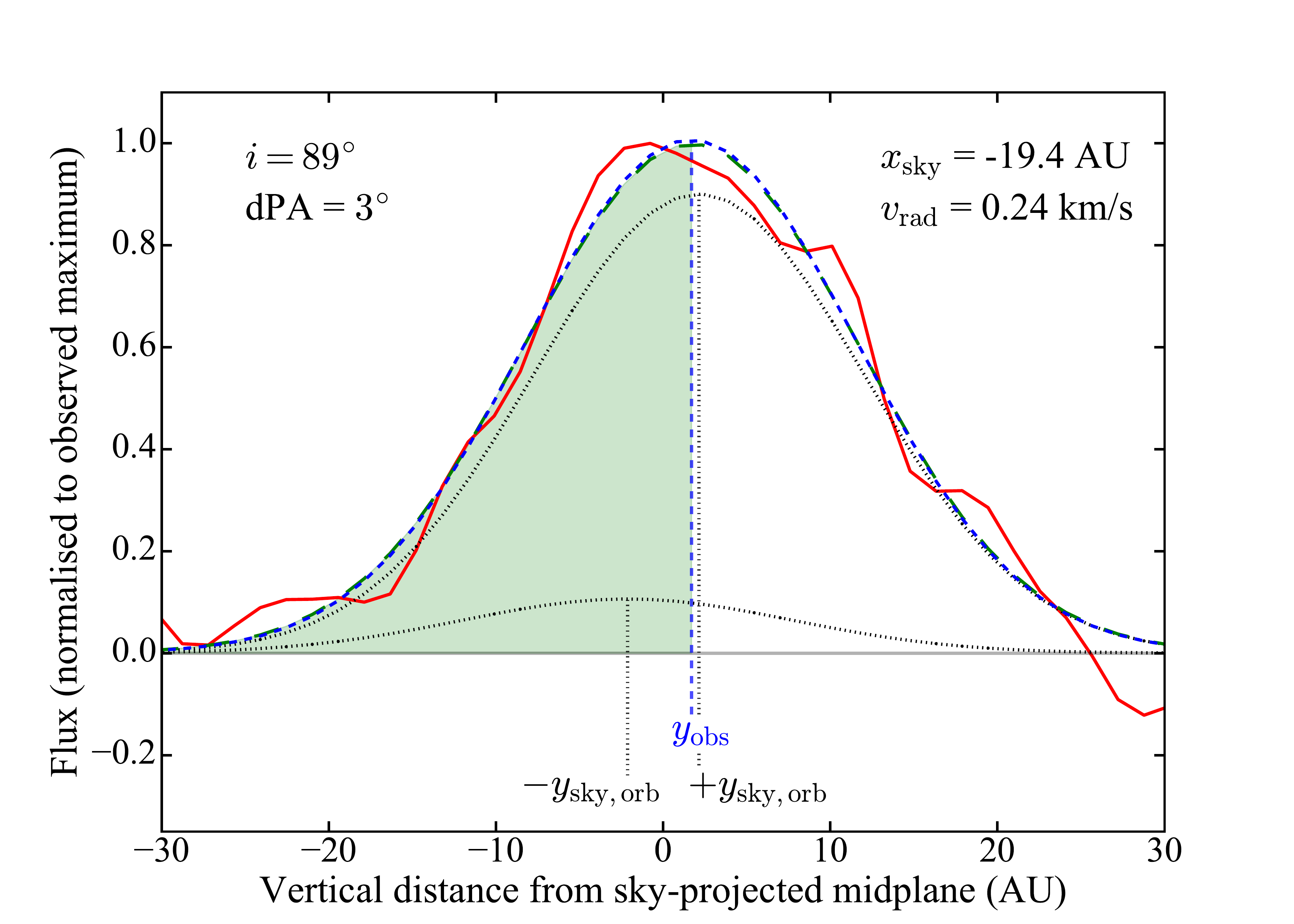}
\vspace{-3mm}
\caption{Graphical representation of the procedure employed to recover the flux fraction originating from above (at $+y_{\rm sky, orb}$) and below the disk on-sky midplane (at $-y_{\rm sky, orb}$). Here we show an example for an orbit inclined by 89$^{\circ}$ to the line of sight, for a disk that is tilted in the plane of the sky by 3$^{\circ}$, at a location of -19.4 AU along the disk sky-projected midplane, and radial velocity of +0.4 km/s with respect to the star. See main text for details of the procedure.}
\vspace{0mm}
\label{fig:fitexample}
\end{figure}

However, if the disk is azimuthally asymmetric, the front and back of the disk will contribute to the observed emission in unequal amounts, causing a vertical shift in the centroid of the observed emission (red solid line in Fig. \ref{fig:fitexample}) and consequently also in the centre of the best-fit vertical Gaussian $y_{\rm obs}$ (blue dashed line) towards either + or $-y_{\rm sky, orb}$, whichever contributes most to the emission. We here aim to use this shift to retrieve the fraction of the flux originating at +$y_{\rm sky, orb}$ and -$y_{\rm sky, orb}$, and hence in front and behind the sky plane at $\pm y$. We note that a degeneracy is still present in that we do not know whether CO above the on-sky midplane is in front or behind the sky plane, and viceversa for CO observed below the midplane. 

For simplicity, we assume that the two flux contributions we are trying to find take the form of Gaussians with the same standard deviation $\sigma_{\rm obs}$ as that of the best-fit vertical Gaussian, but centred at +$y_{\rm sky, orb}$ and -$y_{\rm sky, orb}$ for a given orbit (black dotted lines in Fig. \ref{fig:fitexample}). We then extract their relative flux contribution by imposing that the integral under the sum of the two Gaussians (green area in Fig. \ref{fig:fitexample}) from -$\infty$ to $y_{\rm obs}$ is equal to exactly half of the integral between $\pm\infty$ under the original best-fit Gaussian (blue dashed curve). This reduces to a simple analytical formula for the fraction $f^+$ of flux originating from the +$y_{\rm sky, orb}$ location,
\begin{equation}
f^+=\frac{\rm erf\left(\frac{y_{\rm obs}+y_{\rm sky, orb}}{\sqrt{2}\sigma_{\rm obs}}\right)}{\rm erf\left(\frac{y_{\rm sky, orb}-y_{\rm obs}}{\sqrt{2}\sigma_{\rm obs}}\right)+\rm erf\left(\frac{y_{\rm obs}+y_{\rm sky, orb}}{\sqrt{2}\sigma_{\rm obs}}\right)},
\end{equation}
where erf is the error function, and of course the fraction originating from the -$y_{\rm sky, orb}$ will be $f^-=1-f^+$. We note that due to either noise in the ALMA cubes or a choice of orbit ($i$, dPA) that does not well represent the data, it may happen that at some PV locations $\left(x_{\rm sky},v_{\rm rad}\right)$ $y_{\rm obs}$ does not fall in between $\pm y_{\rm sky, orb}$. In such cases, we assign the entirety of the flux to the nearest of either the $+y_{\rm sky, orb}$ or $-y_{\rm sky, orb}$ locations.

\begin{figure}
 \hspace{-7mm}
  \includegraphics*[scale=0.3]{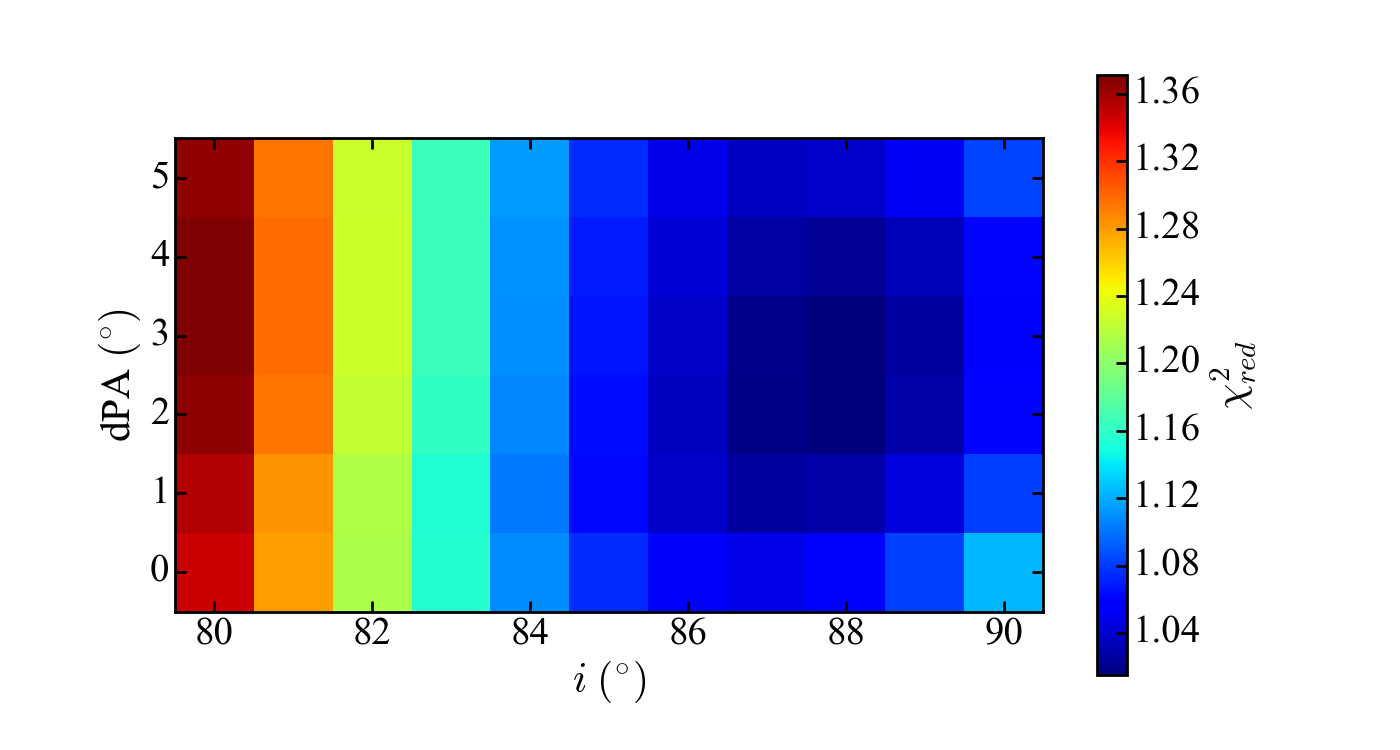}
\vspace{-7mm}
\caption{Reduced $\chi^2$ map obtained by fitting the double-Gaussian vertical profile model to the CO J=2-1 ALMA data cube (as shown by the green long-dashed and red solid lines in Fig. \ref{fig:fitexample} for a specific sky-projected midplane location and radial velocity). The two free parameters for the orbit are the disk inclination $i$ and on-sky tilt angle dPA.}
\vspace{-5mm}
\label{fig:chisqmap}
\end{figure}

By finding the relative contributions of the two Gaussians, we are basically creating a double-Gaussian model (green long-dashed line in Fig. \ref{fig:fitexample}) that we can then compare to the data (red solid line), for each specific PV location. The free parameters of this model orbit are the disk inclination $i$ and the on-sky tilt angle dPA. To attempt constraining these parameters, we therefore construct this model for a grid of $i$ and dPA and in each case evaluate how well it fits the observed data cube. We do this by simply using the $\chi^2$ statistic, showing the resulting reduced $\chi^2$ map in Fig. \ref{fig:chisqmap}. This shows that for any tilt angle dPA our simple model prefers disk inclinations $\geq86^{\circ}$, with a formal best-fit at $i=88^{\circ}$ and dPA $=3^{\circ}$. The results are discussed in more detail in Sect. \ref{sect: structnonedgeon}.


\bsp	
\label{lastpage}
\end{document}